\documentclass[prb,aps,superscriptaddress,floatfix,tightenlines,showpacs,twocolumn]{revtex4-2}
\usepackage{graphicx}
\usepackage{dcolumn}   % needed for some tables
\usepackage{bm}        % for math
\usepackage{amssymb}   % for math
\usepackage{amsmath}
\usepackage{url}
\usepackage{layout}
\usepackage{color}
\usepackage{epsfig}
\usepackage{graphicx}
\usepackage{booktabs}
\usepackage{float}
\usepackage[hyperindex,breaklinks]{hyperref}
\begin{document}

\title{Renormalization group analysis of Dirac fermions with random mass}

\author{Zhiming Pan}
\affiliation{International Center for Quantum Materials, School of Physics, Peking University, Beijing 100871, China}
%\affiliation{Collaborative Innovation Center of Quantum Matter, Beijing 100871, China}%
\author{Tong Wang}
\affiliation{International Center for Quantum Materials, School of Physics, Peking University, Beijing 100871, China}
%\affiliation{Collaborative Innovation Center of Quantum Matter, Beijing 100871, China}%
%
\author{Tomi Ohtsuki}
\affiliation{Physics Division, Sophia University, Chiyoda-ku, Tokyo 102-8554, Japan}%
\author{Ryuichi Shindou}%
\email{rshindou@pku.edu.cn}
\affiliation{International Center for Quantum Materials, School of Physics, Peking University, Beijing 100871, China}%

\date{\today}

\begin{abstract}
Two-dimensional (2D) disordered superconductor (SC) in class D exhibits a 
disorder-induced quantum multicritical phenomenon among diffusive thermal 
metal (DTM), topological superconductor (TS), and conventional localized (AI) 
phases. To characterize the quantum tricritical point where these three phases 
meet, we carry out a two-loop renormalization group (RG) analysis for 2D Dirac 
fermion with random mass in terms of the $\epsilon$-expansion in the spatial 
dimension $d=2-\epsilon$. In 2D ($\epsilon=0$), the random mass is marginally 
irrelevant around a clean-limit fixed point of the gapless Dirac fermion, 
while there exists an IR unstable fixed point at finite disorder strength
that corresponds to the tricritical point. The critical exponent, dynamical 
exponent, and scaling dimension of the (uniform) mass term are evaluated 
around the tricritical point by the two-loop RG analysis. Using a mapping 
between an effective theory for the 2D random-mass Dirac fermion and 
the (1+1)-dimensional Gross-Neveu model, we further deduce the four-loop 
evaluation of the critical exponent, and the scaling dimension of 
the uniform mass around the tricritical point. Both the two-loop and 
four-loop results suggest that criticalities of a AI-DTM transition 
line as well as TS-DTM transition line are controlled by other 
saddle-point fixed point(s) at finite uniform mass. 
\end{abstract}

%\pacs{}

\maketitle
\section{Introduction}

Low-energy fermionic excitation in $p$-wave superconductors 
with broken time-reversal and spin-rotational 
symmetries~\cite{read2000absence} has a `real-valued' 
character of its creation being identical to its annihilation 
(Majorana fermion). Such exotic excitations appear also as 
low-energy fractionalized magnetic excitations in certain 
quantum spin models~\cite{kitaev2006anyons}, 
which could be realized in Mott insulators with heavy magnetic ions~\cite{jackeli2009mott}.  
Experimental realizations of the Majorana particles acquire 
a lot of recent interests in condensed matter experiments, 
while it was suggested that quenched disorders may play 
crucial role in these  experiments~\cite{he17,kayyalha20,wang18,huang18,lian18,yamada20anderson}. 
Two-dimensional (2D) class-D disordered 
superconductor (SC) models~\cite{altland1997} are canonical models of Majorana 
quasiparticles in the presence of the disorder potentials.
The class-D disordered SC models 
have three fundamental phases: topological superconductor (TS) phases with 
quantized thermal Hall conductance $\kappa_{xy}$ in the unit of $\pi^2 k^2_B T/6h$ ($T$ is the temperature, 
$k_B$ is the Boltzmann constant and $h$ is the 
Plank constant)~\cite{cho1997criticality,senthil2000quasiparticle,bocquet00}, 
a diffusive thermal metal (DTM) phase, and conventional Anderson 
localized (AI) phase with $\kappa_{xy}=0$. Natures of quantum 
phase transitions among these three fundamental 
phases have been under active debate for decades, while a 
number of the numerical studies have been carried 
out on network models and  lattice models~\cite{read2000absence,cho1997criticality,chalker2001thermal,mildenberger2006,mildenberger2007density,evers2008,kagalovsky2008universal,kagalovsky2010critical,wimmer2010,medvedyeva2010,laumann2012,yoshioka2018,lian18,fulga2020}.
A phase diagram of the class-D disordered SC models 
has a close connection with a phase diagram 
of a 2D $\pm J$ random bond Ising model (RBIM). 
In an exact mapping between the RBIM and disordered 
fermion models, the diffusive thermal metal phase is absent  
\cite{read2000absence,gruzberg2001random,chalker2001thermal,mildenberger2007density}. 
Cho and Fisher (CF)~\cite{cho1997criticality} found a 
mapping from the RBIM into a Chalker-Coddington network model 
(disordered fermion models)~\cite{chalker1988percolation}. In the 
phase diagram of the CF network model, all the three 
fundamental phases appear and a TS-DTM transition line,
AI-DTM transition line, and AI-TS transition line meet at a quantum 
tricritical point. Critical nature of the tricritical point 
as well as those of the three transition lines have been veiled in mystery. 
Latest numerical studies of the CF model implies a possibility of an 
additional fixed-point structure along the AI-TS transition line, 
which is seemingly related to a Nishimori point in 
the RBIM~\cite{mildenberger2007density,kagalovsky2008universal,kagalovsky2010critical}.

\begin{figure}[t]
\centering
\includegraphics[width=0.9\linewidth]{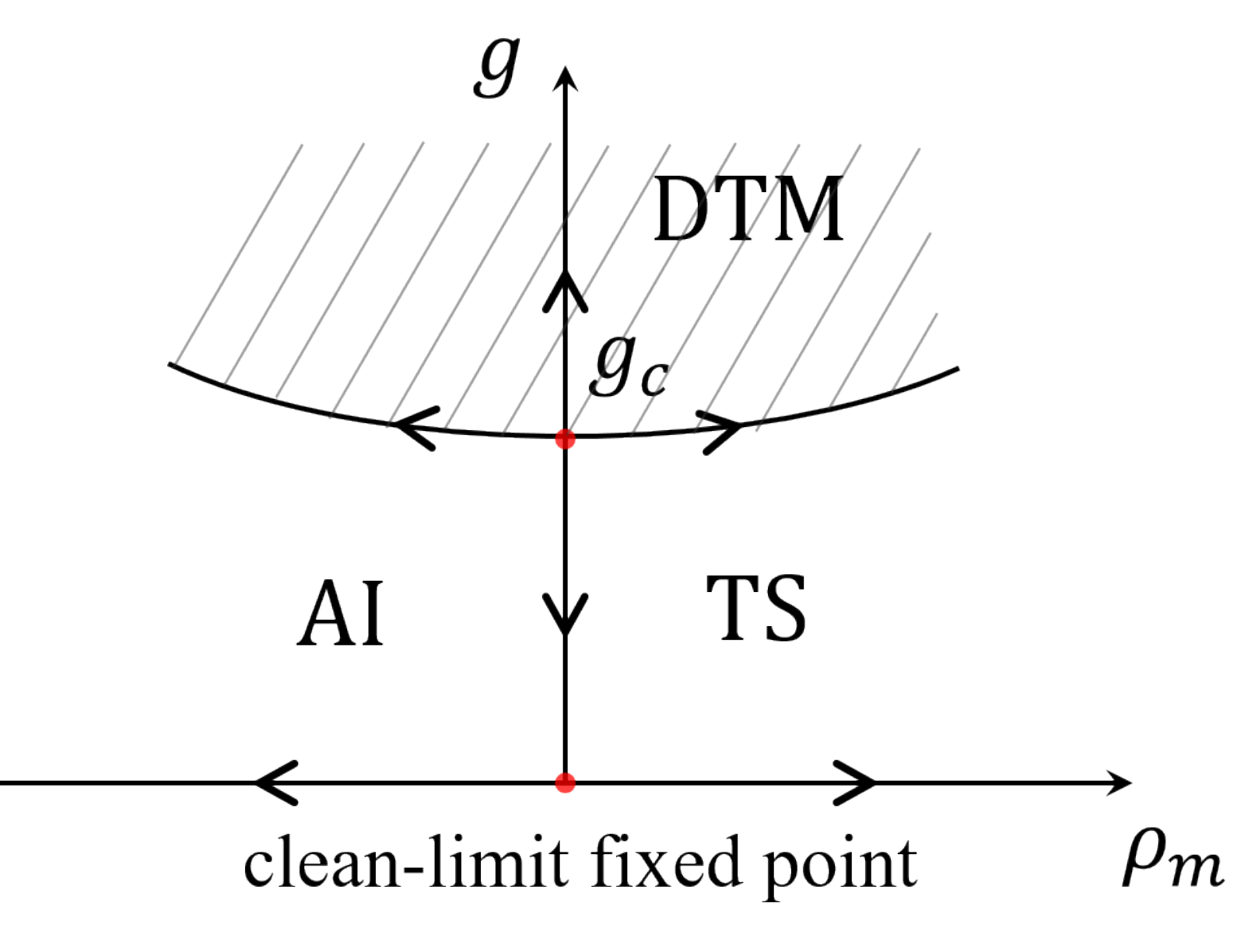}
\caption{Schematic phase diagram for one-component Dirac fermion with 
random Dirac mass. 
DTM is a diffusive thermal metal phase, TS and AI stand a gapped 
phase with the non-zero topological integer ($C=-1$)
and a gapped phase with zero topological 
integer ($C=0$) respectively. 
The topological integers in the gapped phases 
characterize the thermal Hall conductivity as 
$\kappa_{xy}=C \pi^2 k^2_B T/6h$. 
The vertical axis $g$ represents the disorder strength of the 
Dirac-mass type and the horizontal axis represents the uniform 
Dirac mass $m$; $\rho_m$ is a dimensionless renormalized mass. 
A critical point ($g=g_c$) at $\rho_m=0$
is a tricritical point. The arrow stands for the renormalization 
group (RG) flow determined by the two-loop and four-loop 
RG analyses.}
\label{fig:rgflowone}
\end{figure}

In this paper, we clarify universal scaling properties around 
the tricritical point (TCP) and the three phase transition lines
on the basis of the renormalization group (RG) analysis 
of 2D Dirac fermion with random 
Dirac mass. 
The 2D Dirac fermion is an effective theory for a topological phase 
transition between 
two topologically distinct gapped phases in a clean limit, where 
a change of a uniform Dirac mass term $m$ induces the topological 
transition.   
The transition point in the clean limit is described by 
a 2D gapless Dirac fermion. Any on-site 
disorder potentials in the 2D gapless Dirac fermion are marginal 
around the clean-limit fixed point. Particle-hole symmetry 
in the class D symmetry restricts a form of the disorders to 
be of the Dirac-mass type. It is known  
that the Dirac-mass type disorder is marginally irrelevant 
around the clean-limit fixed point~\cite{ludwig1994integer}. 
In this paper, we demonstrate that  an infrared (IR) unstable fixed point 
appears at a finite critical disorder strength, $g=g_c$ (Fig.~\ref{fig:rgflowone}),  
using the two-loop RG analyses and its extension up to the four-loop level. 
%$d=2-\epsilon$~\cite{bondi1990a,schuessler09analytic,roy2014diffusive,roy2016erratum,syzranov2016critical,dudka2016}, we show that 
The IR unstable fixed point corresponds to the TCP, where the 
three phases in the class D systems meet in their phase diagrams. 
From the clean-limit fixed point to the fixed point at $g=g_c$ 
runs a topological phase transition line, 
that intervenes between the two gapped phases; 
AI with $\kappa_{xy}=0$ and TS with $\kappa_{xy} \ne 0$ 
(Fig.~\ref{fig:rgflowone}). We evaluate a scaling dimension 
of the uniform Dirac mass and dynamical exponent around the fixed 
point at $g=g_c$ at the two-loop level, and find that the 
uniform Dirac mass $m$ as well as a deviation of the disorder strength 
from the critical disorder strength, $\delta g \equiv g-g_c$,  
are relevant scaling variables; Fig.~\ref{fig:rgflowone}.  
This determines the renormalization group (RG) flow around the 
TCP as well as the scaling properties of the three 
transition lines. Namely, the criticality of the TS-AI 
transition line is controlled by the clean-limit fixed point, while 
the criticalities of the DTM-TS phase transition and 
the DTM-AI phase transition are controlled by other 
theories with finite uniform mass $m$. % instead of by the TCP. 
Using a mapping between an  
effective model for the random-mass Dirac fermions and 
Gross-Neveu (GN) model together with preceding four-loop RG 
calculation of the GN model, we further discuss the scaling 
dimensions of the uniform mass $m$ and the disorder strength 
$\delta g$ around the TCP~\cite{gracey1991,gracey2016four,choi17question}.
Both two-loop and four-loop results suggest that the 
tricritical point (TCP) is unstable in the IR limit (Fig.~\ref{fig:rgflowone}).   

The organization of this paper is as follows. In the next section, 
we introduce two-dimensional (2D) random-mass Dirac fermions 
as a low-energy theory for Bogoliubov 
excitations in a disordered $p_x+i p_y$ superconductor on a lattice. 
The theory has two controlled parameters; a disorder 
strength of the random-mass type, and the uniform mass that 
induces the topological phase transition between 
two gapped phases with distinct topological numbers. 
In Sec.~\ref{sec:fieldTheory}, we introduce an effective low-energy 
theory for the 2D random-mass Dirac fermions, generalize it in 
general spatial dimension $d$, and discuss the renormalizability 
of the effective theory in $d\le 2$. 
In Sec.~\ref{sec:rg}, we use minimal subtraction method  
in $d=2-\epsilon$ and derive two-loop renormalization group (RG) 
equations for the disorder strength. We obtain an anomalous 
dimension of the uniform mass as well as the dynamical exponent 
up to the two-loop level in Sec.~\ref{sec:rg}. 
In Sec.~\ref{sec:randomMass}, we analyze the RG equation 
and obtain the critical disorder strength for the TCP,
the scaling dimensions of the random mass and 
dynamical exponent around the TCP. In Sec.~\ref{sec:GN}, we discuss a 
relation between the effective theory for the 2D random-mass Dirac 
fermions and (1+1)D SU(N) Gross-Neveu model. The relation gives 
the four-loop evaluations of the critical disorder strength $g_c$ as well as 
scaling dimensions of the uniform mass $m$ and 
disorder strength $\delta g$ around the TCP. 
Sec.~\ref{sec:summary} is devoted to summary and discussion. 

\section{Tight-binding model of spinless $p_x +ip_y$ superconductor}
\label{sec:tbmodel}
A square-lattice model of spinless fermions with 
$p_x +ip_y$ Cooper pairing and the random chemical-potential type 
disorder is considered ~\cite{potter10,bernevig13}, 
\begin{align}
\mathcal{H}/2=& \sum_{\bm{j}} \big( \varepsilon_{\bm{j}} +\mu\big) \hat{c}_{ \bm{j} }^{\dagger} \hat{c}_{\bm{j}} +t \sum_{\bm{j},\mu=x,y} \big[ \hat{c}_{ \bm{j} +\bm{e}_{\mu} }^{\dagger} \hat{c}_{\bm{j}} +\text{h.c.} \big] 	\nonumber\\
+& \Delta \sum_{ \bm{j} } \big[ i \hat{c}_{ \bm{j} +\bm{e}_{x} }^{\dagger} \hat{c}_{\bm{j}}^{\dagger} +\hat{c}_{ \bm{j} +\bm{e}_{y} }^{\dagger} \hat{c}_{\bm{j}}^{\dagger}  +\text{h.c.} \big],	\label{eq:2dTBSC}
\end{align}
with uniform chemical potential $\mu$, 
nearest-neighbor hopping amplitude $t$ and superconducting pairing amplitude 
$\Delta\in\mathbb{R}$. 
Here $\bm{j}\equiv (j_x,j_y)$ denotes 
the lattice vectors, 
and 
$\bm{e}_x$ and $\bm{e}_y$ are primitive unit vectors 
in $x$ and $y$ direction, respectively.
$\varepsilon_{\bm{j}} \in [-W/2,W/2]$ is a short-ranged 
 on-site chemical-potential-type random potential,
\begin{align}
\langle {\varepsilon_{\bm{i}} \rangle_{\text{dis}}} =0,\quad
\langle {\varepsilon_{\bm{i}} \varepsilon_{\bm{j}} \rangle_{\text{dis}}} =\delta_{\bm{i},\bm{j}} W^2/2, 
\end{align}
where $\langle\cdots \rangle_{\text{dis}}$ stands for Gaussian disorder average. $W$ represents 
the disorder strength of 
the random potential.
With a two-components Nambu vector,
\begin{align}
\Psi_{\bm{j}} =\begin{pmatrix}
\hat{c}_{\bm{j}} \\ \hat{c}_{\bm{j}}^{\dagger}
\end{pmatrix} ,\quad 
\Psi_{\bm{j}}^{\dagger} =\begin{pmatrix}
\hat{c}_{\bm{j}}^{\dagger} & \hat{c}_{\bm{j}}
\end{pmatrix},	\label{eq:NambuSpinor}
\end{align}
the Hamiltonian in Eq.~(\ref{eq:2dTBSC}) is written 
as a Bogoliubov-de Gennes (BdG) form,
\begin{align}
& \mathcal{H} 
=\sum_{\bm{j}} 
\Psi_{\bm{j}}^{\dagger}
\big[ \mathbb{H} \big]_{\bm{j},\bm{m}}
\Psi_{\bm{m}} 
= \sum_{\bm{j}} 
\Psi_{\bm{j}}^{\dagger}
\big[ \big( \varepsilon_{\bm{j}} + \mu \big) \sigma_3 \big]
\Psi_{\bm{j}}	\nonumber\\	
& +t \!\ \sum_{\bm{j}} \sum_{\mu=x,y} 
\big[ \Psi_{\bm{j} +\bm{e}_{\mu}}^{\dagger}
\big( \sigma_3 \big)
\Psi_{\bm{j}}+ 
\mathrm{h.c.}
 \big]	\nonumber\\
& +i\Delta \!\ \sum_{\bm{j}}
\big[ \Psi_{\bm{j} +\bm{e}_x}^{\dagger}
\big( \sigma_1 \big)
\Psi_{\bm{j}}
+ 
\Psi_{\bm{j} +\bm{e}_y}^{\dagger}
\big( \sigma_2 \big)
\Psi_{\bm{j}}
+\mathrm{h.c.}
 \big]	\label{eq:2dTBBdG}
\end{align}
with the three Pauli matrices $\sigma_i \, (i=1,2,3)$. 
The BdG Hamiltonian in Eq.~(\ref{eq:2dTBBdG}) satisfies a particle-hole symmetry,
\begin{align}
\mathbb{C} \mathbb{H}^{\rm T}\mathbb{C}^{-1} = -\mathbb{H},	\label{eq:ParticleHoleSym}
\end{align}
with $[\mathbb{C}]_{{\bm j},{\bm m}}=\delta_{{\bm j},{\bm m}} \sigma_1$. The Hamiltonian also breaks the time-reversal symmetry.
The Hamiltonian belongs to the class D in the 10-fold AZ symmetry 
class classification~\cite{altland1997}.

The quasi-particle (Bogoliubov) excitation is a gapped  
excitation, except for $\mu=0,\pm 4t$. 
At $\mu=0, \pm 4 t$, the Bogoliubov excitation forms point nodes at 
high symmetric momentum points. Especially, when $\mu=-4t$, the particle 
and the hole bands form a gapless Dirac-cone dispersion 
around $(k_x,k_y)=(0,0)$, 
\begin{align}
\mathbb{H}(\bm{k}) 
\simeq & \begin{pmatrix}
m -t(k_x^2+k_y^2) & 2\Delta (k_x -ik_y)	\\
2\Delta (k_x +ik_y) & -m +t(k_x^2+k_y^2)
\end{pmatrix}.	\label{eq:BdGBZ0}
\end{align}
A finite $m \equiv \mu+4t$ endows the gapless Dirac fermion 
with a finite gap.
In the gapped phase, the bulk state is 
characterized by a quantized thermal Hall 
conductivity $\kappa_{xy}= C \pi^2 k^2_B T/6h$ 
with $C$ a quantized integer number~\cite{senthil2000quasiparticle}. 
The gapless point $m=0$ separates 
the two gapped phases with the quantized number 
$C=0$ ($m<0$) and $C=-1$ ($m>0$). 
The low-energy effective theory around $m\simeq 0$ and 
around the $\Gamma$ point ($\bm{k}\simeq 0$) is described by a 
Dirac fermion Hamiltonian for a slowly-varying component of the Nambu 
field $\psi({\bm r})$,
\begin{align}
\hat{H}_{0}=& \int d^2\bm{x} \, \psi^{\dagger}(\bm{x}) \big( v \bm{\sigma} \cdot \hat{\bm{p}} +m\sigma_3\big) \psi(\bm{x}),
\end{align}
with $\hat{\bm{p}} \equiv -i\bm{\partial} \equiv (-i\partial_x,-i\partial_y)$, 
$\bm{\sigma} \equiv (\sigma_1,\sigma_2)$,  
a uniform mass term $m$ and a Dirac fermion's  
velocity $v\equiv 2\Delta$. The Dirac fermion Hamiltonian 
generally describes a phase transition between two gapped phases 
with their topological numbers being different from each other by one. 
The gapped phase with $C=0$ stands for a topologically trivial band insulator 
phase or a conventional Anderson insulator (AI) phase that is adiabatically 
connect to the topologically trivial band insulator.  The gapped phase 
with $C=\pm 1$ represents a topological superconductor (TS) phase. 
Whether the positive $m$ corresponds to TS or AI phase depends on 
a global topology of the BdG Hamiltonian in the clean limit; 
the Dirac fermion Hamiltonian can only tell that when the positive 
$m$ corresponds to TS with $C=-1$ (AI) phase, then the negative $m$ 
corresponds to AI (TS with $C=+1$) phase. For simplicity, 
we call the positive (negative) 
$m$ side to be in the TS (AI) phase throughout this paper. 

The chemical-potential type disorder in the tight-binding Hamiltonian Eq.~(\ref{eq:2dTBBdG}) results 
in a Dirac-mass-type disorder potential in the Dirac 
Hamiltonian,  
\begin{align} 
\hat{H}\big[\{V_3({\bm r})\}\big]
= \hat{H}_0+\int d^2\bm{x} V_3(\bm{x})  \psi^{\dagger}(\bm{x}) \sigma_3 \psi(\bm{x}).	\label{eq:2DDiaorderHamil} 
\end{align} 
$V_3(\bm{x})$ is the Dirac-mass-type disorder potential, 
which is short-ranged and obeys the Gaussian distribution 
under a quenched average $\langle \cdots \rangle_{\rm dis}$, i.e.  
$\langle V_3(\bm{r}) \rangle_{\text{dis}} =0$, 
$\langle V_3(\bm{x}) V_3(\bm{x}^{\prime}) \rangle_{\text{dis}} =\Delta_3 \delta^2(\bm{x} -\bm{x}^{\prime})$. 
The disordered Dirac Hamiltonian keeps the particle-hole 
symmetry. %The sign of the uniform mass $m$ separates the 
%two gapped phases with different topological numbers. 
%Around the clean-limit massless point ($\Delta_3=m=0$), 
%the uniform mass is a relevant scaling variable and its 
%scaling dimension is $1$.

\section{Field theory for the random-mass Dirac fermion}
\label{sec:fieldTheory}
Disorder-averaged Green functions for the disordered single-particle Dirac Hamiltonian can be 
systematically treated by a replica method~\cite{altland2010condensed,aharony2018renormalization}.
In the replica method, %we introduce an replicated action that comprises of 
$R$-numbers of the identical free Dirac fermion Hamiltonians 
of $\hat{H}_0$ are replicated together with an elastic-scattering interaction among 
the replicated Dirac fermions,
\begin{align}
& S_{\text{eff}}= \int d\tau\int d^2\bm{x}\, \psi_{\alpha}^{\dagger}(\bm{x},\tau) \big(\partial_{\tau} +v \bm{\sigma} \cdot\hat{\bm{p}} +m\sigma_3 \big) \psi_{\alpha}(\bm{r},\tau)	\nonumber\\
&\ -\frac{\Delta_{3}}{2} \int d\tau d\tau^{\prime}\int d^2\bm{x} \big( \psi_{\alpha}^{\dagger} \sigma_{3} \psi_{\alpha} \big)_{\bm{x},\tau} \big( \psi_{\beta}^{\dagger} \sigma_{3} \psi_{\beta} \big)_{\bm{x},\tau^{\prime}},	\label{eq:2DEffReplica}
\end{align}
with replica indices $\alpha,\beta=1,\cdots R$. Here the summation over 
the replica indices are omitted in Eq.~(\ref{eq:2DEffReplica}), 
and will be omitted in the following unless 
mentioned otherwise. The disordered-averaged connected Green functions 
for the disordered single-particle Hamiltonian are equivalent to 
Green functions for the replicated effective action $S_{\rm eff}$ in a zero-replica limit 
($R\rightarrow 0$); e.g. see the appendix~A. 

Based on the equivalence, 
we will argue renormalizability of $S_{\rm eff}$  around the gapless point ($m=0$) 
and the clean limit ($\Delta_3=0$) in the replica limit $(R\rightarrow 0)$. 
To put it generally, we consider the replica action $S_{\rm eff}$ in 
general spatial dimensions.
The $d$-dimensional action at the gapless point takes a form of,
\begin{align}
S_{\text{eff}}=&S_0+ S_I, \nonumber	\\
S_0
=& \int d^d\bm{x} d\tau \, \psi_{\alpha}^{\dagger}(\bm{x},\tau) \, 
\big( \partial_{\tau}-iv \bm{\gamma} \cdot \bm{\nabla} \big) 
\psi_{\alpha}(\bm{x},\tau) 		\label{eq:DKin}	\\
S_{I}=&-\frac{\Delta_3}{2} \int d^d\bm{x} d\tau d\tau^{\prime} 
\big( \psi_{\alpha}^{\dagger} \hat{\gamma} \psi_{\alpha} \big)_{\bm{x},\tau} 
\big( \psi_{\beta}^{\dagger} \hat{\gamma} \psi_{\beta} \big)_{\bm{x},\tau^{\prime}}	\label{eq:DRepInt}
\end{align}
with $\bm{\nabla}\equiv (\partial_1,\partial_2,\cdots,\partial_d)$, where  
$\bm{\gamma}\equiv (\gamma_i)$ ($i=1,\cdots,d$) is 
a $d$-components vector of matrices that generate a 
$d$-dimensional Clifford algebra satisfying the anti-commutation 
relations $\{ \gamma_i, \gamma_j \}= 2\delta_{ij}$ ($i,j=1,\cdots,d$) 
and $\gamma_0=I$. $\hat{\gamma}$ is a generalization of the 
2D Pauli matrix $\sigma_3$ into the $d$-dimensions, 
with the anti-communication relation 
$\{ \gamma_i, \hat{\gamma} \}= 0$.

\subsection{Renormalizability}
The $d$-dimensional effective field theory 
around the clean-limit fixed point 
has the two spatial dimensions ($d=2$) as its upper critical 
dimension~\cite{ludwig1994integer}.
In a unit of an inverse length (scaling of momentum), dimensions 
of coordinate and derivatives are given by $\dim[x]=\dim[\tau]=-1$ 
and $\dim[\partial_{\tau}] =\dim[\partial_{x}]=1$. To evaluate 
a tree-level scaling dimension of the disorder strength, we take 
the action $S_{\text{eff}}$ to be dimensionless, 
$\dim[S_{\text{eff}}]=0$. In a perturbative renormalization group  
analysis around the clean-limit fixed point, the velocity 
and the coefficient in front of $\partial_{\tau}$ in $S_0$ 
are chosen to be marginal at the tree-level: $\dim[v]=0$. 
Then, a dimension of the field operator is given by 
$\dim[\psi]=d/2$. The disorder strength has a 
dimension $\dim[\Delta_3]=2-d$ from $S_I$ in Eq.~(\ref{eq:DRepInt}).
Thus, the disorder strength is marginal, irrelevant and relevant 
at the tree level in $d=2$, $d>2$ and $d<2$ respectively~\cite{peskin-schroeder}.

The Green functions of the action $S_{\rm eff}$ may have 
ultraviolet (UV) divergences. The UV divergences can 
be renormalizable, 
non-renormalizable and super-renormalizable in $d=2$, $d>2$ and $d<2$, respectively. To explain the UV divergences in the 
Green functions and their renormalizability in general $d$ dimensions, let us put the action $S_{\rm eff} = S_0+S_I$ 
in the momentum-frequency space, 
\begin{align}
S_0
=& \int_{k} \, \psi_{\alpha}^{\dagger}(\bm{k},\omega) \, \big( -i\omega +v \bm{\gamma} \cdot \bm{k} \big) \psi_{\alpha}(\bm{k},\omega),	\label{eq:DKinMoM}\\
S_{I}=&-\frac{\Delta_3}{2} \int_{\omega_1,\omega_2} \int_{\bm{k}_1,\bm{k}_2,\bm{k}_3} \psi_{\alpha}^{\dagger}(\bm{k}_1,\omega_1) \hat{\gamma} \psi_{\alpha}(\bm{k}_3,\omega_1)	\nonumber\\
&\quad \times 
\psi_{\beta}^{\dagger}(\bm{k}_2,\omega_2) \hat{\gamma} \psi_{\beta}(\bm{k}_1+\bm{k}_2-\bm{k}_3,\omega_2) 	\label{eq:DRepIntMom}
\end{align}
with momentum and frequency integrals,
\begin{align}
\int_k \equiv \int_{\bm{k}} \int_{\omega},\quad
\int_{\omega}=\int^{+\infty}_{-\infty} \frac{d\omega}{2\pi},\quad  \int_{\bm{k}}=\int_{|\bm{k}|<\Lambda} \frac{d^d\bm{k}}{(2\pi)^d}. 
\end{align} 
Here $\Lambda$ is a UV momentum cutoff. The interaction in 
Eq.~(\ref{eq:DRepIntMom}) does not exchange energy (frequency) 
or replica index.

Disorder-averaged $2n$-points connected Green functions in the disordered 
single-particle Hamiltonian are identical to $2n$-points Green functions 
for the replica action $S_{\rm eff}$ in the zero replica limit, see    
Eqs.~(\ref{eq:cGreenDis2}) and (\ref{eq:cGreenDis4}) in the Appendix A. 
We thus consider  the renormalization of the $2n$-points Green functions 
of $S_{\rm eff}$ in the limit of $R\rightarrow 0$. 
In the momentum-frequency space, 
they are defined as follows: 
\begin{align}
& (2\pi)^{d+1} {\delta}^{(d)}(\bm{k}- \bm{k}^{\prime}) {\delta}(\omega- \omega^{\prime}) 
G_{\alpha}^{(2)}(\bm{k},\omega)		\nonumber\\
& \equiv 
\frac{1}{Z_{\text{eff}}} \int  {D}\psi_{\gamma}^{\dagger} {D}\psi_{\gamma} \, \psi_{\alpha}(\bm{k},\omega) \psi_{\alpha}^{\dagger}(\bm{k}^{\prime},\omega^{\prime}) \, e^{-S_{\text{eff}}},	\label{eq:2Green} \\
& (2\pi)^{d+2} {\delta}^{(d)}(\bm{k}_1 +\bm{k}_2- \bm{k}_3 - \bm{k}_4) 
{\delta}(\omega_1- \omega_4) 
{\delta}({\omega}_2- {\omega}_3) 	\nonumber\\
&\, \times G^{(4)}_{\alpha\beta}(\bm{k}_1,\bm{k}_2,\bm{k}_3; \omega_1,{\omega}_2) \equiv 
\frac{1}{Z_{\text{eff}}} \int  {D}\psi_{\gamma}^{\dagger} {D}\psi_{\gamma} e^{-S_{\text{eff}}} 	\nonumber\\
& \ \ \ \ 
\psi_{\alpha}(\bm{k}_1,\omega_1) \psi_{\beta}(\bm{k}_2,\omega_2) \psi_{\beta}^{\dagger}(\bm{k}_3,\omega_3) \psi_{\alpha}^{\dagger}(\bm{k}_4,\omega_4),	\label{eq:4Green}
\end{align}
with 
\begin{align}
Z_{\rm eff} \equiv \int D\psi^{\dagger}_{\alpha} D \psi_{\alpha} 
e^{-S_{\rm eff}}.   \label{eq:partition-main} 
\end{align}
According to the standard Dyson-Feynman perturbation theory, 
the Green functions are given by amputated one-particle 
irreducible (1PI) parts of the 
connected Green's functions (vertex functions) ~\cite{peskin-schroeder,altland2010condensed,amit}.
The two-points and four-points vertex functions,   
$\Gamma_{\alpha}^{(2)}$ and $\Gamma_{\alpha\beta}^{(4)}$, are 
related to the respective Green functions, 
\begin{align}
&G_{\alpha}^{(2)}(\bm{k},\omega) \Gamma_{\alpha}^{(2)}(\bm{k},\omega) =1,		\label{eq:2Vertex} \\
& G^{(4)}_{\alpha\beta}(\bm{k}_1,\bm{k}_2,\bm{k}_3; \omega_1,{\omega}_2)
= G_{\alpha}(\bm{k}_1,\omega_1) G_{\beta}(\bm{k}_2,\omega_2) 	\nonumber \\
& \times G_{\beta}(\bm{k}_3,\omega_2)  G_{\alpha}(\bm{k}_1+\bm{k}_2-\bm{k}_3,\omega_1) \Gamma^{(4)}_{\alpha\beta}(\bm{k}_1,\bm{k}_2,\bm{k}_3; \omega_1,{\omega}_2)	\nonumber	\\
&\,\  +G_{\alpha}^{(2)}(\bm{k}_1,\omega_1) G_{\beta}^{(2)}(\bm{k}_2,\omega_2) (2\pi)^d	 \delta^{(d)}(\bm{k}_2-\bm{k}_3). \label{eq:4Vertex}
\end{align}
Similarly, higher-order $2n$-points vertex 
functions can be defined from the $2n$-points connected 
Green functions ~\cite{peskin-schroeder,altland2010condensed,amit}.
Note that those 1PI parts with 
closed internal fermion loops vanish in the limit of the 
zero replica, as every loop gives a factor $R$. Thus, 
frequencies and replica indices of all the internal fermion lines 
in the 1PI parts are fixed by those of external fermion lines. 
The 1PI parts are given only by integrals over the internal 
momenta that depend on the UV momentum cutoff $\Lambda$.

The $2n$-points vertex functions $\Gamma^{(2n)}_{\alpha_1\cdots\alpha_n}$ 
%in Eqs.~(\ref{eq:2Vertex},\ref{eq:4Vertex}) 
could diverge when the UV momentum cutoff $\Lambda$ is taken infinitely large. 
To evaluate how $\Gamma^{(2n)}$ with $n\equiv N_F/2$ 
number of external fermion lines diverges in the limit 
of the large $\Lambda$, let us suppose that an amputated 
Feynman diagram with the $N_F$ external 
fermion points has integrals over $L$-number of internal $d$-dimensional momenta. The integrand is a product 
among $V_F$ number of internal fermion lines and 
$V$ number of the fermion's quartic couplings (vertices). 
From dimensional power counting~\cite{peskin-schroeder,amit}, superficial degree of the UV divergence $M$ of the 
amputated diagram is given by $M=dL-V_F$. 
Each vertex of the quartic coupling connects four fermion 
lines and each fermion line attaches to two vertices or external 
points. Thus, the total number of internal and external fermion 
lines of an unamputated Feynman diagrams is given 
by $V_F+N_F=2V+N_F/2$.  
Each vertex with four fermion lines imposes a momentum conservation 
onto four momenta of the four fermion lines. Besides, the total sum of 
the external momenta is zero. Thus, $L=V_F-(V-1)$. 
Combing them together, we obtain $M$ in terms of $N_F$ and $V$ as follows;
\begin{align}
M=d- \frac{d-1}{2} N_F +(d-2) V.
\end{align}
This shows that the system is renormalizable, non-renormalizable and super-renormalizable at $d=2$, 
$d>2$ and $d<2$, respectively~\cite{peskin-schroeder,amit}.  

At $d=2$, the superficial degree of the UV divergence of the 
amputated 1PI parts only depends on the number of the 
external fermion points, i.e. $M=2- N_F/2$. 
Two-points ($N_F=2$) and four-points ($N_F=4$) vertex functions in 
Eqs.~(\ref{eq:2Vertex}) and (\ref{eq:4Vertex}) have potentially UV divergences 
in the large $\Lambda$ limit, while $\Gamma^{(N_F)}$ ($N_F\geq 6$) has 
no UV divergence.  The dimensional counting shows that 
in the two-points vertex function 
$\Gamma^{(2)}$, the coefficient of $\sigma_3$ has a linear divergence in 
$\Lambda$, while those coefficients of $\omega$ and $\bm{k}$ have the 
logarithmic divergence in $\Lambda$. Note that because of the particle-hole 
symmetry, $\Gamma^{(2)}$ has no $\sigma_0$ term that is linear in $\Lambda$. 
Thus, the most general form of the divergences in the 
two-points vertex is given by, 
\begin{align}
& \Gamma_{\alpha}^{(2)}(\bm{k},\omega) =(\cdots) \cdot \Lambda \sigma_3 + (\cdots) \cdot \ln\Lambda 
\cdot (-i\omega) 	\nonumber \\
&\qquad \quad \qquad +(\cdots) \cdot \ln\Lambda \cdot 
(v \bm{\sigma} \cdot \bm{k}). \label{eqCHap4:vertex2div}
\end{align}
The linear divergence in Eq.~(\ref{eqCHap4:vertex2div}) can be 
absorbed into a shift of the uniform mass, so that 
the theory remains massless. In practice, the $\Lambda$ linear 
term does not appear in the following perturbative renormalization 
calculation (see Sec. IV). The logarithmic UV divergences in the coefficients 
of $\omega$ and $\bm{k}$ can be absorbed into renormalizations of 
field operator amplitude and the single-particle energy (frequency) $\omega$.  
The four-points vertex function is dimensionless and shows the 
logarithmic divergence in $\Lambda$, 
\begin{align}
&\Gamma_{\alpha\beta}^{(4)}(\cdots) = (\cdots) \cdot \ln\Lambda \cdot \sigma_3\otimes  \sigma_3 	\nonumber\\
&+ (\cdots) \cdot \ln\Lambda \cdot \big(\text{other ` $\otimes$ tensor products'} \big)+\mathcal{O}(\omega,k).	\label{eqCHap4:vertex4div}
\end{align}
Under the particle-hole symmetry, the logarithmic UV divergence 
is allowed to take a tensor form of 
$\sigma_3\otimes  \sigma_3$ as well as other tensor forms, 
e.g. $\sigma_0\otimes  \sigma_0$, $\sigma_j\otimes  \sigma_j$ 
with $j=1,2$. The logarithmic divergence with 
$\sigma_3\otimes \sigma_3$ can be absorbed into a 
renormalization of the disorder strength $\Delta_3$. 
When the logarithmic divergence appears in coefficients of the 
other tensor forms, one should also include such tensor 
form of bare interactions into the original 
action $S_{\rm eff}$ to make the theory to be renormalizable. 
In the following two-loop calculation, we will see 
that only the coefficient of $\sigma_3 \otimes \sigma_3$  
has the UV divergence, while the coefficients of the other 
form have no UV divergence. When all the UV divergences  
in the vertex functions are absorbed into the 
renormalizations of field operator amplitude, 
the single-particle energy $\omega$,    
disorder strength $\Delta_3$ and the uniform mass $m$, 
the effective field theory is renormalizable.

\section{Renormalization}
\label{sec:rg}
In the previous section, we introduced 
three kinds of the logarithmic UV divergences 
in vertex functions $\Gamma^{(N_F)}$ ($N_F=2,4$),  Eqs.~(\ref{eqCHap4:vertex2div}) and (\ref{eqCHap4:vertex4div}). In this section, 
we include them into the renormalization of the field operator 
strength $Z_2$ with $\psi\equiv \sqrt{Z_2} \phi$, 
the renormalization of single-particle energy (frequency) 
$\omega$ and renormalization of the effective 
interaction $\Delta_3$. In practice, we use 
the dimensional regularization by putting spatial dimensions into 
$d=2-\epsilon$, where $\ln\Lambda$ in $d=2$ becomes $1/\epsilon$ in 
small $\epsilon$ ~\cite{peskin-schroeder,bondi1990a,schuessler09analytic,roy2014diffusive,syzranov2016critical}. In the following, we will see that 
$\Gamma^{(2)}$ and $\Gamma^{(4)}$ in $d=2-\epsilon$ 
have $1/\epsilon$ divergent terms 
and we shall include them into the renormalizations of the field 
strength, the single-particle energy and the 
interaction strength.

To this end, we use a minimal subtraction method~\cite{peskin-schroeder,syzranov2016critical} and 
separate the action in Eqs.~(\ref{eq:DRepInt}) and (\ref{eq:DKinMoM}) into an effective action 
$S_{E}$ and counterterm part $S_C$,
\begin{align}
& S_{\text{eff}}=S_E+ S_C,	\nonumber\\
& S_E=\int_{\bm{k},\omega} \, \phi_{\alpha}^{\dagger}(\bm{k},\omega) \, \big( -i\Omega +v \bm{\gamma} \cdot \bm{k} \big) \phi_{\alpha}(\bm{k},\omega)	\notag\\
& -\frac{ \Omega^{\epsilon} \kappa }{2} \int d\tau d\tau^{\prime} \int d^d\bm{x} \big( \phi_{\alpha}^{\dagger} \hat{\gamma} \phi_{\alpha} \big)_{\bm{x},\tau} \big( \phi_{\beta}^{\dagger} \hat{\gamma} \phi_{\beta} \big)_{\bm{x},\tau^{\prime}},	
\label{eq:SEeffective}\\
& S_C=\int_{\bm{k},\omega} \, \phi_{\alpha}^{\dagger}(\bm{k},\omega) \, \big( -i\delta\Omega +\delta_2v \bm{\gamma} \cdot \bm{k} \big) \phi_{\alpha}(\bm{k},\omega)	\notag\\
& -\frac{\Omega^{\epsilon} \delta\kappa}{2} \int d\tau d\tau^{\prime} \int d^d\bm{x} \big( \phi_{\alpha}^{\dagger} \hat{\gamma} \phi_{\alpha} \big)_{\bm{x},\tau} \big( \phi_{\beta}^{\dagger} \hat{\gamma} \phi_{\beta} \big)_{\bm{x},\tau^{\prime}}.	
\label{eq:SCcounter}
\end{align}
Here $\phi$ is a renormalized field and it is related with the bare field $\psi$ by a field renormalization $\sqrt{Z_2}$,
\begin{align}
\psi\equiv \sqrt{Z_2} \phi,\quad Z_2\equiv 1+\delta_2, 
\label{eq:Z2}
\end{align}
with a field counterterm $\delta_2$.
$\Omega$ and $\delta \Omega$ are the renormalized 
single-particle energy and its counterterm,
\begin{align}
 \Omega + \delta \Omega = Z_2 \omega. \label{eq:RenorBareCou2}
\end{align}
$\kappa$ and 
$\delta \kappa$ are renormalized {\it dimensionless} interaction 
strength and its counterterm.
Since $\Delta_3$ has a scaling dimension of $2-d$: $\dim[\Delta_3]=\epsilon$,  we normalize $\Delta_3$ by $\Omega^{\epsilon}$ to have dimensionless $\kappa$,
\begin{eqnarray}
\Omega^{\epsilon} \big(\kappa +\delta \kappa\big)  =Z_2^2 \Delta_3.  \label{eq:RenorBareCou}
\end{eqnarray} 
In this renormalization scheme, the renormalized 
single-particle energy $\Omega$ plays a role of 
a renormalization group (RG) scale, where all the 
physical quantities are normalized by a proper power of 
$\Omega$. Taking $\Omega$ to be finite, we can 
also control infrared divergences that could appear in 
momentum integrals for self-energy and 
the vertex function. % (see Eq.~(\ref{eq:vertex-sub})).  
The renormalization of the single-particle energy results in 
an anisotropy in space and time~\cite{aharony2018renormalization}, 
leading to a non-trivial dynamical exponent around 
a non-trivial fixed point [see Eq.~(\ref{eq:z})].

The primary objective of the renormalization is to make the two-points 
and four-points vertex functions of the {\it renormalized} field 
to be free from the UV divergences as functions of the renormalized 
quantities, $\kappa$ and $\Omega$. The vertex functions 
and Green functions of the renormalized field 
(let us call them renormalized vertex and Green functions,
respectively) are defined through the same equations as 
Eqs.~(\ref{eq:2Green}), (\ref{eq:4Green}), (\ref{eq:2Vertex}), and (\ref{eq:4Vertex}) with the same action and partition function 
as $S_{\rm eff}$ and $Z_{\rm eff}$ and 
with the bare fields $\psi$ being replaced by the renormalized 
field $\phi$, e.g. 
\begin{align}
    & \overline{G}_{\alpha}({\bm k},\Omega)
    \overline{\Gamma}_{\alpha}({\bm k},\Omega) = 1, \label{eq:re-2vertex} \\
    & (2\pi)^{d+1} {\delta}^{(d)}(\bm{k}- \bm{k}^{\prime}) {\delta}(\omega- \omega^{\prime}) 
\overline{G}_{\alpha}^{(2)}(\bm{k},\Omega)		\nonumber\\
& \ \ \ \equiv 
\frac{1}{Z_{\text{eff}}} \int  {D}\psi_{\gamma}^{\dagger} {D}\psi_{\gamma} \, 
\phi_{\alpha}(\bm{k},\omega) \phi_{\alpha}^{\dagger}(\bm{k}^{\prime},\omega^{\prime}) \, 
e^{-S_{\text{eff}}}.	\label{eq:re-2Green} 
\end{align}
Here the renormalized frequency $\Omega$ in the 
left hand sides and bare frequency $\omega$ in 
the right hand side of Eq.~(\ref{eq:re-2Green}) 
are related to each other by 
Eq.~(\ref{eq:RenorBareCou2}). The UV divergent 
terms in the vertex functions of the bare field can be then 
absorbed into the counterterms, $\delta_2$, $\delta \kappa$ 
and $\delta \Omega$, in such a way that the renormalized 
vertex functions  
%$\overline{\Gamma}^{(2n)}$ 
have no UV divergence as functions of renormalized 
quantities $\kappa$ and $\Omega$. 

To this end, $\kappa$ in 
Eq.~(\ref{eq:SEeffective}) and 
$S_C$ in Eq.~(\ref{eq:SCcounter}) are treated perturbatively, and two-points and four-points 
renormalized vertex functions are calculated in terms of the standard perturbation theory. 
The $1/\epsilon$ divergent terms and 
the counterterms are set to cancel each other in the renormalized 
vertex functions at every order in the perturbation. 
In the perturbative expansion, the zeroth-order 
renormalized Green function %$\overline{G}_0$ 
is given by the first term in Eq.~(\ref{eq:SEeffective}); 
\begin{align}
\overline{G}_0(\bm{p},\Omega) = \frac{1}{-i\Omega +v\bm{\gamma} \cdot \bm{p}} = \frac{i\Omega +v \bm{\gamma}\cdot \bm{p}} {\Omega^2 +v^2\bm{p}^2 }.	\label{eq:BaiGreen}
\end{align}
The velocity $v$ is free from the renormalization in this RG 
scheme. We henceforth set $v=1$ for simplicity. 

To cancel $1/\epsilon$ 
divergent terms by the counterterms in the two-points 
renormalized vertex function, we require the two-point 
renormalized vertex function 
as a function of $\kappa$ and $\Omega$ 
to be on the order of ${\cal O}(\epsilon^0)$ in the small 
$\epsilon$ limit,
\begin{align}
\overline{\Gamma}^{(2)}_{\alpha}({\bm k},\Omega) 
&\equiv -i(\Omega+\delta\Omega) 
+(1+\delta_2) \bm{\gamma} \cdot \bm{k} 
- \overline{\Sigma}(\bm{k},\Omega) \nonumber \\
&= -i\Omega +  \bm{\gamma} \cdot \bm{k} + 
{\cal O}(1). \label{eq:self-energy-sub}
\end{align}
To cancel $1/\epsilon$ divergent terms by the 
counterterm in the four-points renormalized 
vertex function, 
we require the four-points renormalized vertex function at 
$\Omega_1=\Omega_2=\Omega$ and at ${\bm k}_1={\bm k}_2={\bm k}_3=0$ 
to be finite as a function of $\kappa$ and $\Omega$ in the small 
$\epsilon$ limit,  
\begin{align}
&\overline{\Gamma}^{(4)}_{\alpha\beta}(\bm{k}_1,\bm{k}_2,\bm{k}_3;\Omega_1,\Omega_2)\big|_{\bm{k}_i=0,\Omega_1=\Omega_2=\Omega} \nonumber \\
& \equiv \Omega^{\epsilon} (\kappa+\delta \kappa) \!\ \hat{\gamma} 
\otimes\hat{\gamma} + \overline{\Gamma}^{(4),\prime}_{\alpha\beta}
(\bm{0},\bm{0},\bm{0};\Omega,\Omega) \nonumber \\
& =\Omega^{\epsilon} \kappa \!\ \hat{\gamma} 
\otimes\hat{\gamma} + {\cal O}(1). \label{eq:vertex-sub}
\end{align}
Note that an 
external single-particle energy $\Omega$ is kept 
finite, so that the integrals in the right hand sides 
are free from any infrared divergence associated with 
the momentum integrals. Based on Eqs.~(\ref{eq:self-energy-sub}) and (\ref{eq:vertex-sub}), 
the counterterms, $\delta_2$, 
$\delta\kappa$ and  $\delta\Omega$ in 
Eqs.~(\ref{eq:self-energy-sub}) and (\ref{eq:vertex-sub}),  
are set to   
cancel the divergent contribution (in power of $1/\epsilon$) 
in the self energy $\overline{\Sigma}$ 
and four-points vertex function $\overline{\Gamma}^{(4),\prime}_{\alpha\beta}$ 
order by order in $\kappa$. Being dimensionless, 
$\delta_2$, $\delta\kappa$ and $\delta\Omega/\Omega$ 
thus obtained  
are given as functions only of the dimensionless disorder 
strength $\kappa$ [see, for example, 
Eqs.~(\ref{eq:g-kappa}), (\ref{eq:omegaZogZg}), (\ref{eq:omegaZogZg2}), (\ref{eq:omegaZogZg3}), and (\ref{eq:polynomial})]. 
The bare coupling constant $\Delta_3$ and the bare 
single-particle energy $\omega$ are 
given by $\kappa$ and $\Omega$ through 
Eqs.~(\ref{eq:Z2}), (\ref{eq:RenorBareCou}), and (\ref{eq:RenorBareCou2}).

For simplicity of the following notation, we rescale the 
dimensionless disorder strength $\kappa$ 
by the spherical integral of $d$-dimensional momentum ~\cite{syzranov2015unconventional,syzranov2016critical},
to have another dimensionless disorder strength 
$g$, 
\begin{align}
g=2C_d \kappa,\quad C_d\equiv \int \frac{d S_d}{(2\pi)^d}=\frac{2^{1-d} \pi^{-d/2}}{\Gamma(d/2)}. \label{eq:g-kappa}
\end{align}
We use $g$ instead of $\kappa$ throughout the remaining part of 
the paper.

The vertex functions of $S_{\rm eff}$ are renormalized at 
a finite single-particle energy $\Omega$ through  Eqs.~(\ref{eq:self-energy-sub}) and (\ref{eq:vertex-sub}), where 
high-energy (short-ranged) degrees of freedom in the bare vertex 
functions are renormalized into the counterterms. The 
bare disorder strength $\Delta_3$, bare single-particle 
energy $\omega$ are given as functions of renormalized 
disorder strength $g$ and renormalized  single-particle energy 
(RG scale) $\Omega$, 
 \begin{align}
\Delta_3 &=\Omega^{\epsilon} 
Z^{-2}_{2} (\kappa+\delta \kappa) 
\equiv \frac{\Omega^{\epsilon}}{2C_d} Z_g g, \label{eq:omegaZogZg} \\ 
\omega &= Z^{-1}_2(\Omega + \delta \Omega) \equiv 
Z_{\omega} \Omega,  
%\quad \Gamma^{(2n)} = Z^{-n}_{2} \overline{\Gamma}^{(2n)}
\label{eq:omegaZogZg2} \\
Z_{2} &= 1 + \delta_2. \label{eq:omegaZogZg3}
\end{align}
The divergent contributions in the bare vertex functions 
in the small $\epsilon$ limit are included in the three 
renormalization constants, $Z_{\mu}=Z_2,Z_{\omega},Z_g$ 
($\mu=2,\omega,g$). 
Being dimensionless, these constants must be 
polynomials in the dimensionless disorder 
strength $g$. Namely, they generally take the following 
forms, 
\begin{align}
Z_{\mu}(\epsilon,g)\equiv 1+\delta_{\mu}=  
1+ \sum_{n=1} \frac{a_{n,\mu}(g)}{\epsilon^n},
\label{eq:polynomial}
\end{align}
with $\mu=2,\omega,g$. 
Here $a_{n,\mu}(g)$ is a $m$-th order polynomial of $g$ 
in the $m$-loop perturbative RG calculation.

The renormalized vertex functions in the small $\Omega$ limit 
determine ground-state nature of the action $S_{\rm eff}$~\cite{peskin-schroeder,syzranov2016critical}.
When $\Omega$ 
goes to the zero with a fixed $\Delta_3$, $g$ changes according to Eq.~(\ref{eq:omegaZogZg}). Thereby, 
a limiting value of the renormalized disorder 
strength $g$ in the small $\Omega$ limit
determines the ground state phase diagram of the 
action $S_{\rm eff}$. To determine how 
$g$ changes in the small $\Omega$ limit, let us take an $\Omega$ 
derivative of Eq.~(\ref{eq:omegaZogZg}) with a fixed $\Delta_3$,
\begin{align*}
0 \equiv \Omega \frac{\partial \Delta_3}{\partial \Omega}
&=\Omega^{\epsilon} gZ_g \big(\epsilon +\frac{\Omega}{g} \frac{\partial g}{\partial \Omega}+\frac{\Omega}{Z_g}\frac{\partial Z_g}{\partial \Omega}  \Big) \nonumber \\
&= \Omega^{\epsilon} gZ_g \bigg[\epsilon +\frac{\partial \ln g}{\partial \ln \Omega} \Big( 1 + \frac{d \ln Z_g}{d \ln g}  \Big)\bigg].  
\end{align*} 
The derivative gives out a $\beta$ function of the coupling constant $g$, 
$\beta_{g}$, that tells how $g$ changes in the small $\Omega$ limit,
\begin{align}
\frac{\partial g}{\partial l} =-\beta_g= \epsilon g \big( 1+{g} \frac{d \ln Z_g}{d g} \Big)^{-1},	\label{eq:betag}	
\end{align}
with an RG parameter $l \equiv -\ln\Omega$. The small $\Omega$ limit 
corresponds to the large $l$ limit. $Z_g$ is a polynomial 
function of $g$ [Eq.~(\ref{eq:polynomial})] and so is the $\beta$ function. 
In the next section, we will calculate the $\beta$ function up to the 
two-loop level (the third order in $g$). In Sec.~\ref{sec:GN}, we deduce the 
$\beta$ function up to the four-loop level (the fifth order in $g$) 
using a correspondence between the random Dirac fermion Hamiltonian 
model and an SU(N) Gross-Neveu model.

The renormalized single-particle 
energy $\Omega$ plays a role of the scale parameter in 
this RG scheme~\cite{peskin-schroeder,syzranov2016critical,aharony2018renormalization}.
According to the renormalization 
condition Eq.~(\ref{eq:self-energy-sub}), $\Omega$ has the same scaling as 
the momentum or inverse of the length scale in 
the long wavelength limit. 
%comes from a finite system size 
%through the renormarlization condition Eq.~(\ref{eq:self-energy-sub}). 
%Thereby, the RG scale of a system with a finite linear dimension $L$ can 
%be given by $\ln L$, 
%\begin{align}
% \ln \Omega = - \ln L + {\rm constant}. \label{lislnL}
%\end{align}
When $\Omega$ goes to zero with a fixed $\Delta_3$, the bare 
single-particle energy $\omega$ changes  
according to Eqs.~(\ref{eq:omegaZogZg}) and (\ref{eq:omegaZogZg2}), 
e.g. $\omega$ also goes to zero. According to the 
definitions of the renormalized Green functions, e.g.    
Eq.~(\ref{eq:re-2Green}), the bare 
frequency $\omega$ thus changed is dual to temporal variables in 
the renormalized Green functions. Thus, a scaling 
of the bare frequency with respect to the RG scale 
$\Omega$ under a fixed $\Delta_3$ determines 
how a characteristic 
time is scaled by a characteristic length  
in the infrared (IR) regime. 
To be more specific, we can define the dynamical 
exponent $z$~\cite{syzranov2016critical} 
as a derivative of $\ln \omega$ with respect to $\ln \Omega$ 
with a fixed $\Delta_3$,  
\begin{align}
z \equiv \frac{\partial \ln \omega}{\partial \ln \Omega} = 
1 + \frac{\partial \ln Z_{\omega}}{\partial \ln \Omega} = 
1 + \beta_{g} \frac{d\ln Z_{\omega}}{dg}, \label{eq:z}
\end{align}
where Eq.~(\ref{eq:omegaZogZg2}) was used from the 
second equality in the right hand side. 
Note that $Z_\omega$ is a polynomial function of 
$g$ [Eq.~(\ref{eq:polynomial})] and so is the dynamical exponent. 
In the next section, we will calculate the dynamical 
exponent up to the two-loop level (the second order in $g$).

\subsection{One-loop renormalization}
\begin{figure}[t]
\centering
\includegraphics[width=0.9\linewidth]{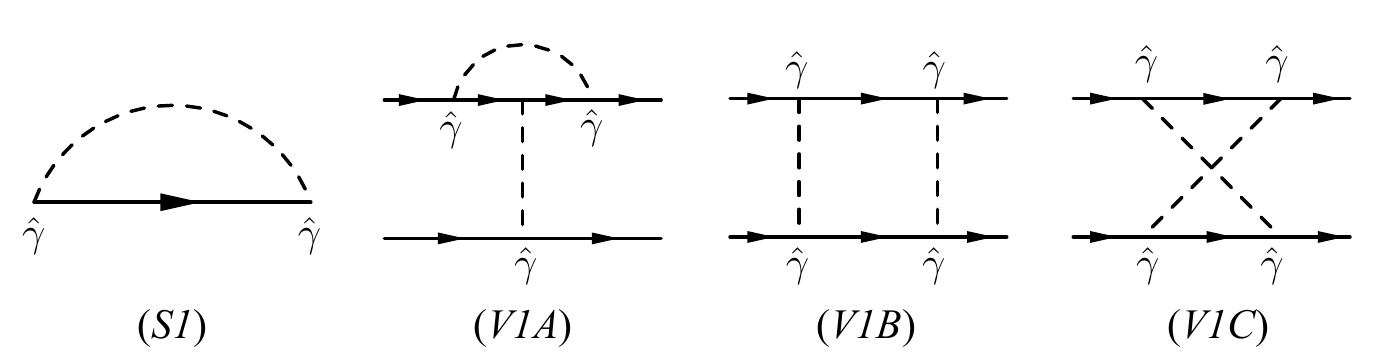}
\caption{One-loop diagrams: ($S1$) self-energy; ($V1$) four-points vertex function. Solid line stands for the zeroth order renormalized Green function given in Eq.~(\ref{eq:BaiGreen}). Dotted line 
stands for the effective interaction with $\kappa$ in $S_E$.}
\label{fig:OneLoop}
\end{figure}

We first calculate the one-loop renormalization.
One-loop diagrams to the right hand sides of  Eqs.~(\ref{eq:self-energy-sub}) and (\ref{eq:vertex-sub}) 
are shown in Fig.~\ref{fig:OneLoop}. 
A diagram $[S1]$ gives the one-loop contribution to 
the self-energy part $\overline{\Sigma}({\bm k},\Omega)$ 
in Eq.~(\ref{eq:self-energy-sub}),
\begin{align}
[S1]= \kappa \Omega^{\epsilon} \hat{\gamma} \int_{\bm{p}} \overline{G}_0(\bm{p}) \hat{\gamma}  
=\big(i\Omega\big) \frac{g}{2\epsilon},	\label{eq:S1}
\end{align}
with the zero-th order renormalized Green function 
$\overline{G}_0({\bm p})$ defined in Eq.~(\ref{eq:BaiGreen}). 
Since the single-particle energies in any internal fermion 
lines in Feynman diagrams are always fixed to be $\Omega$ in the RG conditions, 
Eqs.~(\ref{eq:self-energy-sub}) and (\ref{eq:vertex-sub}),  
we omit the argument $\Omega$ in 
$\overline{G}_0({\bm p},\Omega)$ and simplify it 
by $\overline{G}_0({\bm p})$. 
Eq.~(\ref{eq:S1}) has no linear term in 
$\bm{\gamma}\cdot\bm{p}$. Thus, $\delta_2=0$ at the 
one-loop level. The one-loop contributions to the 
four-points vertex function are shown in the 
three diagrams, $[V1A]$, $[V1B]$ and $[V1C]$. 
The diagram $[V1A]$ takes a tensor 
form of $\hat{\gamma}\otimes \hat{\gamma}$, 
\begin{align}
& [V1A]=\kappa^2 \Omega^{2\epsilon} \int_{\bm{p}}  \overline{G}_0(-\bm{p}) \overline{G}_0(\bm{p}) \hat{\gamma}\otimes \hat{\gamma}	\nonumber\\
=& -\kappa^2 \Omega^{2\epsilon} \int_{ \bm{p}} \frac{1}{ \Omega^2+ p^2} \hat{\gamma}\otimes \hat{\gamma}
= -\frac{\Omega^{\epsilon}}{2C_d} \frac{g^2}{2\epsilon} \hat{\gamma}\otimes \hat{\gamma}.	\label{eq:V1A}
\end{align}
The diagram $[V1B]$ and $[V1C]$ take tensor forms 
of $\gamma_i\otimes \gamma_j$ ($i,j=0,1,\cdots,d$), respectively;
\begin{align*}
[V1B]&
=\kappa^2 \Omega^{2\epsilon} \int_{\bm{p}} \frac{i\Omega -\bm{\gamma}\cdot \bm{p}}{\Omega^2 +p^2} \otimes  \frac{i\Omega +\bm{\gamma}\cdot \bm{p}}{\Omega^2 +p^2},	\\
[V1C]&
= \kappa^2 \Omega^{2\epsilon} \int_{\bm{p}} \frac{i\Omega -\bm{\gamma}\cdot \bm{p}}{\Omega^2 +p^2} \otimes  \frac{i\Omega -\bm{\gamma}\cdot \bm{p}}{\Omega^2 +p^2}.
\end{align*}
However, a sum of these two diagrams is on the order of 
one in the limit of $\epsilon\rightarrow 0$;
\begin{align}
& [V1B]+[V1C]=\kappa^2 \Omega^{2\epsilon} 
\int_{\bm{p}} \frac{i\Omega -\bm{\gamma}\cdot \bm{p}}{\Omega^2 +p^2} \otimes \frac{2i\Omega}{\Omega^2 +p^2}, 	\nonumber\\
& =\kappa^2 \Omega^{2\epsilon} 
\int_{\bm{p}} \frac{-2\Omega^2}{\Omega^2 +p^2} 
\frac{1}{\Omega^2 +p^2} \gamma_0 \otimes \gamma_0 =\mathcal{O}(1).	\label{eq:V1BC}	
\end{align}
Thus, at the one-loop level, no new vertex form other than 
$\hat{\gamma} \otimes \hat{\gamma}$ is generated with the divergence.

The one-loop counterterms in 
Eqs.~(\ref{eq:self-energy-sub}) and (\ref{eq:vertex-sub}) 
should cancel the $1/\epsilon$ divergent terms 
in the self-energy (\ref{eq:S1}) and the vertex function (\ref{eq:V1A}) and (\ref{eq:V1BC}), 
\begin{gather*}
 -i\delta^{(1)} \Omega - [S1]  =\mathcal{O}(1),	\\
\Omega^{\epsilon} \delta^{(1)} \kappa +2[V1A] +[V1B] +[V1C] 
=\mathcal{O}(1),
\end{gather*}
with a proper symmetry factor for $[V1A]$.
We obtain the one-loop counterterms: 
\begin{align}
\delta^{(1)}\Omega= -\Omega \frac{g}{2\epsilon},\quad 
\Omega^{\epsilon} \delta^{(1)}\kappa 
=\frac{\Omega^{\epsilon}}{2C_d} \frac{g^2}{\epsilon}.	\label{eq:1LoopCounter}
\end{align}
Here $\delta^{(m)}\kappa$, $\delta^{(m)}\Omega$, and $\delta^{(m)}_{2}$ 
stand for the $m$-loop contributions to the counterterms, $\delta \kappa$, 
$\delta \Omega$, and $\delta_2$, respectively;
\begin{align}
\delta \Omega \equiv \sum^{\infty}_{m=1} \delta^{(m)}\Omega, \!\ \!\ 
\delta \kappa \equiv \sum^{\infty}_{m=1} \delta^{(m)}\kappa, \!\  \!\ 
\delta_2 \equiv \sum^{\infty}_{m=1} \delta^{(m)}_2. \label{sum-delta-m}
\end{align}

\subsection{Two-loop renormalization}
\begin{figure}[t]
\centering
\includegraphics[width=1.0\linewidth]{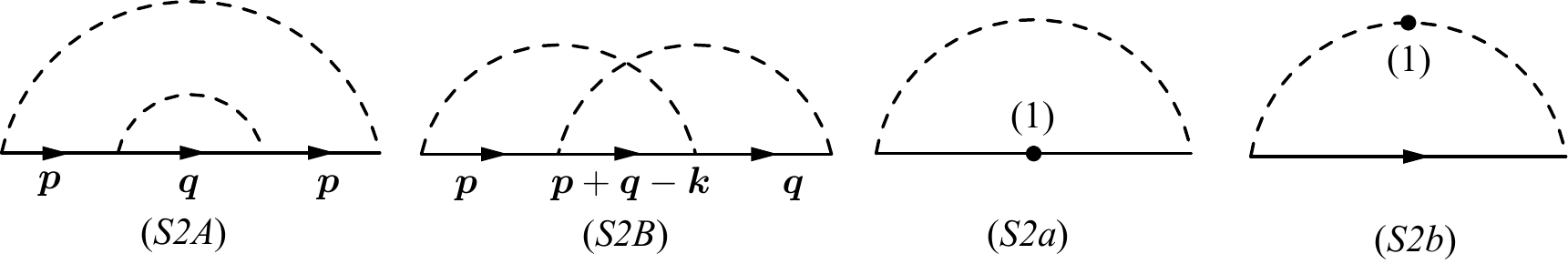}
\caption{Two-loop self-energy diagrams. Black circle on the 
solid line with 
(1) stands for $-i\delta^{(1)}\Omega + \delta^{(1)}_2 v{\bm \gamma}
\cdot {\bm k}$ in $S_C$ where $\delta^{(1)}$ is given by 
Eq.~(\ref{eq:1LoopCounter}) and $\delta^{(1)}_2=0$. Black circle 
on the dotted line with (1) stands for 
$\delta^{(1)}\kappa$ in $S_C$. $\delta^{(1)}\kappa$ is 
given by Eq.~(\ref{eq:1LoopCounter}).}
\label{fig:2loopse}
\end{figure}

We now proceed to the two-loop renormalization. 
The two-loop self-energy diagrams are shown 
in Fig.~\ref{fig:2loopse}. 
Diagrams $[S2A]$, $[S2a]$ and $[S2b]$ are linear in $i\Omega$ 
and they do not have linear terms in $\bm{\gamma}\cdot\bm{k}$. 
Diagram $[S2B]$ has both linear term in $i\Omega$ and 
linear term in  $\bm{\gamma}\cdot\bm{k}$. 
Diagrams $[S2A]$ and $[S2B]$ at $\bm{k}=0$ are given by,
\begin{align}
[S2A] =& \kappa^2 \Omega^{2\epsilon} \int_{\bm{p},\bm{q}} \overline{G}_0(-\bm{p}) \overline{G}_0(\bm{q}) \overline{G}_0(-\bm{p})	\nonumber\\
=& \big( i\Omega \big) \frac{g^2}{4\epsilon^2} \big(1-\epsilon\big) +\mathcal{O}(1),	\label{eq:S2A-1} \\ 
[S2B]({\bm k}=0)=& \kappa^2 \Omega^{2\epsilon} \int_{\bm{p},\bm{q}} \overline{G}_0(-\bm{p}) {G}(\bm{p} + \bm{q}) \overline{G}_0(-\bm{q})	\nonumber\\
= &-(i\Omega) \frac{3g^2}{8\epsilon^2} +\mathcal{O}(1). \label{eq:S2B-1}
\end{align}
The two-loop contributions with the one-loop counterterms are 
shown in $[S2a]$ and $[S2b]$, 
\begin{align}
[S2a]=& \kappa \Omega^{\epsilon} \!\ i\delta^{(1)}\Omega \!\ 
\hat{\gamma} \int_{\bm{p}} \Big( \displaystyle  
\frac{1}{-i\Omega+\bm{\gamma} \cdot \bm{p}} \Big)^2 \hat{\gamma} 	\nonumber\\
=& -\big(i\Omega\big) \frac{g^2}{4\epsilon^2} \big(1-\epsilon\big) +\mathcal{O}(1), 	\\
[S2b]=& \delta^{(1)}\kappa \!\ \Omega^{\epsilon} \hat{\gamma} \int_{\bm{p}} \Big( \displaystyle  \frac{1}{-i\Omega+\bm{\gamma} \cdot \bm{p}} \Big) \hat{\gamma} 		\nonumber\\
=& \big(i\Omega\big) \frac{g^2}{2\epsilon^2}  +\mathcal{O}(1).  
\end{align}
Two-loop contribution to the counterterm of the 
single-particle energy should cancel these $1/\epsilon$ poles in Eq.~(\ref{eq:self-energy-sub}),
\begin{align}
-i\delta^{(2)}\Omega - ([S2A] +[S2B]({\bm k}=0) + [S2a] +[S2b])= \mathcal{O}(1). \label{eq:2-loop-a}
\end{align}
This gives the two-loop contribution to the 
counterterm for the single-particle energy,
\begin{align}
\delta^{(2)}\Omega= -\Omega \frac{g^2}{8\epsilon^2}.	\label{eq:2LCdeltaO}
\end{align}

The two-loop contribution to the field counterterm $\delta_2$ comes from the ${\bm \gamma}\cdot {\bm k}$-linear term in the diagram $[S2B]$. 
The diagram $[S2B]$ for finite ${\bm k}$ is given by 
\begin{align*}
[S2B]=& \kappa^2 \Omega^{2\epsilon} \int_{\bm{p},\bm{q}} \overline{G}_0(-\bm{p}) \overline{G}_0(\bm{p} + \bm{q} -\bm{k}) \overline{G}_0(-\bm{q}). 
\end{align*}
The zero-th order in the small ${\bm k}$ was already 
calculated and included in 
Eqs.~(\ref{eq:S2B-1}) and (\ref{eq:2-loop-a}),
respectively. 
The ${\bm \gamma}\cdot {\bm k}$-linear 
term can be obtained by an 
expansion in ${\bm k}$ of one of the propagators:
\begin{align*}
& \overline{G}_0(\bm{p}+\bm{q}-\bm{k}) =\frac{1}{\overline{G}_0^{-1}(\bm{p}+\bm{q}) 
-\bm{\gamma}\cdot \bm{k}}  	\nonumber \\
&= \overline{G}_0(\bm{p}+\bm{q}) +\overline{G}_0(\bm{p}+\bm{q}) (\bm{\gamma}\cdot \bm{k}) \overline{G}_0(\bm{p}+\bm{q}) + {\cal O}({\bm k}^2). 
\end{align*}
The linear-in-$\bm{k}$ part of $[S2B]$ is
\begin{align}
&[S2B]({\bm \gamma}\cdot\bm{k})	\nonumber \\
=& \kappa^2 \Omega^{2\epsilon} \int_{\bm{p},\bm{q}} \overline{G}_0(-\bm{p}) \overline{G}_0(\bm{p}+\bm{q}) (\bm{\gamma}\cdot \bm{k}) \overline{G}_0(\bm{p}+\bm{q}) \overline{G}_0(-\bm{q})	\nonumber\\
=&\kappa^2 \Omega^{2\epsilon} \frac{1}{4\epsilon} \Big( C_{2-\epsilon} \Omega^{-\epsilon} \Big)^{2} 
\bm{\gamma}
\cdot \bm{k} +\mathcal{O}(1)		\nonumber\\
=&\frac{g^2}{16\epsilon} 
\bm{\gamma}\cdot \bm{k} +\mathcal{O}(1).
\end{align}
The field counterterm 
$\delta_2$ should cancel the $1/\epsilon$-pole from $[S2B](\bm{\gamma}\cdot\bm{k})$ in Eq.~(\ref{eq:self-energy-sub}),  
$\delta_2 \bm{\gamma}\cdot\bm{k} - [S2B](\bm{\gamma}\cdot\bm{k}) 
= {\cal O}(1)$. This gives out the two-loop contribution to 
the field counterterm $\delta_2$ and the field 
renormalization factor $Z_2$ as 
\begin{align}
\delta_2^{(2)}= \frac{g^2}{16\epsilon},\quad 
Z_2=1+\delta_2^{(2)}= 1 +\frac{g^2}{16\epsilon}.	\label{eq:2LFieldR}
\end{align}

The two-loop contributions to the counterterm of the four-points vertex function are calculated 
in the Appendix B.
After the lengthy calculation 
in the Appendix B, 
we obtain the two-loop vertex counterterm as follows,
\begin{align}
\Omega^{\epsilon} \delta^{(2)} \kappa =
\frac{\Omega^{\epsilon}}{2C_d} g \Big( - \frac{g^2}{8\epsilon} +\frac{g^2}{\epsilon^2} \Big).	\label{eq:2Loopdeltak}
\end{align}

\subsection{Two-loop RG equations}
At the two-loop level, the bare single-particle energy $\omega$ 
is given by a sum of the one-loop counterterm 
in Eq.~(\ref{eq:1LoopCounter}) and the two-loop counterterm 
in Eqs.~(\ref{eq:2LCdeltaO}) together with the field renormalization 
factor in Eq.~(\ref{eq:2LFieldR}),
\begin{align*}
\omega=& 
Z_2^{-1} (\Omega +\delta^{(1)}\Omega +\delta^{(2)}\Omega ) 	\nonumber \\
=& \Big( 1- \frac{g}{2\epsilon}  - \frac{g^2}{8\epsilon^2} -\frac{g^2}{16\epsilon} \Big) \Omega.
\end{align*}
The bare interaction is given by a sum of the 
one-loop counterterm in Eq.~(\ref{eq:1LoopCounter}) and 
the two-loop counterterm in Eq.~(\ref{eq:2Loopdeltak}) together 
with the field renormalization factor in Eq.~(\ref{eq:2LFieldR}), 
\begin{align*}
\Delta_3 =& Z_2^{-2} \big(\kappa\Omega^{\epsilon} +\delta^{(1)}\kappa \!\ \Omega^{\epsilon} +\delta^{(2)}\kappa \!\ \Omega^{\epsilon} \big) 	\\
=& \frac{\Omega^{\epsilon}}{2C_{d}} g \Big(1 +\frac{g}{\epsilon} - \frac{g^2}{4\epsilon} +\frac{g^2}{\epsilon^2} \Big).
\end{align*}
Equating these two with the renormalization constants defined in 
Eqs.~(\ref{eq:omegaZogZg}) and (\ref{eq:omegaZogZg2}), 
we obtain the renormalization constants as, 
\begin{align}
Z_{\omega}=& 1- \frac{g}{2\epsilon}  - \frac{g^2}{8\epsilon^2} -\frac{g^2}{16\epsilon},	\label{eq:zomega2loop}\\
Z_g=& 1 +\frac{g}{\epsilon} - \frac{g^2}{4\epsilon} +\frac{g^2}{\epsilon^2}.	\label{eq:zg2loop}
\end{align}
Substituting Eq.~(\ref{eq:zg2loop}) into Eq.~(\ref{eq:betag}) 
and keeping only up to the two-loop order (third order in $g$), 
we finally obtain the $\beta$ function for the renormalized 
dimensionless disorder strength $g$ as,
\begin{align}
\frac{\partial g}{\partial l} =-\beta_g =\epsilon g -g^2 +\frac{1}{2}g^3. \label{eq:2LoopBetag}
\end{align}
Substituting Eq.~(\ref{eq:zomega2loop}) into Eq.~(\ref{eq:z}) 
and keeping only up to the two-loop order (second order in $g$), 
we obtain the dynamical exponent $z$ as:
\begin{eqnarray}
z=  1+ \beta_g \frac{d \ln Z_{\omega}}{dg} =1 + \frac{g}{2}  +\frac{g^2}{8}.	\label{eq:2LoopDyz} 
\end{eqnarray}

\subsection{Scaling dimension of the uniform mass}
\begin{figure}[t]
\centering
\includegraphics[width=1.0\linewidth]{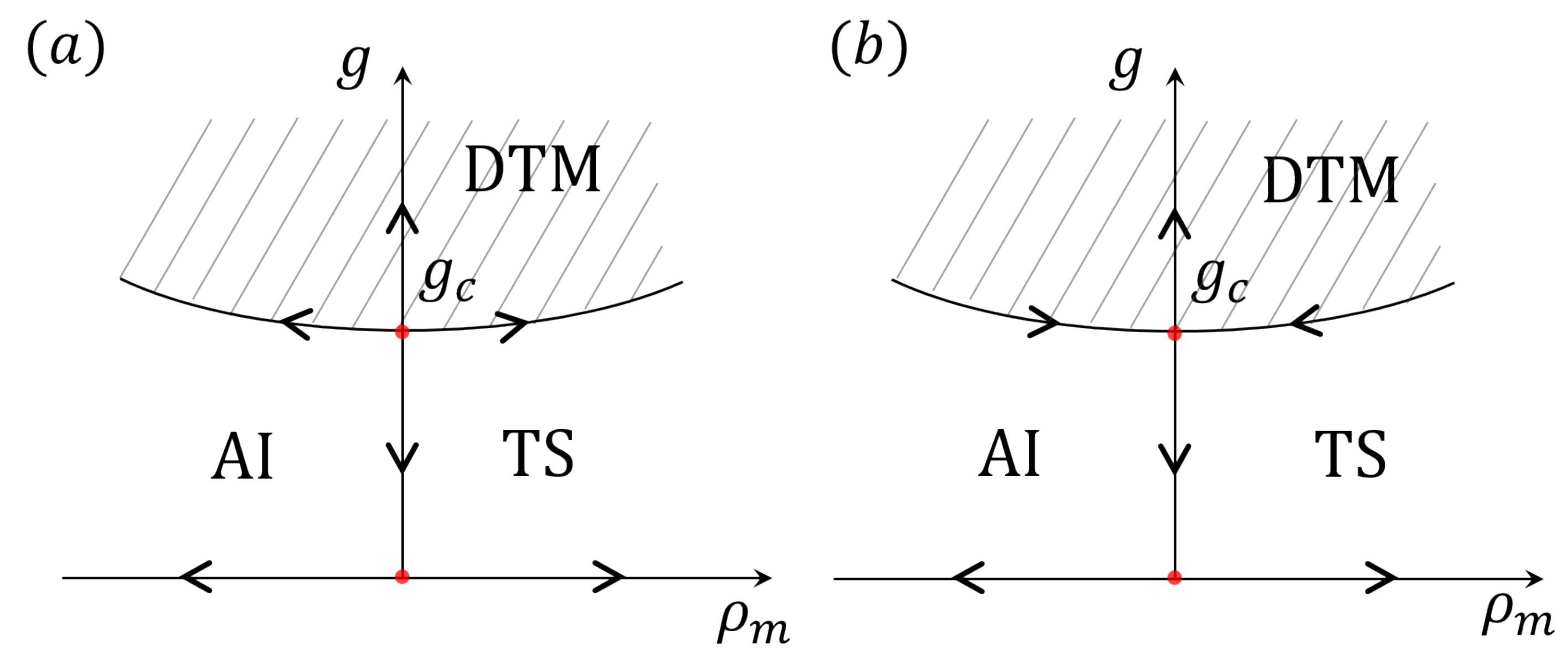}
\caption{Possible RG-flow phase diagram for 2D Dirac fermion with random mass.
(a) uniform mass is a relevant 
scaling variable around the fixed
point at $g=g_c$;
(b) uniform mass is an irrelevant scaling variable around the fixed
point at $g=g_c$.
}
\label{Fig:RGFlowAB}
\end{figure}

In the previous subsection, we derived the $\beta$ function of the 
Dirac-mass type disorder strength in the absence of the uniform 
mass. The $\beta$ function thus obtained is a function 
only of the renormalized disorder strength $g$, which has an 
infrared (IR) unstable fixed point at finite disorder 
strength ($g=g_c$) in 2D ($\epsilon=0$). 
An important question remains; {\it Is the 
uniform mass operator relevant or irrelevant around the fixed 
point at $g=g_c$?} If the uniform mass $m$ is an irrelevant 
scaling variable around the fixed point with the finite critical 
disorder strength, the quantum criticality of DTM-AI transition as 
well as DTM-TS transition are controlled 
by the fixed point at $g=g_c$; see Fig.~\ref{Fig:RGFlowAB}b. 
If the uniform mass $m$ is another relevant scaling variable 
around the fixed point with the finite critical disorder strength 
(Fig.~\ref{Fig:RGFlowAB}a),
 the quantum criticalities of 
these two transition lines are controlled by another 
saddle-point fixed point(s) at finite uniform mass, 
which  may not be captured by $S_{\rm eff}$ studied in 
this paper.  

To clarify the scaling property of the uniform mass, we 
use the same perturbative renormalization theory as in the 
previous section and calculate the scaling dimension of the 
uniform mass operator up to the two-loop order (second order 
in $g$). To this end, we treat the uniform mass operator as
an external perturbation and expand the vertex 
functions in terms of the uniform mass $m$~\cite{peskin-schroeder,schwartz2014quantum,amit,aharony2018renormalization}.
Since the 
uniform mass has a dimension of $1$ at the clean-limit 
fixed point, ${\rm dim}[m]=1$, the $m$-linear term 
of the two-points vertex function has the UV 
logarithmic divergence in $d=2$. Namely, following the 
same line of the argument in Sec.~\ref{sec:fieldTheory}A, we expand the vertex 
functions at finite $m$ in terms of ${\bm k}$, $\omega$, and $m$ in $d=2$,
\begin{align}
\Gamma^{(2)}_{\alpha}({\bm k},\omega,m) 
= &\Gamma^{(2)}_{\alpha}({\bm k},\omega,m=0) \nonumber \\ 
& \ \  + (\cdots) \cdot \ln \Lambda \sigma_3 \cdot m +{\cal O}(m^2). \label{eq:gamma2m} 
\end{align}
Here a $m$-linear term in the four-points vertex function 
as well as the higher-order terms in $m$ in all the vertex functions 
have no UV divergence in the two dimensions. Thus, the bare theory 
in the presence of finite uniform mass have {\it one additional} 
logarithmic divergent term compared to the bare  massless theory. 
The new UV divergent term can be absorbed into a renormalization 
of the mass $m$. We do this mass renormalization by using the same $\epsilon$-expansion as in the previous section, where $\ln \Lambda$ 
in the two dimensions is replaced by $1/\epsilon$ in the 
$d=2-\epsilon$ dimensions. 

The replicated action with the uniform 
mass is given by an addition of the uniform mass term into 
$S_{\rm eff}$ and $Z_{\rm eff}$ in 
Eqs.~(\ref{eq:DKinMoM}), (\ref{eq:DRepInt}), and (\ref{eq:partition-main}),  
\begin{align*}
S_{\rm eff} &= \cdots + 
\int_{{\bm k},\omega} \psi^{\dagger}_{\alpha}({\bm k},\omega) 
m \hat{\gamma} \psi_{\alpha}({\bm k},\omega). \nonumber
\end{align*}
Using the same minimal subtraction method 
as in the previous section, we rewrite the 
bare field in the mass term by the 
renormalized field,
\begin{align}
S_{\rm eff}&= \cdots + \int_{{\bm k},\omega} 
\phi^{\dagger}_{\alpha}({\bm k},\omega) \!\ 
(1+\delta_M) M \!\ \hat{\gamma} \!\ \phi_{\alpha}({\bm k},\omega), \label{SeffwithM}
\end{align} 
where the omitted parts in $S_{\rm eff}$ are already given in Eqs.~(\ref{eq:SEeffective}) and (\ref{eq:SCcounter}). Namely, we put 
the added mass term as 
\begin{align}
 m\psi^{\dagger}_{\alpha} \hat{\gamma} \psi_{\alpha} 
 = Z_{2} Z^{-1}_{\cal O} M \phi^{\dagger}_{\alpha} 
 \hat{\gamma} \phi_{\alpha}. 
\end{align}
Here a renormalized mass $M$ and its counterterm $\delta_M$ are 
related to the bare mass $m$ as;
\begin{align}
     m \equiv Z^{-1}_{\cal O} M, \quad Z_2 Z^{-1}_{\cal O} 
     \equiv 1 + \delta_{M}, 
     \label{mass-renormalization-constant}
\end{align}
with $Z_2$ the field renormalization defined in Eq.~(\ref{eq:Z2}). 
The renormalized vertex functions are given as functions 
of ${\bm k}$, $\Omega$ and the renormalized mass $M$. We determine 
the previous counterterms ($\delta \Omega$, $\delta \kappa$, $\delta_2$) 
including $\delta_M$ in such a way that the following 
RG conditions are satisfied by the renormalized vertex functions;
\begin{align}
    &\overline{\Gamma}^{(2)}_{\alpha}({\bm k},\Omega,M)\Big|_{M=0} = -i\Omega + {\bm \gamma}\cdot {\bm k} + {\cal O}(1), \label{eq:cond1} \\
    &\partial_M \overline{\Gamma}^{(2)}_{\alpha}({\bm k},\Omega,M)\Big|_{{\bm k}=0,M=0} = \hat{\gamma} + {\cal O}(1), \label{eq:cond2} \\ 
    & \overline{\Gamma}^{(4)}_{\alpha\beta}({\bm k}_1,{\bm k}_2,{\bm k}_3:\Omega_1,\Omega_2)\Big|_{{\bm k}_i=0,\Omega_i=\Omega,M=0} \nonumber \\
    & \ \ = \Omega^{\epsilon} \kappa \hat{\gamma} \otimes \hat{\gamma} 
    + {\cal O}(1). \label{eq:cond3} 
\end{align}
Namely, such vertex functions are free from the UV divergence as 
functions of renormalized single-particle energy, renormalized 
uniform mass and renormalized disorder strength, e.g. 
\begin{align}
   \overline{\Gamma}^{(2)}({\bm k},\Omega,M) 
    = - i\Omega + {\bm \gamma}\cdot {\bm k} 
    + M \hat{\gamma} + {\cal O}(1). \label{eq:82} 
\end{align}
The first and the third conditions, Eq.~(\ref{eq:cond1}) and (\ref{eq:cond3}), are 
already satisfied by $\delta_2$, $\delta \kappa$ 
and $\delta \Omega$ determined in the previous section. 
Thus, we have only to determine $\delta_M$ 
together with these counterterms such that 
the second condition Eq.~(\ref{eq:cond2}) is 
satisfied. 

\begin{figure}[t]
\centering
\includegraphics[width=0.9\linewidth]{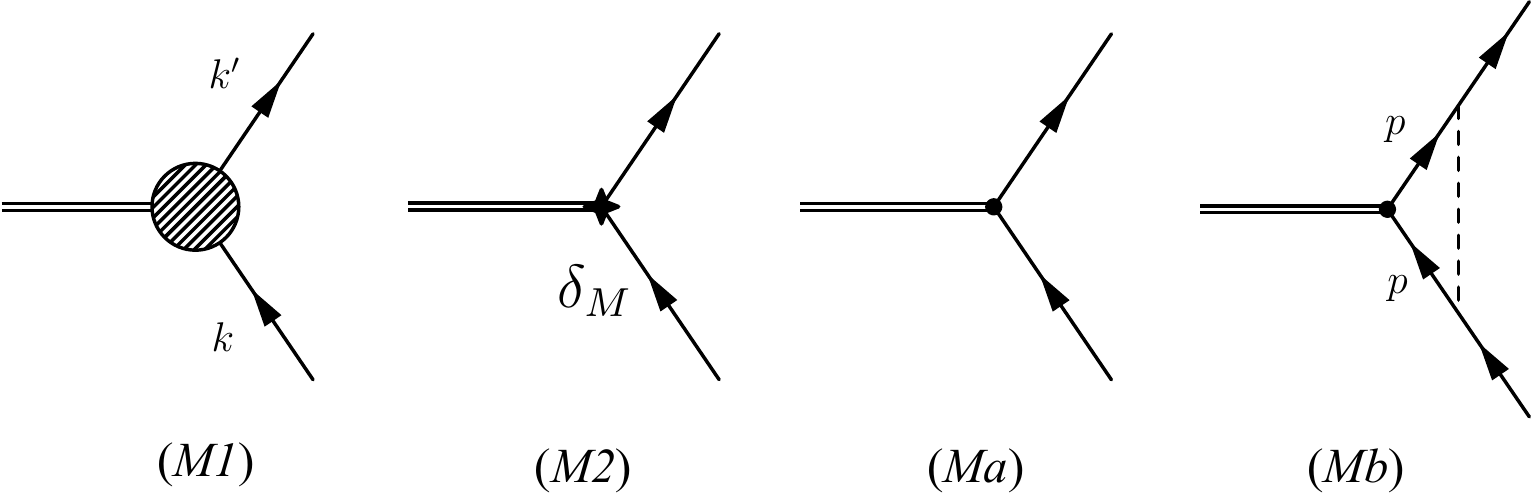}
\caption{(M1) Feynman diagrams of the 1-particle irreducible (1PI) part of $(1+\delta_M)\overline{\Gamma}^{(2,1)}_{\alpha}({\bm k},\Omega)$. ($M2$) The zero-th order in $g$ for $\delta_M\overline{\Gamma}^{(2,1)}_{\alpha}({\bm k},\Omega)$. ($Ma$) The zero-th order in $g$ for $\overline{\Gamma}^{(2,1)}_{\alpha}({\bm k},\Omega)$.
($Mb$) The first order in $g$ for $\overline{\Gamma}^{(2,1)}_{\alpha}({\bm k},\Omega)$.}
\label{Fig:massgreen}
\end{figure}

$\partial_{M}\overline{\Gamma}^{(2)}_{\alpha}({\bm k},\Omega,M)$ evaluated 
at $M=0$ is nothing but an amputated 
one-particle irreducible (1PI) part of a composite 
Green function at the massless point (Fig.~\ref{Fig:massgreen})~\cite{peskin-schroeder,schwartz2014quantum,amit,aharony2018renormalization}.
To see this, let us take the derivative of Eq.~(\ref{eq:re-2vertex}) 
with respect to the renormalized mass,
\begin{align}
&\partial_M \overline{\Gamma}^{(2)}_{\alpha}({\bm k},\Omega,M)\Big|_{{\bm k}=0,M=0} \nonumber \\
&\ \ = -\overline{G}^{-1}_{\alpha} \cdot \partial_M \overline{G}^{(2)}_{\alpha}({\bm k},\Omega,M)\Big|_{{\bm k}=0,M=0} \cdot \overline{G}^{-1}_{\alpha}. \label{eq:derGamma2derM}
\end{align}
From Eqs.~(\ref{eq:re-2Green}) and (\ref{SeffwithM}), 
the derivative of the two-points Green function 
in the right hand side is given by the following composite Green function, 
\begin{align}
   & (2\pi)^{d+1}\delta^{d+1}(k-k^{\prime}) \!\ 
    \partial_M \overline{G}^{(2)}_{\alpha}({\bm k},\Omega,M)\Big|_{M=0} \nonumber \\
    &\ \ = 
     - (1+\delta_M) 
     \int_{k^{\prime\prime}} 
     \langle \phi_{\alpha}(k) 
     \phi^{\dagger}_{\epsilon}(k^{\prime\prime}) \hat{\gamma} 
     \phi_{\epsilon}(k^{\prime\prime})\phi^{\dagger}_{\alpha}(k^{\prime})\rangle_{{\rm eff},c}, \nonumber \\
     & \ \ \equiv - (1+\delta_M) (2\pi)^{d+1}\delta^{d+1}(k-k^{\prime}) 
     \!\  \nonumber \\
    & \hspace{2.cm} \times \overline{G}_{\alpha}({\bm k},\Omega) \overline{\Gamma}^{(2,1)}_{\alpha}({\bm k},\Omega) \overline{G}_{\alpha}({\bm k},\Omega), \label{eq:derG2derM} 
\end{align}
with $\langle \cdots \rangle_{{\rm eff},c} 
\equiv \langle \phi_{\alpha} \phi^{\dagger}_{\epsilon}\hat{\gamma}\phi_{\epsilon} \phi^{\dagger}_{\alpha} \rangle_{\rm eff} - \langle \phi_{\alpha}\phi^{\dagger}_{\alpha} \rangle_{\rm eff} \langle  \phi^{\dagger}_{\epsilon}\hat{\gamma}\phi_{\epsilon}  \rangle_{\rm eff}$, 
$k \equiv ({\bm k},\omega)$, $k^{\prime} \equiv 
({\bm k}^{\prime},\omega^{\prime})$ and $k^{\prime\prime} \equiv 
({\bm k}^{\prime\prime},\omega^{\prime\prime})$. 
Here $\langle \cdots\rangle_{\rm eff}$ in the right hand side 
is taken over $S_{\rm eff}$ with the zero uniform mass, $M=0$. 
The two inverse Green functions in  Eq.~(\ref{eq:derGamma2derM}) amputate one-particle 
reducible parts of the composite Green function;
\begin{align}
    \partial_{M} \Gamma^{(2)}_{\alpha}({\bm k},\Omega,M)\Big|_{M=0} 
    = (1+\delta_M) \overline{\Gamma}^{(2,1)}_{\alpha}({\bm k},\Omega). 
    \label{eq:amputated21}
\end{align}

In the following, the 1PI part of the amputated composite 
Green function $\overline{\Gamma}^{(2,1)}_{\alpha}({\bm k},\Omega)$ 
will be calculated at ${\bm k}=0$ perturbatively in $g$ in $d=2-\epsilon$ 
with the renormalized massless theory $S_{\rm eff}$ of Eqs.~(\ref{eq:SEeffective}) and (\ref{eq:SCcounter}). 
The 1PI part thus obtained contains $1/\epsilon$ 
divergent terms. $\delta_M$ in the right hand 
side of Eq.~(\ref{eq:amputated21}) 
shall be chosen in such a way that all the $1/\epsilon$ 
divergent terms in $\overline{\Gamma}^{(2,1)}_{\alpha}({\bm k}=0,\Omega)$ are cancelled 
by $\delta_M$, satisfying Eq.~(\ref{eq:cond2}).

The zero-th and the first order contributions to the 
right hand side of Eq.~(\ref{eq:amputated21}) are shown in Fig.~\ref{Fig:massgreen}. $[Ma]$ and $[M2]$ 
comprise the zero-th order contribution to the 1PI part of 
$(1+\delta_M)\overline{\Gamma}^{(2,1)}_{\alpha}({\bm k},\Omega)$;
\begin{align}
    [Ma] = \hat{\gamma}, \quad [M2]=\delta_M^{(1)} \hat{\gamma}. 
    \nonumber 
\end{align}
As in Eq.~(\ref{sum-delta-m}), $\delta^{(m)}_{M}$ stands for the $m$-loop contribution to 
the counterterm $\delta_M$; $\delta_{M} \equiv \sum_{m} \delta^{(m)}_M$. 
A one-loop diagram $[Mb]$ is the first 
order contribution to $\overline{\Gamma}^{(2,1)}_{\alpha}({\bm k},\Omega)$; 
\begin{align}
& [Mb]= \kappa\Omega^{\epsilon}  \int_{\bm{p}} \frac{1}{-i\Omega -\bm{\gamma}\cdot\bm{p}} \cdot \frac{1}{-i\Omega +\bm{\gamma}\cdot\bm{p} } \hat{\gamma}	\nonumber\\ 
=&\kappa\Omega^{\epsilon} \cdot \Big(  -\frac{C_{2-\epsilon}}{\epsilon} {\Omega}^{-\epsilon} \Big) \hat{\gamma} +\mathcal{O}(1) 
=-\frac{g}{2\epsilon} \hat{\gamma} +\mathcal{O}(1).	
\end{align}
The one-loop counterterm $\delta_M^{(1)}$ should cancel this $1/\epsilon$ divergence,
\begin{align} 
\delta^{(1)}_{M} =\frac{g}{2\epsilon}, \label{eq:1loop-dM}
\end{align}

\begin{figure}[t]
\centering
\includegraphics[width=0.9\linewidth]{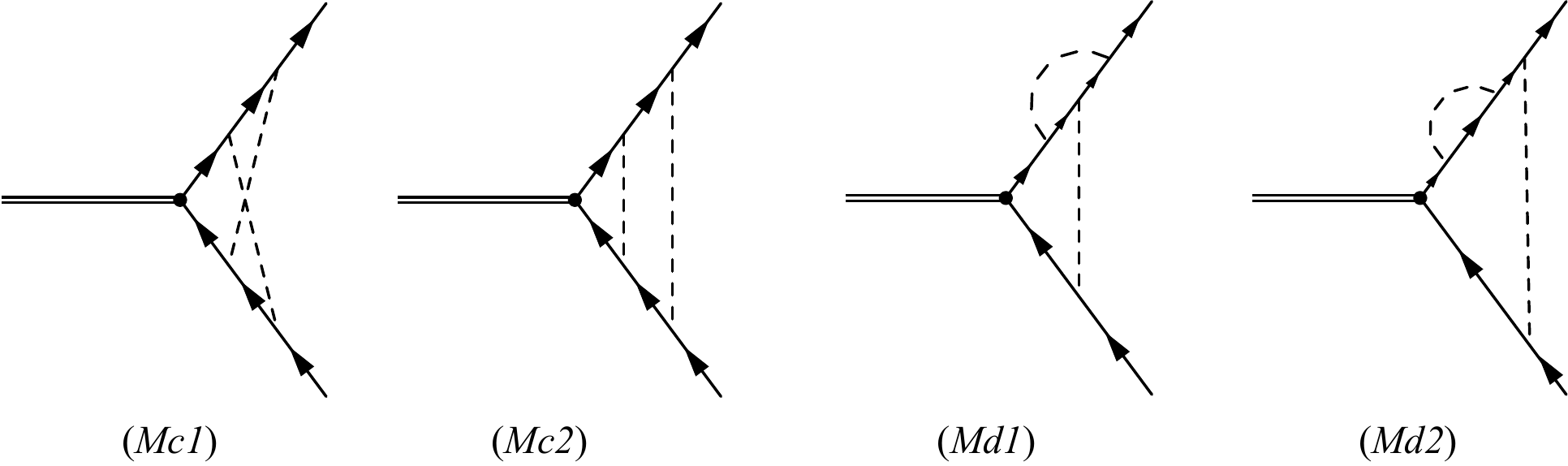}
\caption{Two-loop Feynman diagrams of the 1PI part of $(1+\delta_M)\overline{\Gamma}^{(2,1)}_{\alpha}({\bm k},\Omega)$.}
\label{Fig:massgreen2}
\end{figure}

The two-loop contributions to the 1PI part of 
$(1+\delta_M)\overline{\Gamma}^{(2,1)}_{\alpha}({\bm k},\Omega)$ are shown in Fig.~\ref{Fig:massgreen2} and 
Fig.~\ref{Fig:massgreen3}. All these diagrams are evaluated 
at ${\bm k}=0$, and take a 
form of $\hat{\gamma}$ (see below). We will 
evaluate their coefficients of $\hat{\gamma}$ 
in the following. The coefficient of $\hat{\gamma}$ in 
the diagram $[Mc1]$ is calculated as follows;
\begin{align}
& [Mc1]=\kappa^2 \Omega^{2\epsilon} \int_{ \bm{p}, \bm{q}} \frac{1}{-i\Omega +\bm{\gamma}\cdot\bm{q}} \cdot\frac{1}{-i\Omega -\bm{\gamma}\cdot \bm{p}} \nonumber \\
&\times\frac{1}{-i\Omega +\bm{\gamma}\cdot \bm{p} } \cdot\frac{1}{-i\Omega -(\bm{\gamma}\cdot \bm{p} -\bm{\gamma}\cdot\bm{q} )}	\nonumber\\
=& -\kappa^2 \Omega^{2\epsilon} \int_{ \bm{p}, \bm{q}} \frac{\bm{p}\bm{q} +q^2 -\Omega^2}{ (\Omega^2 +\bm{q}^2) (\Omega^2 +\bm{p}^2) \big(\Omega^2 +(\bm{p} +\bm{q} )^2\big)}	\nonumber\\
=& \kappa^2 \Omega^{2\epsilon} \Big[ \frac{1}{2} \Big( \frac{C_d}{\epsilon} \Omega^{-\epsilon} \Big)^2  - \Big( \frac{C_d}{\epsilon} \Omega^{-\epsilon} \Big)^2  \Big] +\mathcal{O}(1)   \nonumber\\
=&-\frac{1}{2} \frac{g^2}{4\epsilon^2} +\mathcal{O}(1) .
\end{align}
The coefficients of $\hat{\gamma}$ in 
the diagrams $[Mc2]$, $[Md1]$ and $[Md2]$ are calculated as follows;
\begin{align}
[Mc2]=&\kappa^2 \Omega^{2\epsilon} \int_{ \bm{p}, \bm{q}} \frac{1}{-i\Omega +\bm{\gamma}\cdot\bm{q}} 
\cdot\frac{1}{-i\Omega -\bm{\gamma}\cdot \bm{p}} 	\nonumber \\
&  \times\frac{1}{-i\Omega +\bm{\gamma}\cdot \bm{p}  } \cdot\frac{1}{-i\Omega -\bm{\gamma}\cdot\bm{q}} 	\nonumber\\
=& \kappa^2 \Omega^{2\epsilon} \Big( \frac{C_{2-\epsilon}}{\epsilon} {\Omega}^{-\epsilon} +\mathcal{O}(\epsilon) \Big)^2  
=\frac{g^2}{4\epsilon^2} +\mathcal{O}(1),  \\ 
%\end{align}
%The coefficient of $\hat{\gamma}$ in the diagram $[Md1]$ is calculated as
%Eq.~(\ref{eq:diaV2C}),
%\begin{align}
[Md1]%=& \kappa^2 \mu^{2\epsilon} \int_{\bm{p},\bm{q}} \overline{G}_0(-\bm{p}) \overline{G}_0(\bm{p}) \overline{G}_0(-\bm{q}) \overline{G}_0(\bm{q}-\bm{p})	\nonumber\\
=& \kappa^2 \Omega^{2\epsilon} \int_{\bm{p},\bm{q}}  \overline{G}_0(\bm{q}) \overline{G}_0(-\bm{q}-\bm{p}) \overline{G}_0(\bm{p}) \overline{G}_0(-\bm{p})\nonumber \\ 
=&\kappa^2 \Omega^{2\epsilon} \frac{1}{2} \Big( \frac{C_d}{\epsilon} \Omega^{-\epsilon} \Big)^2 +\mathcal{O}(1)
=\frac{g^2}{8\epsilon^2} +\mathcal{O}(1) , \\ 
%\end{align}
%The coefficient of $\hat{\gamma}$ in 
%the diagram $[Md2]$ is calculated as follows, 
%similar to the diagram $[V2D]$ Eq.~(\ref{eq:diaV2D}),
%\begin{align}
[Md2]=& \kappa^2 \Omega^{2\epsilon} \int_{\bm{p},\bm{q}} \overline{G}_0(\bm{p}) \overline{G}_0(-\bm{q}) \overline{G}_0(\bm{p}) \overline{G}_0(-\bm{p})  \nonumber\\
=&\frac{g^2}{8\epsilon} +\mathcal{O}(1).
\end{align}

The two-loop contributions to $(1+\delta_M) \overline{\Gamma}^{(2,1)}_{\alpha}({\bm k},\Omega)$ 
that contain one-loop counterterms are shown in Fig.~\ref{Fig:massgreen3}. The diagram $[Me]$ 
contains $\delta \kappa$'s one-loop counterterm,  
\begin{align}
& [Me]=\Omega^{\epsilon} \delta^{(1)}\kappa 
\int_{\bm{p}} \frac{1}{-i\Omega -\bm{\gamma}\cdot\bm{p}} \cdot \frac{1}{-i\Omega +\bm{\gamma}\cdot\bm{p} } 	\nonumber \\
=&\frac{\Omega^{\epsilon}}{2C_d} \frac{g^2}{\epsilon} \Big( -\frac{C_{2-\epsilon}}{\epsilon} {\Omega}^{-\epsilon} +\mathcal{O}(\epsilon) \Big) 
=-\frac{g^2}{2\epsilon^2} +\mathcal{O}(1).
\end{align}
The diagrams $[Mf1]$ and $[Mf2]$ are the $\kappa$'s first-order terms that 
contain the $\delta \Omega$'s one-loop counterterm. 
A sum of these two is calculated as follows,
\begin{align}
& [Mf1]+[Mf2]
=2\kappa \Omega^{\epsilon} \!\ i\delta^{(1)}\Omega \!\  \int_{\bm{p}} \overline{G}_0(\bm{p}) \overline{G}_0(\bm{p}) \overline{G}_0(-\bm{p}) 	\nonumber\\
=& 2\kappa \Omega^{\epsilon} \big(-i\Omega\big)^2 \frac{g}{2\epsilon}  \Big( \frac{1}{2} C_{2-\epsilon} \Omega^{-2-\epsilon} +\mathcal{O}(\epsilon) \Big)    \nonumber\\
=&- \frac{g^2}{4\epsilon} +\mathcal{O}(1).
\end{align}
The diagram $[Mg]$ is the $\kappa$'s first-order contribution to  
$\delta_M\overline{\Gamma}^{(2,1)}_{\alpha}
({\bm  k},\Omega)$. Up to the second order in $g$, we use 
Eq.~(\ref{eq:1loop-dM}) as $\delta_M$, 
\begin{align}
[Mg]
=&\delta^{(1)}_{M} \cdot \kappa \Omega^{\epsilon}  \int_{\bm{p}} \frac{1}{-i\Omega -\bm{\gamma}\cdot\bm{p}} \cdot \frac{1}{-i\Omega +\bm{\gamma}\cdot \bm{p}} 	\nonumber \\
=&\frac{g}{2\epsilon} \cdot \kappa \Omega^{\epsilon}  \int_{\bm{p}} \frac{-1}{\Omega^2 +p^2} 
=-\frac{g^2}{4\epsilon^2} +\mathcal{O}(1).
\end{align}
By taking into account appropriate symmetry factors in each diagram, we let the two-loop 
counterterm $\delta^{(2)}_M$ cancels the $1/\epsilon$ divergent terms in Eq.~(\ref{eq:amputated21}):
\begin{align}
\delta^{(2)}_{M} 
&=-\big([Mc1] +[Mc2] +2[Md1] +2[Md2] 	\nonumber\\
&\quad +[Me] +[Mf1] + [Mf2] +[Mg] \big)
= \frac{3g^2}{8\epsilon^2}. \label{eq:2loop-dM}
\end{align}
Combining Eqs.~(\ref{eq:1loop-dM}) and (\ref{eq:2loop-dM}), 
we finally obtain the mass renormalization 
constant $Z_{\cal O}$ up to the two-loop 
level as follows,
\begin{align}
Z_{\mathcal{O}}^{-1} 
=& Z^{-1}_2 (1+\delta^{(1)}_M + \delta^{(2)}_M) 
= 1+ \frac{g}{2\epsilon}- \frac{g^2}{16\epsilon} + \frac{3g^2}{8\epsilon^2}.	\label{eq:ZO2loop}
\end{align}
\begin{figure}[t]
\centering
\includegraphics[width=1.0\linewidth]{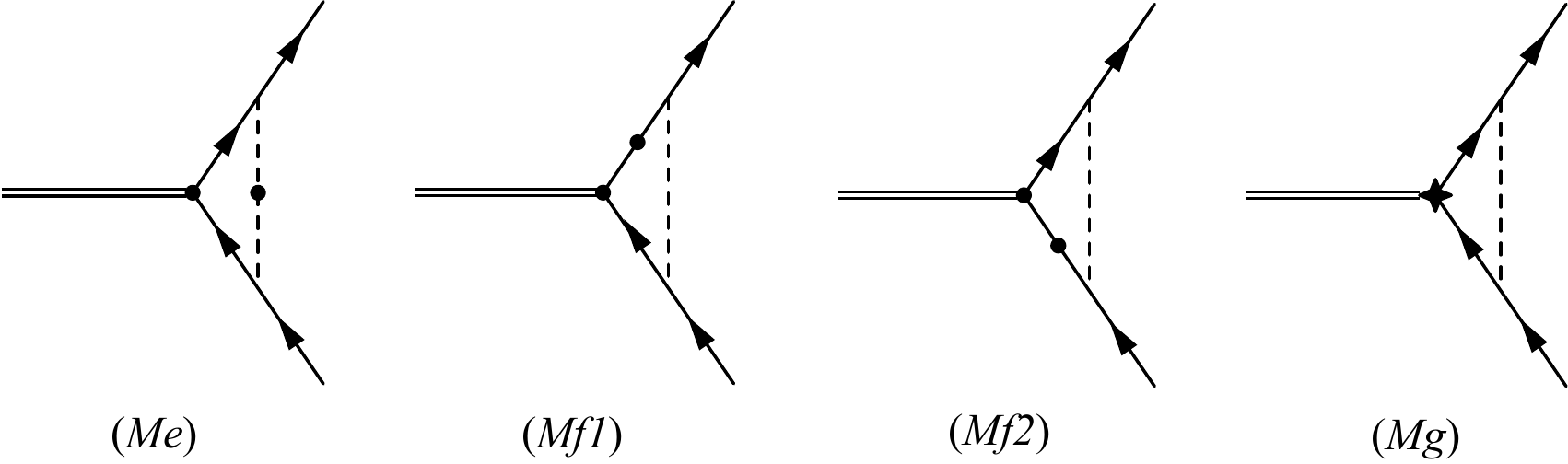}
\caption{Two-loop contributions to the 1PI part that 
contains one-loop counterterms, such as 
$\delta^{(1)} \kappa$, $\delta^{(1)}\Omega$, 
and $\delta^{(1)}_M$. Note that as shown in the previous section, 
$\delta_2$ starts from the second order in $g$, so that it does not contribute to these diagrams.}
\label{Fig:massgreen3}
\end{figure}

The two-points vertex function has been renormalized 
up to the linear order in the uniform mass, Eq.~(\ref{eq:82}). 
The UV divergent terms of the bare two-points 
vertex function are included in the renormalization constants at the finite 
RG scale $\Omega$. The renormalized 
vertex functions in the small $\Omega$ limit determine 
ground-state property of $S_{\rm eff}$ with 
a finite uniform mass. When the RG scale $\Omega$ 
goes to zero with fixed bare uniform mass $m$ and a fixed 
bare disorder strength $\Delta_3$, the renormalized 
mass $M$ changes its value according to Eq.~(\ref{mass-renormalization-constant}).  
The renormalized $M$ thus changed determines the 
ground-state property of $S_{\rm eff}$ with a finite (but small) 
uniform mass (see below). To determine how $M$ changes in the small 
$\Omega$ limit, let us take an $\Omega$ derivative of 
Eq.~(\ref{mass-renormalization-constant}) 
with a fixed $m$ and $\Delta_3$,
\begin{align}
0 \equiv & \frac{\partial \ln m}{\partial \ln \Omega}\Big|_{\Delta_3,m} = 
 \frac{\partial \ln M}{\partial \ln\Omega}\Big|_{\Delta_3,m} -\frac{\partial \ln Z_{\mathcal{O}}}{\partial \ln\Omega}\Big|_{\Delta_3,m}. \label{0=dmdOmega}
\end{align}
This gives an anomalous dimension of the uniform mass, $\gamma_{\cal O}$,
\begin{align}
    \frac{\partial \ln M}{\partial \ln\Omega}\Big|_{\Delta_3,m} 
    &= \frac{\partial g}{\ln \Omega}\Big|_{\Delta_3,m} 
    \frac{d \ln Z_{\mathcal{O}}}{d g} \nonumber \\
    &= \beta_g \frac{d \ln Z_{\mathcal{O}}}{d g} 
    \equiv \gamma_{\cal O}. 
\end{align}
From Eqs.~(\ref{eq:ZO2loop}) and (\ref{eq:2LoopBetag}), we 
obtain the two-loop evaluation of the anomalous dimension of 
the uniform mass:
\begin{align}
\gamma_{\mathcal{O}} \equiv \beta_g \frac{\partial \ln Z_{\mathcal{O}}}{\partial  g} = \frac{g}{2} -\frac{g^2}{8}.
\end{align}
In the clean limit ($g=0$), $\gamma_{\mathcal{O}}=0$, while $\gamma_{\mathcal{O}}$ can be non-zero around a fixed point 
with finite $g$. With an introduction of a renormalized 
dimensionless uniform mass $\rho_m$ as $M \equiv \Omega \rho_m$, 
a scaling property of $\rho_m$ around 
the massless theory is obtained as follows,  
\begin{align}
\frac{\partial \rho_m}{\partial l} 
=(1- \gamma_{\mathcal{O}}) \rho_m \equiv d_m \rho_m, 
\end{align}
with $l \equiv -\ln \Omega$. 
The two-loop scaling dimension of the renormalized dimensionless uniform mass 
is finally obtained around a massless fixed 
point ($\rho_m=0$) as 
\begin{align}
d_m=1- \frac{g}{2} +\frac{g^2}{8}.	\label{eq:dm2Loop}
\end{align}

\section{RG phase diagram of two-dimensional Dirac fermion with random mass}
\label{sec:randomMass}
In the previous section, the two-loop RG equations 
for the disorder strength and uniform mass as well as the 
dynamical exponent 
are evaluated  
perturbatively in the disorder strength around the clean-limit 
fixed point in $d=2-\epsilon$, 
Eqs.~(\ref{eq:2LoopBetag}), (\ref{eq:dm2Loop}), and (\ref{eq:2LoopDyz}). 
In this section, we set $\epsilon=0$, and study a structure of a RG phase 
diagram for the random-mass Dirac fermions in the two dimensions,  
\begin{align}
&\begin{cases}
\displaystyle \frac{\partial g}{\partial l}=-\beta_g= \, 
-g^2 +\frac{g^3}{2}, \phantom{\frac{}{\phantom{\displaystyle \frac{1}{}}}}	\\
\displaystyle \frac{\partial \rho_m}{\partial l}=\,d_m  \rho_m = \Big( 1- \frac{g}{2} +\frac{g^2}{8} \Big) \rho_m,
\end{cases}		\label{eq:2LoopRGE}\\
&\quad z=\, 1 + \frac{g}{2}  +\frac{g^2}{8}.	\label{eq:2LoopDymz}	
\end{align}
Here $g$, $\rho_m$ and $z$ stand for the 
dimensionless random-mass-type disorder strength, the 
dimensionless uniform mass, and the dynamical 
exponent respectively. 

In the low-energy limit ($l\rightarrow +\infty$), the two coupling constants 
flow in the $g$-$\rho_m$ parameter space, forming a 
phase diagram as shown in Fig.~\ref{fig:rgflowone}. 
The RG phase diagram comprises of three phases, 
a diffusive thermal metal (DTM) with larger $g$, a 
topological superconductor (TS) with $g=0$ and $\rho_m>0$, 
and a conventional Anderson insulator (AI) with $g=0$ 
and $\rho_m<0$. The three phases 
meet at an IR unstable fixed point at $(g,\rho_m)=(g_c,0)$, that 
can be regarded as the multicritical (tricritical) point in the preceding 
numerical phase diagram of the CF model ~\cite{cho1997criticality,chalker2001thermal}.

In the massless case ($\rho_m=0$), the fixed point at $g=g_c$ corresponds 
to a semimetal-metal (SM-M) quantum phase transition point. 
For $g<g_c$ and $\rho_m=0$, the finite disorder strength is 
renormalized to zero in the low-energy limit, where the ground state is 
characterized by the clean-limit massless Dirac-fermion 
fixed point at $(g,\rho_m)=(0,0)$ (semimetal phase). 
For $g>g_c$, the disorder 
strength grows up into a larger value. The ground state for $g>g_c$ 
is in a diffusive metal phase, which is presumably 
described by another stable fixed point at larger $g$. 
The criticality of the SM-M quantum phase 
transition point at $\rho_m=0$ 
is controlled by the fixed point at $g=g_c$. Since the weak 
disorder strength ($g<g_c$) is renormalized to zero, 
universality class of a phase transition between AI and TS 
phases is determined by the clean-limit massless Dirac-fermion 
fixed point. According to the mapping between the random-mass 
Dirac fermion model and the random-bond Ising model (RBIM), 
the massless Dirac-fermion fixed point corresponds to the 
clean-limit Ising fixed point in the RBIM ~\cite{cho1997criticality}. 

The renormalized uniform mass is a relevant operator not only around the clean-limit 
massless Dirac-fermion fixed point but also around the fixed point at $g=g_c$. 
Accordingly, universality classes of the phase transition(s) between 
AI and DTM as well as that between TS and DTM must be determined  
by other saddle-point fixed point(s) at finite $\rho_m$. Exploring these 
fixed points at finite $\rho_m$ goes beyond the scope of this paper and 
we leave it for future study. Two-loop evaluations of  
the critical disorder strength, the scaling dimensions 
of the disorder strength and the uniform mass, and the dynamical 
exponent around the unstable fixed point are $g_c=2$, 
${\rm dim}[g-g_c]=2$, ${\rm dim}[\rho_m]=1/2$, and $z(g_c)=5/2$,
respectively.  
 
A correlation-length critical exponent for the SM-M  
quantum phase transition at the massless case is obtained from 
the following relation 
\begin{align}
\frac{\partial\ln (g-g_c)}{\partial l} =\nu^{-1}
\end{align} 
with $\nu(g_c)=1/2$ at the two-loop level. 
This critical exponent  violates the Chayes inequality, that 
dictates $\nu d\geq 2$ with the spatial dimension $d$~\cite{chayes1986finite}. 
To see whether the inequality holds or not at the higher order in $g$, we 
evaluate in the next section the correlation-length critical 
exponent up to the four-loop level, using a relation between 
the effective theory $S_{\rm eff}$ and SU(N) Gross-Neveu (GN) 
model. 

\section{higher-loop results deduced from mappings to other models}
\label{sec:GN}
\subsection{Transformation between random-mass and random-chemical-potential 
in Dirac fermions in two dimensions}
The 2-loop RG equation [Eq.~(\ref{eq:2LoopRGE})]
is consistent 
with the previous studies of random Dirac fermion with 
chemical-potential-type disorder, under a transformation between random mass 
and random chemical potential~\cite{ludwig1994integer,schuessler09analytic}.
To explain this transformation, we 
start from the 2D replicated action for 
the random-mass Dirac fermion Eq.~(\ref{eq:2DEffReplica}),
\begin{align}
S_0=& \int d\tau d^2\bm{x}\, \psi_{\alpha}^{\dagger}\Big\{ \partial_{\tau} -iv(\partial_1\sigma_1+\partial_2\sigma_2) +m\sigma_3 \Big\}\psi_{\alpha},	\nonumber\\
S_1=& -\frac{\Delta_3}{2} \! \int \! d\tau d\tau^{\prime} d^2\bm{x} \big( \psi_{\alpha}^{\dagger} \sigma_3 \psi_{\alpha} \big)_{\bm{x},\tau} \big( \psi_{\beta}^{\dagger} \sigma_3 \psi_{\beta}\big)_{\bm{x},\tau^{\prime}}.	\label{eq:2DEffReplica-a}
\end{align}
In the path integral formulation, 
$\psi^{\dagger}$ and $\psi$ can be considered as independent integral 
variables. Thus, we can define a new set of integral variables 
$\{\bar{\psi},\psi\}$ as ~\cite{dudka2016},
\begin{align}
\psi^{\dagger} \rightarrow \bar{\psi}= -i\psi^{\dagger} \sigma_3,\quad \psi^{\dagger} =i\bar{\psi} \sigma_3
\end{align}
while keeping $\psi$ unchanged. The action of Eq.~(\ref{eq:2DEffReplica-a})  
with $\{\bar{\psi},\psi\}$ is given by, 
\begin{align}
S_0=& \int d\tau d^2\bm{x}\, \bar{\psi}_{\alpha} \Big\{ (i\sigma_3) \partial_{\tau} +iv(\sigma_2 \partial_1- \sigma_1 \partial_2) +im \Big\}\psi_{\alpha},	\nonumber\\
S_1=& +\frac{\Delta_3}{2}  \int d\tau d\tau^{\prime} d^2\bm{x} \, \big( \bar{\psi}_{\alpha} \psi_{\alpha} \big)_{\bm{x},\tau} \big( \bar{\psi}_{\beta} \psi_{\beta}\big)_{\bm{x},\tau^{\prime}}.	\label{eq:2DEffReplica-b}
\end{align}
Under an SU(2) rotation by the 90 degree, 
\begin{align*}
\psi\rightarrow e^{i\frac{\pi}{4}\sigma_3} \psi,\quad \bar{\psi}\rightarrow  \bar{\psi} e^{-i\frac{\pi}{4}\sigma_3},
\end{align*}
the action with $\{\bar{\psi},\psi\}$ can be put into the following form,
\begin{align}
S_0=& \int d\tau d^2\bm{x}\, \bar{\psi}_{\alpha}  \Big\{ (i\sigma_3) \partial_{\tau} -iv(\sigma_1 \partial_1+ \sigma_2 \partial_2) +im \Big\}\psi_{\alpha},	\nonumber\\
S_1=& +\frac{\Delta_3}{2} \int d\tau d\tau^{\prime} d^2\bm{x} \, \big( \bar{\psi}_{\alpha} \psi_{\alpha} \big)_{\bm{x},\tau} \big( \bar{\psi}_{\beta} \psi_{\beta}\big)_{\bm{x},\tau^{\prime}}.	\label{eq:2DEffReplica-c}
\end{align}
This action is related to a replicated effective theory for the random Dirac fermion 
with the chemical-potential-type disorder potential,~\cite{ludwig1994integer,schuessler09analytic,dudka2016}
\begin{align}
S_0^{\prime}=& \int d\tau d^2\bm{x}\, \bar{\psi}_{\alpha}  \Big\{ \partial_{\tau}  -iv(\sigma_1 \partial_1+ \sigma_2 \partial_2) +\tilde{m} \sigma_3 \Big\} \psi_{\alpha},	\nonumber\\
S_1^{\prime}=& -\frac{\Delta_0}{2} \int d\tau d\tau^{\prime} d^2\bm{x} \, \big( \bar{\psi}_{\alpha} \psi_{\alpha} \big)_{\bm{x},\tau} \big( \bar{\psi}_{\beta} \psi_{\beta}\big)_{\bm{x},\tau^{\prime}}. 	\label{eq:2DChemiEffReplica}
\end{align}
Here $\tilde{m}$ and $\Delta_0$ stand for bare uniform mass and chemical-potential-type 
disorder strength. The two theories are mapped to each other, where 
$i\partial_\tau$, $im$ and $\Delta_3$ in Eq.~(\ref{eq:2DEffReplica-c}) correspond to 
$\tilde{m}$ and $\partial_\tau$ and $-\Delta_0$ in Eq.~(\ref{eq:2DChemiEffReplica}), respectively. In the 
latter theory, the preceding work employed the renormalized 
single-particle energy as the RG scale and derive two-loop RG equation 
for the disorder strength $\Delta_0$~\cite{schuessler09analytic,roy2014diffusive,roy2016erratum,syzranov2016critical}. 
One could also use the renormalized uniform mass as 
the RG scale, to derive the same two-loop RG equation. Thanks to 
the correspondence between Eq.~(\ref{eq:2DEffReplica-c}) 
and Eq.~(\ref{eq:2DChemiEffReplica}), such two-loop 
RG equation must be identical to Eq.~(\ref{eq:2LoopBetag}) 
under the sign change of the disorder strength, 
$\Delta_0 \leftrightarrow -\Delta_3$. 
In fact, this is the case up to the two-loop 
level~\cite{schuessler09analytic,roy2014diffusive,roy2016erratum,syzranov2016critical}.

\subsection{Relation to (1+1) dimensional SU(N) Gross-Neveu model} 
%For 2D random mass Dirac fermion (\ref{eq:2DEffReplica-b}),
%\begin{align*}
%S_0=& \int d\tau\int d^2\bm{x}\, \bar{\psi}_{\alpha}(\bm{x},\tau)  \Big\{ (i\sigma_3) \partial_{\tau} -iv(\sigma_1 \partial_1+ \sigma_2 \partial_2) +im \Big\}\psi_{\alpha}(\bm{x},\tau),	\nonumber\\
%S_1=& +\frac{\Delta_3}{2} \int d\tau d\tau^{\prime}\int d^2\bm{x} \, \big( \bar{\psi}_{\alpha} \psi_{\alpha} \big)_{\bm{x},\tau} \big( \bar{\psi}_{\beta} \psi_{\beta}\big)_{\bm{x},\tau^{\prime}},
%\end{align*}
%

The replicated 2D Dirac fermion theory with mass-type disorder  
Eq.~(\ref{eq:2DEffReplica-a}) as well as that with chemical-potential-type 
disorder  Eq.~(\ref{eq:2DChemiEffReplica}) are related to ($d$+1)D 
SU(N) Gross-Neveu (GN) model ~\cite{gross74,roy2016erratum,louvet2016disorder},
\begin{align}
S_{\text{GN}}= \int d\tau d^d{\bm x} \bigg\{
\bar{\psi}_a i\big( \partial_{\mu} \gamma^{\mu} -m\big) \psi_a+ \frac{\bar{g}}{2} \big(\bar{\psi}_a \psi_a \big)^2 \bigg\},	\label{eq:GNModel}
\end{align}
with $a=1,2,\cdots, N$. 
The Dirac matrices $\{\gamma^{\mu}=\gamma_{\mu}\}$ ($\mu=0,1,\cdots,d$) satisfy 
the relation $\{\gamma_{\mu},\gamma_{\nu}\}=2\delta_{\mu\nu}$ in  $(d+1)$ Euclidean 
space time. The conjugate of the Dirac filed is defined as 
$\bar{\psi}=i\psi^{\dagger} \gamma^0$. In the $(1+1)$-dimension, a representation 
of $\gamma^{\mu}$ matrices reduces to the two by two Pauli matrices, 
\begin{align}
\gamma^0= \sigma_1,\quad \gamma^1= \sigma_2,\quad \gamma^0\gamma^1=i\sigma_3.
\end{align}
Note that the quenched disorder in Eq.~(\ref{eq:2DEffReplica-c}) 
does not change the frequency of fermion lines, so that the 
fermion frequencies can be regarded as an external parameter. 
Accordingly, Eq.~(\ref{eq:2DEffReplica-c}) 
in the two-spatial dimension in the zero replica limit ($R\rightarrow 0$) 
and Eq.~(\ref{eq:GNModel}) in the $(1+1)$-dimension in the limit 
of $N\rightarrow 0$ share the same renormalization group theory, 
where $\partial_1$, $\partial_2$, $\Delta_3$, $m$ and $v$ in Eq.~(\ref{eq:2DEffReplica-c}) 
are replaced by $\partial_0$, $\partial_1$, $\bar{g}$, $m$ 
and $1$ in Eq.~(\ref{eq:GNModel}), respectively~\cite{louvet2016disorder}.

\begin{figure}[t]
\centering
\includegraphics[width=0.95\linewidth]{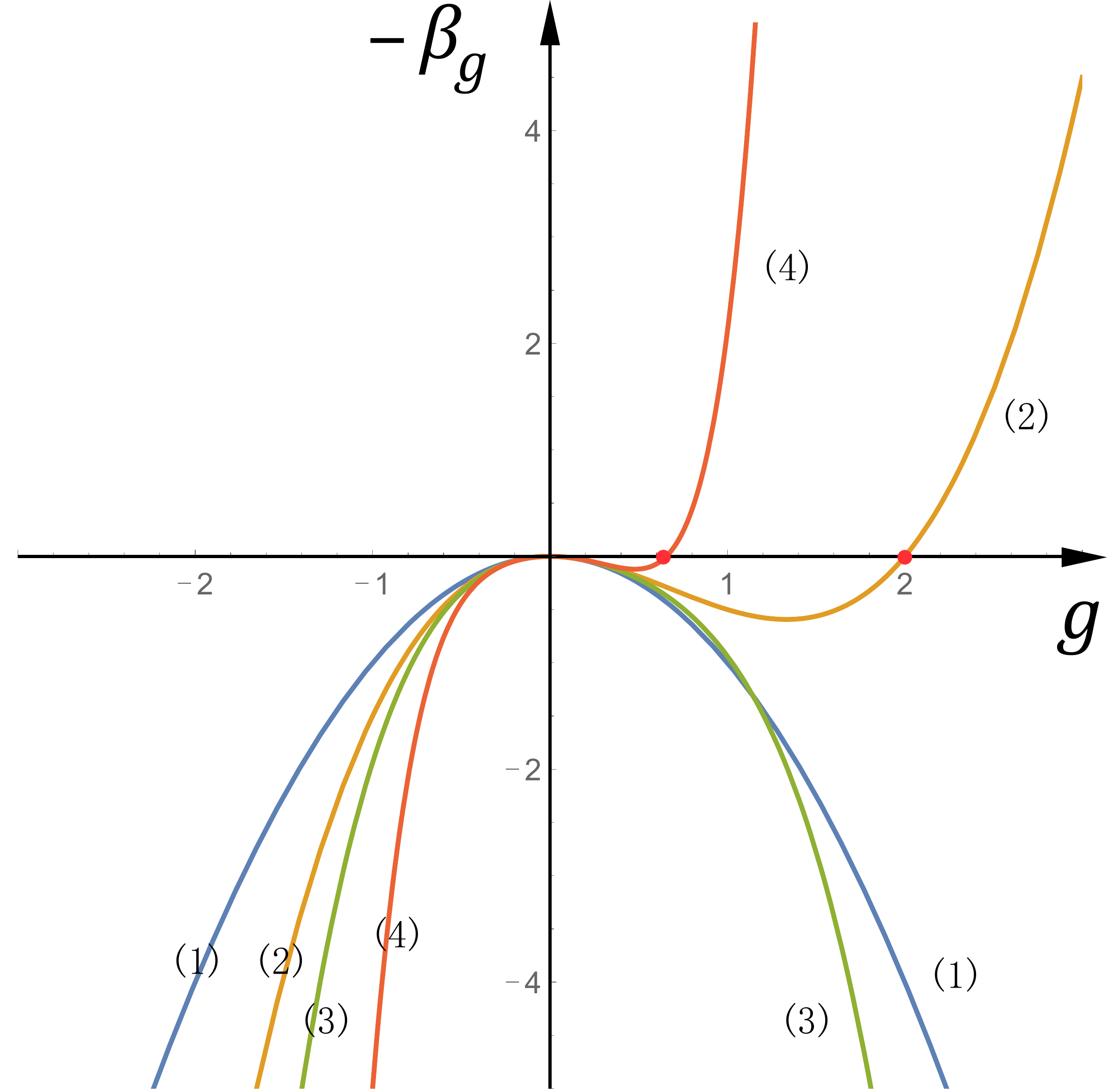}
\caption{The $m$-loop $\beta$ function for coupling constant $g$ 
with $m=1,2,3,4$. The two-loop and four-loop $\beta$ function 
show a zero point at positive values of $g$.} 
\label{fig:rgbetaloop}
\end{figure}
The $\beta$ function
for the GN coupling constant $\beta_g$ 
and the anomalous dimension $\gamma_m$ of the uniform mass $m$ 
have been calculated up to the four-loop 
order ~\cite{gracey1991,gracey2016four}. In the limit of 
$N\rightarrow 0$, they are given by  
\begin{align}
& \beta_{\bar{g}}\big|_{N=0}
=(d-2) \bar{g} +\frac{\bar{g}^2}{\pi} -\frac{\bar{g}^3}{2\pi^2} +\frac{7\bar{g}^4}{16\pi^3} -\frac{17\zeta_3+4}{8\pi^4} \bar{g}^5,	\nonumber\\
& \gamma_m\big|_{N=0}=\frac{\bar{g}}{2\pi} -\frac{ \bar{g}^2}{8\pi^2} +\frac{3\bar{g}^3}{32\pi^3} - \frac{[52 \zeta_3 +19]}{128\pi^4} \bar{g}^4, 	\label{eq:GNfourGamma}
\end{align}
where $\overline{g}$ corresponds to $\Delta_3$ in 
Eq.~(\ref{eq:2DEffReplica-c}). $\zeta_3$ is a 
value of Riemann zeta function $\zeta_3=\zeta(3)$ (Ap\'ery's constant). 
To compare these equations with Eq.~(\ref{eq:2LoopRGE}), we 
normalize $\overline{g} \equiv \Delta_3$ 
by $\pi$ in the two dimensions from Eqs.~(\ref{eq:g-kappa}) and (\ref{eq:RenorBareCou})
\begin{align}
g= 2C_2 \Delta_3 =\frac{\overline{g}}{\pi}. \label{eq:g-gg}
\end{align}
With this normalization, Eq.~(\ref{eq:GNfourGamma}) is translated into  
the following $\beta$ functions for $g$ and the uniform mass 
$\rho_m$ for the 2D random-mass Dirac fermions,
\begin{align}
&\begin{cases}
\displaystyle \frac{\partial g}{\partial l}= \, 
-g^2 +\frac{g^3}{2} -\frac{7}{16} g^4 +\frac{17\zeta_3+4}{8} g^5, \phantom{\frac{}{\phantom{\displaystyle \frac{1}{}}}}	\\
\displaystyle \frac{\partial \rho_m}{\partial l}=\, \Big( 1- \frac{g}{2} +\frac{g^2}{8} 
-\frac{3g^3}{32} + \frac{52 \zeta_3 +19}{128} g^4 \Big) \rho_m. 
\end{cases}	\label{eq:RGfourLoop}
\end{align}

At the two-loop level, Eq.~(\ref{eq:2LoopRGE}) has 
the IR unstable fixed point at $g_c=2$, 
which corresponds to the multicritical 
(tricritical) point. 
At the four-loop level, this critical disorder strength 
moves to a smaller value, $g_c \approx 0.655$. 
Fig.~\ref{fig:rgbetaloop} shows the $\beta$ function of $g$ 
for the $1,2,3,4$-loop order.   

The four-loop evaluation 
of the scaling dimension of $g-g_c$ and the uniform 
mass around the fixed point are 
${\rm dim}[g-g_c]\approx 1.66$
and ${\rm dim}[\rho_m]\approx 0.817$ 
respectively. The critical exponent associated with the SM-M 
quantum phase transition is evaluated as $\nu\approx 0.60$, which 
also violates the Chayes inequality.
%The existence of such infrared fixed point for GN model at finite $N$ is studied with the four-loop result and Pad\'e approximant in Ref~\cite{choi17question}.
%It may be hard to draw a conclusion from these perturbative renormalization. 
%Further non-perturbative method is necessary.

\vspace{0.4cm}

\section{Summary and Discussion}
\label{sec:summary}
In summary, we have performed renormalization 
group analyses for the 2D Dirac fermion with the random-mass 
type disorder. We obtained the two-loop $\beta$ function for 
the disorder strength and observed an infrared unstable 
fixed point at a finite disorder strength. The fixed point 
corresponds to the quantum tricritical point that intervenes 
diffusive metal phase, and two topologically distinct 
gapped phases in 2-dimensional class D models. Two phase 
transition lines between the diffusive metal and the two gapped 
phases and a transition line between the two topologically-distinct 
gapped phases meet at the tricritical point. The dynamical exponent 
and scaling dimension of the uniform mass are also evaluated at 
the tricritical point up to the two-loop level.  

The two-loop evaluations of the scaling dimensions of the 
uniform mass and disorder strength shows that (i) 
the transition line between the two gapped phases is controlled 
by the clean-limit massless-Dirac-fermion fixed point with 
the Ising criticality, (ii) the transition lines between 
the diffusive metal phase and gapped phases are controlled {\it not} 
by the tricritical point {\it but} by another saddle-point fixed point 
with finite uniform Dirac mass. Using a mapping between the effective theory 
for the 2D random-mass Dirac fermion and (1+1)D SU(N) Gross-Neveu model 
in the limit of $N\rightarrow 0$, 
we also obtained the four-loop evaluations of the scaling dimensions 
of the uniform mass and disorder strength around the tricritical point. 
The four-loop result gives the same conclusion as the two-loop 
result about the criticality of the three transition lines. We found 
that the critical exponent $\nu$ for the semimetal-metal(SM-M) 
quantum phase transition breaks the Chayes inequality ($\nu d>1$) both at the 
two-loop level and at the four-loop level, indicating an unusual 
aspect of the disorder-driven SM-M quantum phase transition 
in 2D class D models. 

\noindent
\begin{acknowledgments}
The authors appreciate Ilya Gruzberg for helpful discussions.  Zhiming Pan, Tong Wang 
and Ryuichi Shindou were  
supported by the National Basic Research 
Programs of China (No. 2019YFA0308401) and by National Natural Science 
Foundation of China (No.11674011 and No. 12074008). Tomi Ohtsuki 
was supported  by JSPS KAKENHI Grants 19H00658. 
\end{acknowledgments}

\appendix
\onecolumngrid
\section{disordered Dirac Hamiltonian and replicated Dirac-fermions action}
In the main text, the renormalization of the Green functions of replicated Driac fermions 
is intensively studied. The study leads to the renormalization group equation for the 
2D Dirac fermions with random mass.  In this appendix, we review an equivalence between 
averaged Green functions for disordered Dirac Hamiltonian and Green functions for the 
replicated Dirac fermions action $S_{\rm eff}$.  A partition function for 
the disordered Dirac Hamiltonian is considered,
\begin{align}
Z\big[\{V_3({\bm r})\}\big] 
\equiv \int {D}\psi^{\dagger} {D}\psi\, e^{-S\big[\{V_3({\bm r})\}\big] },	\ \ 
S\big[\{V_3({\bm r})\}\big] 
\equiv \int_{-\beta/2}^{\beta/2} d\tau \,\big( \int d^2\bm{x} \psi^{\dagger} \partial_{\tau} \psi +\hat{H}\big[\{V_3({\bm r})\}\big] \big),	\label{eq:disorderedS}	
\end{align}
where an integral over an imaginary time $\tau$ 
ranges from $-\beta/2$ to $\beta/2$ with an inverse temperature 
$\beta \equiv 1/k_BT$. In the main text, we always take $\beta\rightarrow +\infty$ ($T=0$). 
A time-ordered $2n$-points Green function is given by a 
trace over the action in Eq.~(\ref{eq:disorderedS}),
\begin{align}
\langle \psi(1) \cdots \psi^{\dagger}(2n)\rangle 
=\frac{1}{Z} \int {D}\psi^{\dagger} {D}\psi\, \psi(1) \cdots \psi^{\dagger}(2n) e^{-S},	\label{eq:2nGreenS}
\end{align}
with space-time coordinates $i\equiv (\tau_i,\bm{x}_i)$ 
($i=1,\cdots,2n$). The Green function is averaged over 
different disorder realization through the Gaussian distribution,
\begin{align}
\langle (\cdots) \rangle_{\text{dis}} \equiv  \frac{1}{\mathcal{N}} \int DV_3\, \big(\cdots\big) \!\ 
\exp\big( -\frac{1}{2\Delta_3} \int d^2\bm{x} V_3(\bm{x})^2 \Big).	\label{eq:disAver}
\end{align}
where $\mathcal{N}$ is a normalization factor such that $\langle 1\rangle_{\text{dis}}=1$. 
The averaged Green function is given by,
\begin{align}
\langle\langle \psi(1) \cdots \psi^{\dagger}(2n)\rangle\rangle_{\text{dis}}
=\frac{1}{\mathcal{N}} \int DV_3\, \langle \psi(1) \cdots \psi^{\dagger}(2n)\rangle \, \exp\big( -\frac{1}{2\Delta_3} \int V_3^2 \big).	\label{eq:2nGreenSdis}
\end{align}

To treat the disorder-averaged $2n$-points 
Green functions systematically, we use a 
replica method throughout the paper~\cite{altland2010condensed,aharony2018renormalization}.
In the replica method, we introduce a replicated action that comprises of 
$R$-numbers of the identical free Dirac-fermion Hamiltonians 
of $\hat{H}_0$ and an elastic-scattering interaction between 
the replicated Dirac fermions, $S_{\rm eff}$ in Eq.~(\ref{eq:2DEffReplica}). 
The averaged $2n$-points {\it connected} Green functions are 
given by the replica-limit of $2n$-points Green 
functions for the replicated Dirac-fermion action 
(\ref{eq:2DEffReplica})~\cite{aharony2018renormalization}.
The averaged two-points Green function is given by,
\begin{align}
&\langle \langle \psi(\bm{x},\tau) \psi^{\dagger}(\bm{x}^{\prime},\tau^{\prime})\rangle \rangle_{\text{dis}}	
=\lim_{R\rightarrow 0} 
\langle \psi_{\alpha}(\bm{x},\tau) \psi_{\alpha}^{\dagger}(\bm{x}^{\prime},\tau^{\prime}) \rangle_{\text{eff}},	\label{eq:cGreenDis2}
\end{align}
where the summation over the replica index is {\it not} assumed in the right hand side. 
$\langle \cdots\rangle_{\text{eff}}$ in the right hand side stands for a trace over 
the replicated action with the elastic interaction,
\begin{align}
\langle \cdots \rangle_{\text{eff}} \equiv 
\frac{1}{Z_{\text{eff}}} \int {D}\psi_{\alpha}^{\dagger} {D}\psi_{\alpha} \, \big( \cdots \big)\, e^{-S_{\text{eff}}}, 
\!\ \!\ 
Z_{\text{eff}}\equiv \int  {D}\psi_{\alpha}^{\dagger} {D}\psi_{\alpha} \, e^{-S_{\text{eff}}}. \label{eq:partition}
\end{align}
The averaged 4-points connected Green function is 
given by the four-points Green function of the 
replicated action in the replica limit,
\begin{align}
\langle \langle \psi(1) \psi(2) \psi^{\dagger}(2^{\prime}) \psi^{\dagger}(1^{\prime}) \rangle_c \rangle_{\text{dis}} 
&= \lim_{R\rightarrow 0} \frac{1}{R} \sum_{\alpha_1,\alpha_2,\beta_2,\beta_1}	\!\ 
\int {D}\psi_{\alpha}^{\dagger} {D}\psi_{\alpha} \, \psi_{\alpha_1}(1) \psi_{\alpha_2}(2) \psi_{\beta_2}^{\dagger}(2^{\prime}) \psi_{\beta_1}^{\dagger}(1^{\prime}) \, e^{-S_{\text{eff}}}, \nonumber \\	
&= \lim_{R\rightarrow 0} \frac{1}{R} \sum_{\alpha_1,\alpha_2,\beta_2,\beta_1} 
\langle 
\psi_{\alpha_1}(1) \psi_{\alpha_2}(2) \psi_{\beta_2}^{\dagger}(2^{\prime}) \psi_{\beta_1}^{\dagger}(1^{\prime}) \rangle_{\rm eff}. 
\label{eq:cGreenDis4}
\end{align}
Similarly, the equivalence between the disorder-averaged 
$2n$-points connected Green function of the disordered single-particle 
Hamiltonian and $2n$-points Green function of the 
replicated action in the replica limit holds 
true for the higher order.   
\section{Two-loop renormalization to the four--points vertex function}
\begin{figure}[t]
\centering
\includegraphics[width=1.0\linewidth]{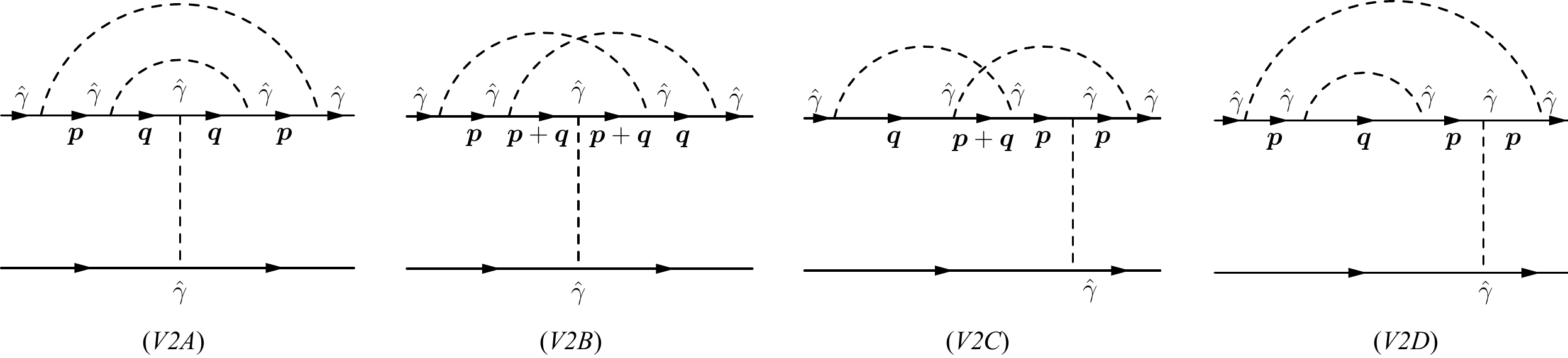}
\caption{Two-loop vertex diagrams.}
\label{fig:2loopver1}
\end{figure}

\begin{figure}[t]
\centering
\includegraphics[width=0.4\linewidth]{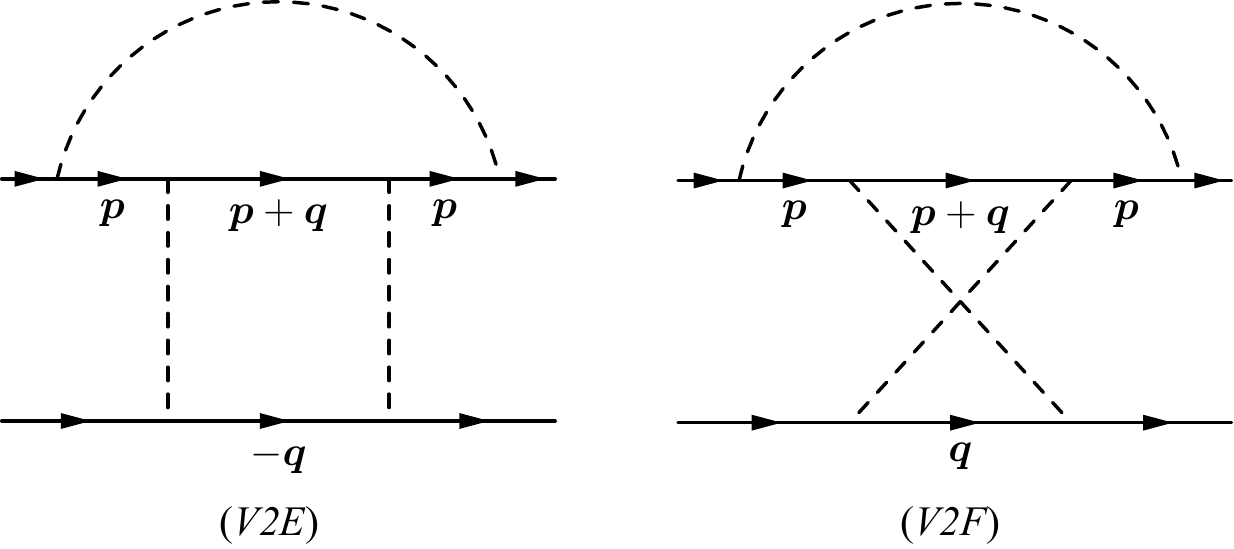}
\caption{Two-loop vertex diagrams}
\label{fig:2loopver2}
\end{figure}

\begin{figure}[t]
\centering
\includegraphics[width=1.0\linewidth]{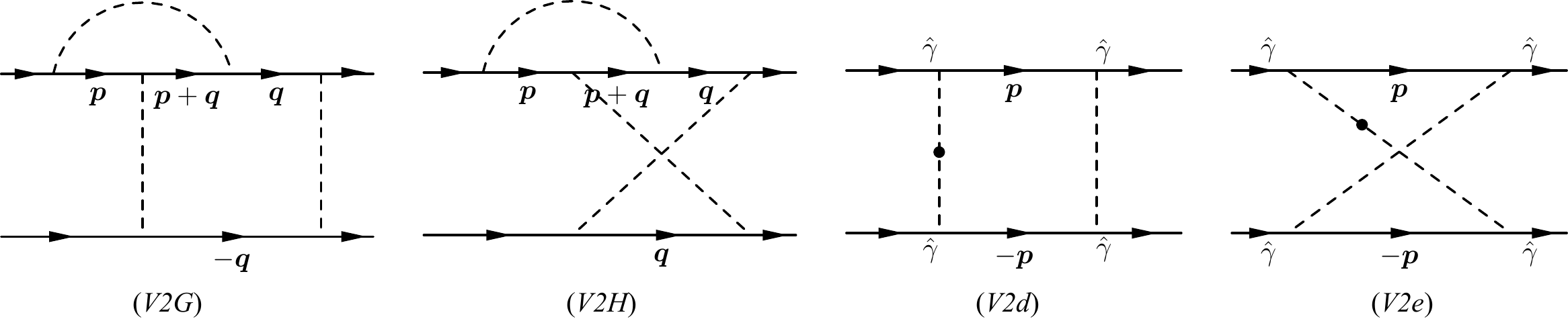}
\caption{Two-loop vertex diagrams}
\label{fig:2loopver21}
\end{figure}

\begin{figure}[t]
\centering
\includegraphics[width=1.0\linewidth]{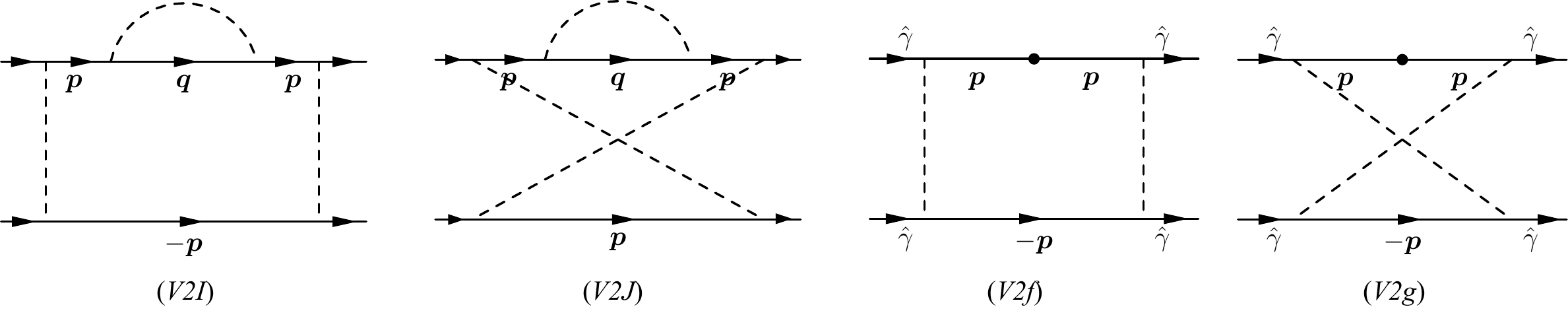}
\caption{Two-loop vertex diagrams}
\label{fig:2loopver22}
\end{figure}

\begin{figure}[t]
\centering
\includegraphics[width=1.0\linewidth]{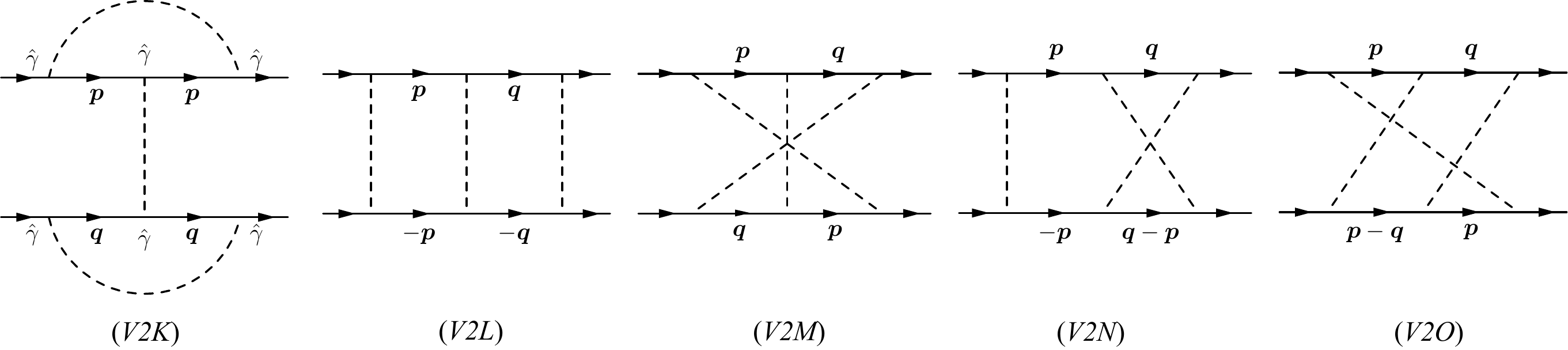}
\caption{Two-loop vertex diagrams.}
\label{fig:2loopver3}
\end{figure}

\begin{figure}[t]
\centering
\includegraphics[width=0.6\linewidth]{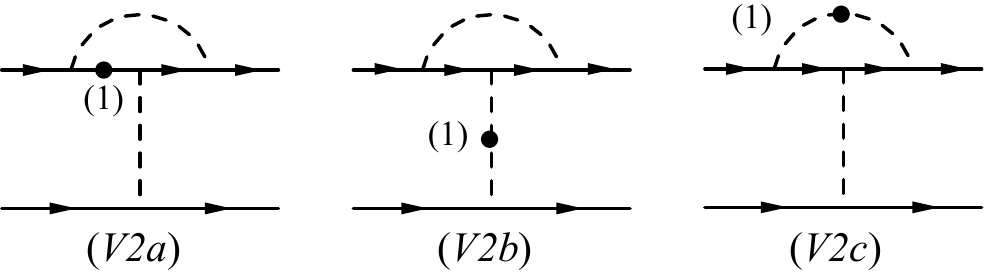}
\caption{Two-loop diagrams within one-loop counterterms}
\label{fig:2loopver4}
\end{figure}

In this Appendix, the two-loop contribution to  
the four-points vertex counterterm is 
calculated in details. The two-loop contributions to the right hand side of Eq.~(\ref{eq:vertex-sub}) 
are given by Feynman diagrams in Figs.~\ref{fig:2loopver1},  
\ref{fig:2loopver2}, \ref{fig:2loopver21}, \ref{fig:2loopver22}, \ref{fig:2loopver3}, and 
\ref{fig:2loopver4}. Note that each of  
them could have divergent terms with tensor forms 
other than $\hat{\gamma}\otimes \hat{\gamma}$.  We will 
show in this appendix that such divergent terms cancel exactly in the 
summation: the sum only gives divergent terms with the 
tensor form of $\hat{\gamma}\otimes \hat{\gamma}$. As 
in Eq.~(\ref{eq:vertex-sub}), the vertex function is evaluated 
with zero external momenta and with the two external frequencies 
set to $\Omega$. Accordingly, all the internal fermion 
lines carry the same frequency $\Omega$ and we thus 
abbreviate $\overline{G}_0({\bm p},\Omega)$ for internal fermion lines 
as $\overline{G}_0({\bm p})$ in the following. 

All the diagrams in Fig.~\ref{fig:2loopver1} give the divergent 
terms with $\hat{\gamma}\otimes \hat{\gamma}$.
%The momentum integral in these diagrams goes along on external fermion line, 
\begin{align}
[V2A]=& \kappa^3 \Omega^{3\epsilon} \int_{\bm{p},\bm{q}} \hat{\gamma} \overline{G}_0(\bm{p}) \hat{\gamma} \overline{G}_0(\bm{q}) \hat{\gamma} \overline{G}_0(\bm{q}) \hat{\gamma} \overline{G}_0(\bm{p}) \hat{\gamma}\otimes \hat{\gamma},	\\
[V2B]=& \kappa^3 \Omega^{3\epsilon} \int_{\bm{p},\bm{q}} \hat{\gamma} \overline{G}_0(\bm{p}) \hat{\gamma} \overline{G}_0(\bm{p}+\bm{q}) \hat{\gamma} \overline{G}_0(\bm{p}+\bm{q}) \hat{\gamma} \overline{G}_0(\bm{q}) \hat{\gamma}\otimes \hat{\gamma}, \\
[V2C]=& \kappa^3 \Omega^{3\epsilon} \int_{\bm{p},\bm{q}}  \hat{\gamma} \overline{G}_0(\bm{q}) \hat{\gamma} \overline{G}_0(\bm{p}+\bm{q}) \hat{\gamma} \overline{G}_0(\bm{p}) \hat{\gamma} \overline{G}_0(\bm{p}) \hat{\gamma}\otimes \hat{\gamma},		\label{eq:diaV2C}\\
[V2D]=& \kappa^3 \Omega^{3\epsilon} \int_{\bm{p},\bm{q}} \hat{\gamma} \overline{G}_0(\bm{p}) \hat{\gamma} \overline{G}_0(\bm{q}) \hat{\gamma} \overline{G}_0(\bm{p}) \hat{\gamma} \overline{G}_0(\bm{p}) \hat{\gamma}\otimes \hat{\gamma}.	\label{eq:diaV2D}
\end{align}
These integrals always take the tensor-form of 
$\hat{\gamma}\otimes \hat{\gamma}$ and their coefficients are 
calculated as follows;
\begin{align}
[V2A]=& \kappa^3 \Omega^{3\epsilon} \int_{\bm{p}} \overline{G}_0(\bm{p}) \frac{-1}{\Omega^2 +q^2} \overline{G}_0(-\bm{p}) =\kappa^3 \Omega^{3\epsilon} 
\Big( \int_{\bm{p}} \frac{-1}{\Omega^2 +p^2} \Big)^2	\nonumber\\
=&\kappa^3 \Omega^{3\epsilon} \Big( -\frac{C_{2-\epsilon}}{\epsilon} {\Omega}^{-\epsilon}  \Big)^2 +\mathcal{O}(1) =\frac{\Omega^{\epsilon}}{2C_d}g \frac{g^2}{4\epsilon^2} +\mathcal{O}(1),	\label{eq:V2A}\\
[V2B] =& \kappa^3 \Omega^{3\epsilon} \int_{\bm{p},\bm{q}} 
\frac{i\Omega- (\bm{\gamma}\cdot\bm{p})}{\Omega^2 +p^2}\cdot \frac{-1}{\Omega^2 +(\bm{p}+\bm{q})^2}\cdot \frac{i\Omega+(\bm{\gamma}\cdot\bm{q})}{ \Omega^2 +q^2}
=\kappa^3 \Omega^{3\epsilon} \int_{\bm{p},\bm{q}}  
\frac{ \Omega^2+(\bm{\gamma}\cdot\bm{q}) (\bm{\gamma}\cdot \bm{p}) }{\big(\Omega^2 +p^2\big) \big[\Omega^2 +(\bm{p}+\bm{q})^2\big] \big(\Omega^2 +q^2\big)} 	\nonumber\\
=& -\kappa^3 \Omega^{3\epsilon} \frac{1}{2} 
\Big( \frac{C_{2-\epsilon}}{\epsilon} \Omega^{-\epsilon} \Big)^2 +\mathcal{O}(1) = -\frac{\Omega^{\epsilon}}{ 2C_d} g \frac{g^2}{8\epsilon^2} +\mathcal{O}(1),	\label{eq:V2B}
\end{align}
and
\begin{align}
[V2C]= & \kappa^3 \Omega^{3\epsilon} \int_{\bm{p},\bm{q}} \frac{i\Omega+ (\bm{\gamma}\cdot\bm{q})}{\Omega^2+q^2} \cdot \frac{i\Omega- \bm{\gamma}(\bm{p}+\bm{q}) }{\Omega^2+(\bm{p}+\bm{q})^2} \cdot\frac{-1}{\Omega^2+p^2} 	\nonumber\\
=& \kappa^3 \Omega^{3\epsilon} \int_{\bm{p},\bm{q}}  \frac{1}{\big(\Omega^2 +p^2\big) \big[\Omega^2 +(\bm{p}+\bm{q})^2\big] } +\kappa^3 \Omega^{3\epsilon} \int_{\bm{p},\bm{q}}  \frac{ (\bm{\gamma}\cdot\bm{q}) (\bm{\gamma}\cdot\bm{p}) }{\big(\Omega^2 +p^2\big) \big[\Omega^2 +(\bm{p}+\bm{q})^2\big] \big(\Omega^2 +q^2\big)} 	\nonumber\\
=& \kappa^3 \Omega^{3\epsilon} \Big( \frac{C_{2-\epsilon}}{\epsilon} \Omega^{-\epsilon} +\mathcal{O}(\epsilon) \Big)^2 -\kappa^3 \Omega^{3\epsilon} \frac{1}{2} \Big( \frac{C_{2-\epsilon}}{\epsilon} \Omega^{-\epsilon} \Big)^2 +\mathcal{O}(1) 
=\frac{\Omega^{\epsilon} }{2C_d} g\frac{g^2}{8\epsilon^2} +\mathcal{O}(1),	\label{eq:V2C}\\
[V2D]=& \kappa^3 \Omega^{3\epsilon} \int_{\bm{p},\bm{q}}  \frac{i\Omega -(\bm{\gamma}\cdot\bm{p})}{\Omega^2 +p^2}\cdot \frac{i\Omega +(\bm{\gamma}\cdot\bm{q})}{\Omega^2 +q^2} \cdot\frac{-1}{\Omega^2+p^2} 
=\kappa^3 \Omega^{3\epsilon} \int_{\bm{p},\bm{q}}  \frac{\Omega^2}{(\Omega^2 +p^2)^2 (\Omega^2 +q^2)}	\nonumber\\ 
=&\kappa^3 \Omega^{3\epsilon}\Big(\frac{1}{2} C_{2-\epsilon} \Omega^{-\epsilon} +\mathcal{O}(\epsilon) \Big) \Big( \frac{C_{2-\epsilon}}{\epsilon} \Omega^{-\epsilon} +\mathcal{O}(\epsilon)\Big)  
=\frac{\Omega^{\epsilon}}{2C_d}g  \frac{g^2}{8\epsilon} +\mathcal{O}(1).	\label{eq:V2D}
\end{align}

The second class of the two-loop contributions to 
the vertex function is shown in 
Figs.~\ref{fig:2loopver2}, \ref{fig:2loopver21}, \ref{fig:2loopver22}.
These diagrams always take a form of $\gamma_0\otimes \gamma_0$; 
they do not contribute to the 
$\hat{\gamma}\otimes \hat{\gamma}$-type vertex. 
We will see that a sum of them is finite 
in the limit of $\epsilon \rightarrow 0$. 
%A sum of these diagrams will not contribute to 
%They do not contribute $\hat{\gamma}\otimes \gamma$ vertex. 
A sum of diagram $[V2E]$ and diagram 
$[V2F]$ gives a finite order in small $\epsilon$ limit,
\begin{align}
&[V2E]+[V2F]= \kappa^3\Omega^{3\epsilon} \int_{\bm{p},\bm{q}} \overline{G}_0(-\bm{p}) \overline{G}_0(\bm{p}+\bm{q}) \overline{G}_0(-\bm{p}) \otimes \big( \overline{G}_0(\bm{q}) +\overline{G}_0(-\bm{q}) \big)	\nonumber\\
=& \kappa^3\Omega^{3\epsilon} \int_{\bm{p},\bm{q}}  \frac{i\Omega- (\bm{\gamma}\cdot\bm{p})}{\Omega^2 +p^2} \cdot\frac{i\Omega+ (\bm{\gamma}\cdot\bm{p} +\bm{\gamma}\cdot\bm{q})}{\Omega^2 +(\bm{p} +\bm{q})^2} \cdot \frac{i\Omega- (\bm{\gamma}\cdot\bm{p})}{\Omega^2 +p^2} \otimes \frac{2i\Omega}{\Omega^2+q^2}	\nonumber\\
=& \kappa^3\Omega^{3\epsilon} \int_{\bm{p},\bm{q}}  \frac{ 
2\Omega^4 + 2\Omega^2 (p^2 + 2\bm{p}\bm{q})  }{(\Omega^2 +p^2)^2 \big[\Omega^2 +(\bm{p} +\bm{q})^2 \big] (\Omega^2+q^2)} \gamma_0\otimes \gamma_0 =\mathcal{O}(1)\cdot \gamma_0\otimes \gamma_0.	\label{eq:V2EF}
\end{align}
%whose coefficient of $\gamma_0\otimes \gamma_0$ is finite. 
A sum of diagrams $[V2G]$ and $[V2H]$ has a 
$1/\epsilon$ singularity in the small $\epsilon$ limit,
\begin{align}
&[V2G] +[V2H]= \kappa^3\Omega^{3\epsilon} \int_{\bm{p},\bm{q}} \overline{G}_0(-\bm{p}) \overline{G}_0(\bm{p}+\bm{q}) \overline{G}_0(-\bm{q}) \otimes \big( \overline{G}_0(\bm{q}) +\overline{G}_0(-\bm{q}) \big)	\nonumber\\
=& \kappa^3\Omega^{3\epsilon} \int_{\bm{p},\bm{q}}  \frac{i\Omega- (\bm{\gamma}\cdot\bm{p})}{\Omega^2 +p^2} \cdot\frac{i\Omega+ (\bm{\gamma}\cdot\bm{p} +\bm{\gamma}\cdot\bm{q})}{\Omega^2 +(\bm{p} +\bm{q})^2} \cdot \frac{i\Omega- (\bm{\gamma}\cdot\bm{q})}{\Omega^2 +q^2} \otimes \frac{2i\Omega}{\Omega^2+q^2}	\nonumber\\
=& \kappa^3\Omega^{3\epsilon} \int_{\bm{p},\bm{q}} \frac{
2\Omega^4 + 2\Omega^2 
\big( p^2+q^2 +(\bm{\gamma}\cdot\bm{p}) (\bm{\gamma}\cdot\bm{q})\big)   }{(\Omega^2 +p^2) (\Omega^2 +q^2)^2 \big(\Omega^2 +(\bm{p} +\bm{q})^2 \big)} \gamma_0 \otimes \gamma_0	\nonumber\\
=&\kappa^3\Omega^{3\epsilon}  \int_{\bm{q}} \frac{2\Omega^2}{ (\Omega^2 +q^2)^2} \int_{\bm{p}} \frac{1}{\Omega^2 +{p}^2} \gamma_0\otimes \gamma_0 + {\cal O}(1) 
= \kappa^3\Omega^{\epsilon} \frac{C_{2-\epsilon}}{\epsilon}  C_{2-\epsilon}  \gamma_0\otimes \gamma_0 +\mathcal{O}(1).	\label{eq:V2GH}
\end{align}
The singularity is cancelled by the two-loop contributions  
$[V2d]$ and $[V2e]$ with the one-loop counterterm 
$\delta\kappa$: 
\begin{align}
& [V2d]+[V2e]= \kappa \Omega^{\epsilon} \delta^{(1)}\kappa \cdot \Omega^{\epsilon} \int_{\bm{p}} \overline{G}_0(-\bm{p}) \otimes (\overline{G}_0(\bm{p}) +\overline{G}_0(-\bm{p}))	\nonumber\\
=& \kappa \Omega^{\epsilon} \cdot  2\Omega^{\epsilon} \kappa^2 \frac{C_{2-\epsilon}}{\epsilon} \cdot \int_{\bm{p}} \frac{i\Omega -\bm{\gamma}\cdot\bm{p}}{\Omega^2 +p^2} \otimes \frac{2i\Omega }{\Omega^2 +p^2} = -2\kappa^3 \Omega^{\epsilon} \frac{C_{2-\epsilon}}{\epsilon} C_{2-\epsilon} \gamma_0 \otimes \gamma_0.	\label{eq:V2de}
\end{align}
Namely, with proper symmetry factors taken into account, 
the sum of these four diagrams gives a finite contribution 
in the small $\epsilon$ limit, 
\begin{align}
4[V2G]+4[V2H] +2[V2d] +2[V2e]= \mathcal{O}(1) \cdot 
\gamma_0\otimes \gamma_0.	\label{eq:V2HIde}
\end{align}
Diagrams $[V2I]$ and $[V2J]$ cancel each other,
\begin{align}
& [V2I]+[V2J]= \kappa^3\Omega^{3\epsilon} \int_{\bm{p},\bm{q}} \overline{G}_0(-\bm{p}) \overline{G}_0(\bm{q}) \overline{G}_0(-\bm{p}) \otimes \big( \overline{G}_0(\bm{p}) +\overline{G}_0(-\bm{p}) \big)	\nonumber\\
=& \kappa^3\Omega^{3\epsilon} \int_{\bm{p},\bm{q}}  \frac{i\Omega- (\bm{\gamma}\cdot\bm{p})}{\Omega^2 +p^2}\cdot \frac{i\Omega+ (\bm{\gamma}\cdot\bm{q})}{\Omega^2 +q^2} \cdot \frac{i\Omega- (\bm{\gamma}\cdot\bm{p})}{\Omega^2 +p^2} \otimes \frac{2i\Omega}{\Omega^2+p^2}
= \kappa^3\Omega^{3\epsilon} \int_{\bm{p},\bm{q}} \frac{\Omega^2- p^2 }{(\Omega^2 +p^2)^2 (\Omega^2 +q^2) }  \otimes \frac{2\Omega^2}{\Omega^2+p^2}	\nonumber\\
=&  \kappa^3\Omega^{3\epsilon} 2\Omega^2 \int_{\bm{p}} \frac{\Omega^2- p^2 }{(\Omega^2 +p^2)^3 } \int_{\bm{q}} \frac{1}{ \Omega^2 +q^2 } \gamma_0 \otimes \gamma_0.	\label{eq:V2IJ}
\end{align}
Here the momentum integral over $\bm{p}$ in the right hand side 
reduces to zero,
\begin{align*}
\int_{\bm{p}} \frac{2\Omega^2}{(\Omega^2 +p^2)^3}- \int_{\bm{p}} \frac{1}{(\Omega^2 +p^2)^2}= 2\Omega^2 \frac{1}{4} C_{2-\epsilon} \Omega^{-4-\epsilon} -\frac{1}{2} C_{2-\epsilon} \Omega^{-2-\epsilon}=0.
\end{align*}
Similarly, diagrams $(V2f)$ and $(V2g)$ cancel each other exactly:
\begin{align}
& [V2f]+[V2g]= i\delta^{(1)} \Omega \!\ 
\int_{\bm{p}} \hat{\gamma} \overline{G}_0(\bm{p}) \overline{G}_0(\bm{p}) \hat{\gamma} \otimes (\overline{G}_0(\bm{p}) +\overline{G}_0(-\bm{p}))	\nonumber\\
=&  i\delta^{(1)} \Omega \!\ 
\int_{\bm{p}} \frac{-\Omega^2 +p^2}{(\Omega^2 +p^2)^2 } 
\otimes \frac{2i\Omega }{\Omega^2 +p^2} 
= -2\delta^{(1)}\Omega \!\ 
\Omega \int_{\bm{p}} 
\frac{-\Omega^2 +p^2 }{(\Omega^2 +p^2)^3 } 
\gamma_0 \otimes \gamma_0 =0.	\label{eq:V2fg} 
\end{align}
We conclude that a sum of all the diagrams in Figs.~\ref{fig:2loopver2}, \ref{fig:2loopver21}, and 
\ref{fig:2loopver22} generate a $\gamma_0\otimes \gamma_0$-type 
term, while it has no divergence in the limit of 
$\epsilon\rightarrow 0$. 

The third class of the two-loop diagrams for the vertex function 
is shown in Fig.~\ref{fig:2loopver3}. 
They give divergent terms with the 
$\hat{\gamma}\otimes \hat{\gamma}$ tensor form. 
Diagram $[V2K]$ is calculated as follow:
\begin{align}
[V2K]=& \kappa^3 \Omega^{3\epsilon} \int_{\bm{p},\bm{q}} \hat{\gamma} \overline{G}_0(\bm{p}) \hat{\gamma} \overline{G}_0(\bm{p}) \hat{\gamma} \otimes\hat{\gamma} \overline{G}_0(\bm{p}) \hat{\gamma} \overline{G}_0(\bm{p}) \hat{\gamma}
= \kappa^3 \Omega^{3\epsilon} \int_{\bm{p},\bm{q}}\frac{-1}{\Omega^2 +p^2}\cdot  \frac{-1}{\Omega^2 +q^2} \hat{\gamma} \otimes\hat{\gamma}	\nonumber\\
=& \kappa^3 \Omega^{3\epsilon} \Big( \frac{C_{2-\epsilon}}{\epsilon} \Omega^{-\epsilon} +\mathcal{O}(\epsilon) \Big)^2 
= \kappa \Omega^{\epsilon} \frac{g^2}{4\epsilon^2} +\mathcal{O}(1)  
=\frac{\Omega^{\epsilon}}{2C_d} g \frac{g^2}{4\epsilon^2} +\mathcal{O}(1).	\label{eq:V2K}
\end{align}
It is convenient to calculate a sum of $[V2L]$ and $[V2M]$;
\begin{align*}
&[V2L] +[V2M] =\kappa^3\Omega^{3\epsilon} \int_{\bm{p},\bm{q}}  \overline{G}_0(-\bm{p}) \overline{G}_0(\bm{q}) \hat{\gamma} \otimes \big( \overline{G}_0(\bm{p}) \overline{G}_0(-\bm{q}) +\overline{G}_0(-\bm{q}) \overline{G}_0(\bm{p}) \Big) \hat{\gamma}	\\
=& \kappa^3\Omega^{3\epsilon} \int_{\bm{p},\bm{q}} \frac{i\Omega- (\bm{\gamma}\cdot\bm{p})}{\Omega^2 +p^2}\cdot \frac{i\Omega+ (\bm{\gamma}\cdot\bm{q})}{\Omega^2 +q^2} \hat{\gamma} \otimes 
\big( \frac{i\Omega+ (\bm{\gamma}\cdot\bm{p})}{\Omega^2 +p^2} \cdot\frac{i\Omega- (\bm{\gamma}\cdot\bm{q})}{\Omega^2 +q^2} + \frac{i\Omega- (\bm{\gamma}\cdot\bm{q})}{\Omega^2 +q^2} \cdot\frac{i\Omega+ (\bm{\gamma}\cdot\bm{p})}{\Omega^2 +p^2} \Big) \hat{\gamma}. 
\end{align*}
In the right-hand side, odd terms in $\bm{p}$ or $\bm{q}$ vanish:
\begin{align}
[V2L] +[V2M] =& 2\kappa^3\Omega^{3\epsilon} \int_{\bm{p},\bm{q}} \frac{-\Omega^2 - (\bm{\gamma}\cdot\bm{p}) (\bm{\gamma}\cdot\bm{q}) }{(\Omega^2 +p^2) (\Omega^2 +q^2)}  \hat{\gamma} \otimes \frac{-\Omega^2 -\bm{p}\bm{q} }{(\Omega^2 +p^2) (\Omega^2 +q^2)} \hat{\gamma}  \label{eq:V2LV2M1}	\\
+&2 \kappa^3\Omega^{3\epsilon} \int_{\bm{p},\bm{q}} \frac{-i\Omega(\bm{\gamma}\cdot\bm{p} -\bm{\gamma}\cdot\bm{q} ) }{(\Omega^2 +p^2) (\Omega^2 +q^2)}  \hat{\gamma} \otimes \frac{ i\Omega(\bm{\gamma}\cdot\bm{p} -\bm{\gamma}\cdot\bm{q} ) }{(\Omega^2 +p^2) (\Omega^2 +q^2)} \hat{\gamma}.  \label{eq:V2LV2M2}
\end{align}
Under an exchange between ${\bm p}$ and ${\bm q}$ in the integrand,  
Eq.~(\ref{eq:V2LV2M1}) reduces to:
\begin{align*}
(\ref{eq:V2LV2M1})=& 2 \kappa^3\Omega^{3\epsilon} \int_{\bm{p},\bm{q}} \frac{(\Omega^2 + \bm{p}\bm{q})^2 }{(\Omega^2 +p^2)^2 (\Omega^2 +q^2)^2} \hat{\gamma} \otimes \hat{\gamma} 
= 2 \kappa^3\Omega^{3\epsilon} \sum_{i,j} \int_{\bm{p},\bm{q}} \frac{p_i q_i p_jq_j }{(\Omega^2 +p^2)^2 (\Omega^2 +q^2)^2} \hat{\gamma} \otimes \hat{\gamma} +\mathcal{O}(1). 
\end{align*}
Those summands with $i\neq j$ vanish under the integral over $\bm{p}$ 
or ${\bm q}$. Those summands with $i=j$ give out 
\begin{align*}
(\ref{eq:V2LV2M1})
=& 2\kappa^3 \Omega^{3\epsilon} 
\sum_{i} \int_{\bm{p},\bm{q}} 
\frac{ p_i^2 }{(\Omega^2 +p^2)^2 } \frac{q_i^2 }{(\Omega^2 +q^2)^2} \hat{\gamma} \otimes \hat{\gamma} 
=2\kappa^3 \Omega^{3\epsilon} 
\sum_{i} \int_{\bm{p}} \frac{ p_i^2 }{(\Omega^2 +p^2)^2 } \frac{1}{d} \int_{\bm{q}} \frac{q^2 }{(\Omega^2 +q^2)^2} \hat{\gamma} \otimes \hat{\gamma}	\\
=&\kappa^3 \Omega^{3\epsilon} \frac{2}{d} \int_{\bm{p}} \frac{ p^2 }{(\Omega^2 +p^2)^2 } \int_{\bm{q}}  \frac{q^2 }{(\Omega^2 +q^2)^2} \hat{\gamma} \otimes \hat{\gamma}.
\end{align*}
The integral in the right-hand side is calculated as follow:
\begin{align*}
\int_{\bm{p}} \frac{ p^2 }{(\Omega^2 +p^2)^2 } =\int_{\bm{p}} \frac{ 1 }{\Omega^2 +p^2 } -\int_{\bm{p}} \frac{\Omega^2 }{(\Omega^2 +p^2)^2 } =\frac{C_{2-\epsilon}}{\epsilon}\Omega^{-\epsilon} 
\Big(1 -\frac{\epsilon}{2} \Big)+\mathcal{O}(\epsilon).
\end{align*}
Thus, we finally have Eq.~(\ref{eq:V2LV2M1}) as 
\begin{align*}
(\ref{eq:V2LV2M1})=& \kappa^3 \Omega^{3\epsilon} \frac{2}{d}  
\Big( \frac{C_{2-\epsilon}}{\epsilon}\Omega^{-\epsilon} \Big(1 -\frac{\epsilon}{2} \Big)+\mathcal{O}(\epsilon) \Big)^2 \hat{\gamma} \otimes \hat{\gamma} \\
=& \bigg( 
\kappa^3 \Omega^{\epsilon} \frac{C_{2-\epsilon}^2}{\epsilon^2} -\kappa^3 \Omega^{\epsilon} \frac{C_{2-\epsilon}^2}{2\epsilon} +\mathcal{O}(1)\bigg) 
\hat{\gamma} \otimes \hat{\gamma} 
=\bigg(\frac{\Omega^{\epsilon}}{2C_d} \frac{g^3}{4\epsilon^2} -\frac{\Omega^{\epsilon}}{2C_d} \frac{g^3}{8\epsilon}  +\mathcal{O}(1) \bigg) 
\hat{\gamma} \otimes \hat{\gamma}.  
\end{align*}
For Eq.~(\ref{eq:V2LV2M2}), those 
terms proportional to $p_iq_j$ ($i=j$ or $i\neq j$) or to 
$p_ip_j$ ($i\neq j$) vanish under the integral over $\bm{p}$ or $\bm{q}$,
\begin{align*}
(\ref{eq:V2LV2M2})
=&2 \kappa^3 \Omega^{3\epsilon} \int_{\bm{p},\bm{q}} \frac{-i\Omega(\bm{\gamma}\cdot\bm{p} -\bm{\gamma}\cdot\bm{q} ) }{(\Omega^2 +p^2) (\Omega^2 +q^2)}  \hat{\gamma} \otimes \frac{ i\Omega(\bm{\gamma}\cdot\bm{p} -\bm{\gamma}\cdot\bm{q} ) }{(\Omega^2 +p^2) (\Omega^2 +q^2)} \hat{\gamma}
=2\kappa^3 \Omega^{3\epsilon} \int_{\bm{p},\bm{q}} \frac{\Omega^2 (\bm{p} \bm{\gamma} \hat{\gamma} \otimes  \bm{p} \bm{\gamma} \hat{\gamma} + \bm{q} \bm{\gamma} \hat{\gamma} \otimes \bm{q} \bm{\gamma} \hat{\gamma}) }{(\Omega^2 +p^2)^2 (\Omega^2 +q^2)^2},	\\
=&2\kappa^3 \Omega^{3\epsilon} \sum_i \int_{\bm{p},\bm{q}} \frac{\Omega^2 (p_i^2 +q_i^2) }{(\Omega^2 +p^2)^2 (\Omega^2 +q^2)^2} {\gamma}_i \hat{\gamma} \otimes {\gamma}_i \hat{\gamma}
=2\kappa^3 \Omega^{3\epsilon} \sum_i \frac{1}{d} \int_{\bm{p},\bm{q}} \frac{\Omega^2 (p^2 +q^2) }{(\Omega^2 +p^2)^2 (\Omega^2 +q^2)^2} {\gamma}_i \hat{\gamma} \otimes {\gamma}_i \hat{\gamma}	\\ 
=&\kappa^3 \Omega^{3\epsilon} \sum_i \frac{1}{d} \int_{\bm{p}} \frac{1}{\Omega^2 +p^2} \int_{\bm{q}} \frac{4\Omega^2 }{(\Omega^2 +q^2)^2} {\gamma}_i \hat{\gamma} \otimes {\gamma}_i \hat{\gamma} +\mathcal{O}(1)	\\
=& 2\kappa^3 \Omega^{\epsilon}  \frac{1}{d}  \frac{C_{2-\epsilon}}{\epsilon} C_{2-\epsilon} \sum_i {\gamma}_i \hat{\gamma} \otimes {\gamma}_i \hat{\gamma} +\mathcal{O}(1).
\end{align*}
Thus, the sum of $[V2L]$ and $[V2M]$ generates a divergent term with a tensor-form of $\gamma_i\hat{\gamma}\otimes \gamma_i \hat{\gamma}$. However, such divergent term is cancelled by a sum 
of $[V2N]$ and $[V2O]$. The sum of $[V2N]$ and $[V2O]$ is 
calculated as follows:
\begin{align*}
&[V2N] +[V2O] = \kappa^3 \Omega^{3\epsilon} \int_{\bm{p},\bm{q}}  \overline{G}_0(-\bm{p}) \overline{G}_0(\bm{q}) \hat{\gamma} \otimes \Big( \overline{G}_0(\bm{p}) \overline{G}_0(\bm{q} -\bm{p}) +\overline{G}_0(\bm{q} -\bm{p}) \overline{G}_0(\bm{p})  \Big) \hat{\gamma}	\\
=&\int_{\bm{p},\bm{q}} \frac{i\Omega- (\bm{\gamma}\cdot\bm{p})}{\Omega^2 +p^2}\cdot \frac{i\Omega+ (\bm{\gamma}\cdot\bm{q})}{\Omega^2 +q^2} \hat{\gamma} \otimes \Big( \frac{i\Omega+ (\bm{\gamma}\cdot\bm{p})}{\Omega^2 +p^2} \cdot\frac{i\Omega+ (\bm{\gamma}\cdot\bm{q} -\bm{\gamma}\cdot\bm{p})}{\Omega^2 +(\bm{q}-\bm{p})^2} + \frac{i\Omega +(\bm{\gamma}\cdot\bm{q} -\bm{\gamma}\cdot\bm{p})}{\Omega^2 +(\bm{q}-\bm{p})^2} \cdot\frac{i\Omega+ (\bm{\gamma}\cdot\bm{p})}{\Omega^2 +p^2} \Big) \hat{\gamma}.
\end{align*}
In the right-hand side, odd terms in $\bm{p}$ or $\bm{q}$ vanish 
under the momentum integrals;
\begin{align}
[V2N] +[V2O] =& 2\kappa^3 \Omega^{3\epsilon}  \int_{\bm{p},\bm{q}} \frac{-\Omega^2 - (\bm{\gamma}\cdot\bm{p}) (\bm{\gamma}\cdot\bm{q}) }{(\Omega^2 +p^2) (\Omega^2 +q^2)}  \hat{\gamma} \otimes \frac{-\Omega^2 +\bm{p}(\bm{q} -\bm{p}) }{(\Omega^2 +p^2) \big(\Omega^2 +(\bm{q}-\bm{p})^2 \big)} \hat{\gamma}	\label{eq:V2NV2O1}	\\
+& 2 \kappa^3 \Omega^{3\epsilon} \int_{\bm{p},\bm{q}} \frac{ -i\Omega(\bm{\gamma}\cdot\bm{p} -\bm{\gamma}\cdot\bm{q} ) }{(\Omega^2 +p^2) (\Omega^2 +q^2)}  \hat{\gamma} \otimes \frac{ i\Omega (\bm{\gamma}\cdot\bm{q} ) }{(\Omega^2 +p^2) \big(\Omega^2 +(\bm{q}-\bm{p})^2 \big)} \hat{\gamma}.	\label{eq:V2NV2O2}
\end{align}
Eq.~(\ref{eq:V2NV2O1}) takes the tensor form of 
$\hat{\gamma}\otimes \hat{\gamma}$, whose coefficient  
is calculated as follows,
\begin{align*}
&(\ref{eq:V2NV2O1}) = 2 \kappa^3 \Omega^{3\epsilon} \int_{\bm{p},\bm{q}} \frac{\Omega^2 + (\bm{\gamma}\cdot\bm{p}) (\bm{\gamma}\cdot\bm{q})  }{(\Omega^2 +p^2) (\Omega^2 +q^2)} \cdot \frac{\Omega^2 -\bm{p}(\bm{q} -\bm{p}) }{(\Omega^2 +p^2) \big(\Omega^2 +(\bm{q}-\bm{p})^2 \big)} \hat{\gamma}\otimes \hat{\gamma}		\\
= & \kappa^3 \Omega^{3\epsilon} \int_{\bm{p},\bm{q}} \frac{\Omega^2 - (\bm{\gamma}\cdot\bm{p}) (\bm{\gamma}\cdot\bm{q})  }{(\Omega^2 +p^2) }  \cdot\frac{2\Omega^2 +p^2 -q^2 +(\bm{q}+\bm{p})^2 }{(\Omega^2 +p^2) (\Omega^2 +q^2) \big(\Omega^2 +(\bm{q}+\bm{p})^2 \big)} \hat{\gamma}\otimes \hat{\gamma}	\\
= & \kappa^3 \Omega^{3\epsilon} \int_{\bm{p},\bm{q}} \Big( \frac{ - (\bm{\gamma}\cdot\bm{p}) (\bm{\gamma}\cdot\bm{q}) }{(\Omega^2 +p^2) (\Omega^2 +q^2) \big(\Omega^2 +(\bm{q}+\bm{p})^2 \big)} + \frac{\Omega^2 }{(\Omega^2 +p^2)^2 (\Omega^2 +q^2) } -\frac{1}{(\Omega^2 +p^2) \big(\Omega^2 +{q}^2 \big)} \Big) \hat{\gamma}\otimes \hat{\gamma} 
+ {\cal O}(1). 
\end{align*} 
After the momentum integrals, each of the three integrands
in the right hand side are diverging with respect to small $\epsilon$,
\begin{align*}
(\ref{eq:V2NV2O1})=& \kappa^3 \Omega^{3\epsilon} 
\bigg\{ \frac{1}{2} 
\Big( \frac{C_{2-\epsilon}}{\epsilon} \Omega^{-\epsilon} \Big)^2 + \Big( \frac{C_{2-\epsilon} \Omega^{-\epsilon}}{2} +\mathcal{O}(\epsilon) \Big) 
\Big( \frac{C_{2-\epsilon}}{\epsilon} \Omega^{-\epsilon} +\mathcal{O}(\epsilon) \Big) -  
\Big( \frac{C_{2-\epsilon}}{\epsilon} \Omega^{-\epsilon} +\mathcal{O}(\epsilon) \Big)^2 +{\cal O}(1)\bigg\} \hat{\gamma}\otimes \hat{\gamma} \\
=& \kappa^3 \Omega^{\epsilon}  
\bigg\{\frac{1}{2} \frac{C_{2-\epsilon}^2}{\epsilon}  -
\frac{1}{2} \frac{C_{2-\epsilon}^2}{\epsilon^2} +{\cal O}(1) 
\bigg\}\hat{\gamma}\otimes \hat{\gamma} = \frac{\Omega^{\epsilon}}{2C_d} \bigg\{  \frac{g^3}{8\epsilon} - \frac{g^3}{8\epsilon^2} +\mathcal{O}(1)\bigg\} \hat{\gamma}\otimes \hat{\gamma}.
\end{align*}
Eq.~(\ref{eq:V2NV2O2}) takes the tensor form of 
$\gamma_{i}\hat{\gamma}\otimes \gamma_{i}\hat{\gamma}$, 
whose coefficient is calculated as follows,
\begin{align*}
(\ref{eq:V2NV2O2})=& 2 \kappa^3 \Omega^{3\epsilon} \int_{\bm{p},\bm{q}} \frac{ -i\Omega(\bm{\gamma}\cdot\bm{p} -\bm{\gamma}\cdot\bm{q} ) }{(\Omega^2 +p^2) (\Omega^2 +q^2)}  \hat{\gamma} \otimes \frac{ i\Omega (\bm{\gamma}\cdot\bm{q} ) }{(\Omega^2 +p^2) \big(\Omega^2 +(\bm{q}-\bm{p})^2 \big)} \hat{\gamma}	\\
=& 2\kappa^3 \Omega^{3\epsilon} \Omega^2 \frac{1}{d} \int_{\bm{p},\bm{q}} \frac{ {\bm q}\cdot{\bm p} -{\bm q}^2  }{(\Omega^2 +p^2)^2 (\Omega^2 +q^2) \big(\Omega^2 +(\bm{q}-\bm{p})^2 \big)}  \sum_{i}{\gamma}_i \hat{\gamma} \otimes {\gamma}_i \hat{\gamma}	\\
=& 2\kappa^3 \Omega^{3\epsilon} \Omega^2 \frac{1}{d} \int_{\bm{p},\bm{q}} \frac{ -\bm{q}^2  }{(\Omega^2 +p^2)^2 (\Omega^2 +q^2) \big(\Omega^2 +(\bm{q}-\bm{p})^2 \big)}  \sum_{i} 
{\gamma}_i \hat{\gamma} \otimes {\gamma}_i \hat{\gamma} + {\cal O}(1).
\end{align*}
The first term in the right hand side diverges 
in the small $\epsilon$ limit, taking the similar 
form as (\ref{eq:V2LV2M2}) with the opposite sign:
\begin{align*}
(\ref{eq:V2NV2O2})=& 2\kappa^3 \Omega^{3\epsilon} \Omega^2  
\frac{1}{d}  \int_{\bm{p},\bm{q}} \frac{ -1  }{(\Omega^2 +p^2)^2 \big(\Omega^2 +(\bm{q}-\bm{p})^2 \big)}  \sum_i  {\gamma}_i \hat{\gamma} \otimes {\gamma}_i \hat{\gamma} +\mathcal{O}(1)	\\
=& - \kappa^3 \Omega^{3\epsilon} \frac{1}{d} \int_{\bm{p}} \frac{2\Omega^2}{(\Omega^2 +p^2)^2 } \int_{\bm{q}} \frac{1}{\Omega^2 +\bm{q}^2}  \sum_i  {\gamma}_i \hat{\gamma} \otimes {\gamma}_i \hat{\gamma} +\mathcal{O}(1)	\\
=& -\kappa^3 \Omega^{\epsilon}  \frac{1}{d}  \frac{C_{2-\epsilon}}{\epsilon} C_{2-\epsilon} \sum_i {\gamma}_i \hat{\gamma} \otimes {\gamma}_i \hat{\gamma} +\mathcal{O}(1).
\end{align*}
Accordingly, the divergent terms in Eqs.~(\ref{eq:V2LV2M2}) and (\ref{eq:V2NV2O2}) 
cancel each other in the sum of all the diagrams in Fig.~\ref{fig:2loopver3} with proper symmetry 
factors. As result, the sum of all the diagrams 
gives out only the divergent term with the 
tensor-form of $\hat{\gamma}\otimes \hat{\gamma}$;  
\begin{align}
& [V2L] +[V2M] +2 \big([V2N] +[V2O]\big) 	\nonumber\\
=&(\ref{eq:V2LV2M1})  +(\ref{eq:V2LV2M2})+2(\ref{eq:V2NV2O1}) +2(\ref{eq:V2NV2O2}) 
=\frac{\Omega^{\epsilon}}{2C_d}\frac{g^3}{8\epsilon} \hat{\gamma} \otimes \hat{\gamma}.	\label{eq:V2LMNO}
\end{align}

The last class of the two-loop contributions to 
the vertex function is shown in Fig.~\ref{fig:2loopver4}. They take the tensor-form of $\hat{\gamma}\otimes \hat{\gamma}$. Their respective coefficients are calculated as follows;
\begin{align}
& [V2a]=\kappa^2 \Omega^{2\epsilon} \!\ i\delta^{(1)}\Omega \!\ \int_{ \bm{p}} \overline{G}_0(-\bm{p}) \overline{G}_0(+\bm{p}) \overline{G}_0(+\bm{p}) \hat{\gamma}\otimes \hat{\gamma}
=\kappa^2 \Omega^{2\epsilon}  i\Omega \frac{g}{2\epsilon} \int_{\bm{p}} \frac{i\Omega}{(\Omega^2+p^2)^2}	\hat{\gamma}\otimes \hat{\gamma}\nonumber	\\
=& \kappa^2 \Omega^{2\epsilon} \big(i\Omega\big)^2 \frac{g}{2\epsilon} \Big( \frac{1}{2} C_{2-\epsilon} \Omega^{-2-\epsilon} +\mathcal{O}(\epsilon) \Big)  \hat{\gamma}\otimes \hat{\gamma}
= -\frac{\Omega^{\epsilon}}{2C_d} g \frac{g^2}{8\epsilon} \hat{\gamma}\otimes \hat{\gamma} +\mathcal{O}(1) ,	\label{eq:V2sc}	\\
&[V2b] =[V2c]= \kappa\Omega^{\epsilon} \!\ 
\Omega^{\epsilon}\delta^{(1)}\kappa \!\ \int_{ \bm{p}} \overline{G}_0(+\bm{p}) \overline{G}_0(-\bm{p}) \hat{\gamma}\otimes \hat{\gamma} 
=-\kappa \Omega^{\epsilon} \frac{\Omega^{\epsilon}}{2C_d} \frac{g^2}{\epsilon} \int_{\bm{p}} \frac{1}{\Omega^2 +p^2} \hat{\gamma}\otimes \hat{\gamma}	\nonumber \\
=&  -\kappa \Omega^{\epsilon} \frac{\Omega^{\epsilon}}{2C_d} \frac{g^2}{\epsilon} \Big( \frac{C_{2-\epsilon}}{\epsilon} \Omega^{-\epsilon} +\mathcal{O}(\epsilon) \Big) \hat{\gamma}\otimes \hat{\gamma} 
= - \frac{\Omega^{\epsilon}}{2C_d} \frac{g^3}{2\epsilon^2}  \hat{\gamma}\otimes \hat{\gamma} +\mathcal{O}(1) .	\label{eq:V2sab}
\end{align}

We have calculated all the two-loop vertex diagrams in Fig.~\ref{fig:2loopver2}-\ref{fig:2loopver4} and shown that the 
divergent terms take the tensor form of $\hat{\gamma}\otimes\hat{\gamma}$.
A sum of all these divergent terms, Eq.~(\ref{eq:V2A}) - Eq.~(\ref{eq:V2sab}), 
gives the followings;
\begin{align}
[V2]&=2[V2A] +2[V2B] +4[V2C] +4[V2D] +[V2K] + [V2L] +[V2M] +2 ([V2N] +[V2O]) +4[V2a] 	\nonumber\\
& +2[V2b] +2[V2c]	
=\frac{\Omega^{\epsilon}}{2C_d} g\Big( \frac{g^2}{8\epsilon} -\frac{g^2}{\epsilon^2} \Big) \hat{\gamma}\otimes\hat{\gamma} 
+ {\cal O}(1).	\label{eq:V2Sum}
\end{align}
The two-loop vertex counterterm should cancel the divergence 
in Eq.~(\ref{eq:V2Sum}): 
\begin{align*}
\Omega^{\epsilon} \!\ \delta^{(2)} \kappa \!\ \hat{\gamma}\otimes\hat{\gamma} +[V2]= \mathcal{O}(1)
\end{align*}
From this, we obtain the two-loop vertex counterterm as follows,
\begin{align}
\Omega^{\epsilon} \delta^{(2)} \kappa =\frac{\Omega^{\epsilon}}{2C_d} g \Big( - \frac{g^2}{8\epsilon} +\frac{g^2}{\epsilon^2} \Big).	\label{eq:2Ldeltak}
\end{align}

\twocolumngrid
\bibliography{RMDF}

\begin{thebibliography}{47}%
\makeatletter
\providecommand \@ifxundefined [1]{%
 \@ifx{#1\undefined}
}%
\providecommand \@ifnum [1]{%
 \ifnum #1\expandafter \@firstoftwo
 \else \expandafter \@secondoftwo
 \fi
}%
\providecommand \@ifx [1]{%
 \ifx #1\expandafter \@firstoftwo
 \else \expandafter \@secondoftwo
 \fi
}%
\providecommand \natexlab [1]{#1}%
\providecommand \enquote  [1]{``#1''}%
\providecommand \bibnamefont  [1]{#1}%
\providecommand \bibfnamefont [1]{#1}%
\providecommand \citenamefont [1]{#1}%
\providecommand \href@noop [0]{\@secondoftwo}%
\providecommand \href [0]{\begingroup \@sanitize@url \@href}%
\providecommand \@href[1]{\@@startlink{#1}\@@href}%
\providecommand \@@href[1]{\endgroup#1\@@endlink}%
\providecommand \@sanitize@url [0]{\catcode `\\12\catcode `\$12\catcode
  `\&12\catcode `\#12\catcode `\^12\catcode `\_12\catcode `\%12\relax}%
\providecommand \@@startlink[1]{}%
\providecommand \@@endlink[0]{}%
\providecommand \url  [0]{\begingroup\@sanitize@url \@url }%
\providecommand \@url [1]{\endgroup\@href {#1}{\urlprefix }}%
\providecommand \urlprefix  [0]{URL }%
\providecommand \Eprint [0]{\href }%
\providecommand \doibase [0]{https://doi.org/}%
\providecommand \selectlanguage [0]{\@gobble}%
\providecommand \bibinfo  [0]{\@secondoftwo}%
\providecommand \bibfield  [0]{\@secondoftwo}%
\providecommand \translation [1]{[#1]}%
\providecommand \BibitemOpen [0]{}%
\providecommand \bibitemStop [0]{}%
\providecommand \bibitemNoStop [0]{.\EOS\space}%
\providecommand \EOS [0]{\spacefactor3000\relax}%
\providecommand \BibitemShut  [1]{\csname bibitem#1\endcsname}%
\let\auto@bib@innerbib\@empty
%</preamble>
\bibitem [{\citenamefont {Read}\ and\ \citenamefont
  {Ludwig}(2000)}]{read2000absence}%
  \BibitemOpen
  \bibfield  {author} {\bibinfo {author} {\bibfnamefont {N.}~\bibnamefont
  {Read}}\ and\ \bibinfo {author} {\bibfnamefont {A.~W.~W.}\ \bibnamefont
  {Ludwig}},\ }\bibfield  {title} {\bibinfo {title} {Absence of a metallic
  phase in random-bond ising models in two dimensions: Applications to
  disordered superconductors and paired quantum hall states},\ }\href
  {https://doi.org/10.1103/PhysRevB.63.024404} {\bibfield  {journal} {\bibinfo
  {journal} {Phys. Rev. B}\ }\textbf {\bibinfo {volume} {63}},\ \bibinfo
  {pages} {024404} (\bibinfo {year} {2000})}\BibitemShut {NoStop}%
\bibitem [{\citenamefont {Kitaev}(2006)}]{kitaev2006anyons}%
  \BibitemOpen
  \bibfield  {author} {\bibinfo {author} {\bibfnamefont {A.}~\bibnamefont
  {Kitaev}},\ }\bibfield  {title} {\bibinfo {title} {Anyons in an exactly
  solved model and beyond},\ }\href@noop {} {\bibfield  {journal} {\bibinfo
  {journal} {Annals of Physics}\ }\textbf {\bibinfo {volume} {321}},\ \bibinfo
  {pages} {2} (\bibinfo {year} {2006})}\BibitemShut {NoStop}%
\bibitem [{\citenamefont {Jackeli}\ and\ \citenamefont
  {Khaliullin}(2009)}]{jackeli2009mott}%
  \BibitemOpen
  \bibfield  {author} {\bibinfo {author} {\bibfnamefont {G.}~\bibnamefont
  {Jackeli}}\ and\ \bibinfo {author} {\bibfnamefont {G.}~\bibnamefont
  {Khaliullin}},\ }\bibfield  {title} {\bibinfo {title} {Mott insulators in the
  strong spin-orbit coupling limit: From heisenberg to a quantum compass and
  kitaev models},\ }\href {https://doi.org/10.1103/PhysRevLett.102.017205}
  {\bibfield  {journal} {\bibinfo  {journal} {Phys. Rev. Lett.}\ }\textbf
  {\bibinfo {volume} {102}},\ \bibinfo {pages} {017205} (\bibinfo {year}
  {2009})}\BibitemShut {NoStop}%
\bibitem [{\citenamefont {He}\ \emph {et~al.}(2017)\citenamefont {He},
  \citenamefont {Pan}, \citenamefont {Stern}, \citenamefont {Burks},
  \citenamefont {Che}, \citenamefont {Yin}, \citenamefont {Wang}, \citenamefont
  {Lian}, \citenamefont {Zhou}, \citenamefont {Choi}, \citenamefont {Murata},
  \citenamefont {Kou}, \citenamefont {Chen}, \citenamefont {Nie}, \citenamefont
  {Shao}, \citenamefont {Fan}, \citenamefont {Zhang}, \citenamefont {Liu},
  \citenamefont {Xia},\ and\ \citenamefont {Wang}}]{he17}%
  \BibitemOpen
  \bibfield  {author} {\bibinfo {author} {\bibfnamefont {Q.~L.}\ \bibnamefont
  {He}}, \bibinfo {author} {\bibfnamefont {L.}~\bibnamefont {Pan}}, \bibinfo
  {author} {\bibfnamefont {A.~L.}\ \bibnamefont {Stern}}, \bibinfo {author}
  {\bibfnamefont {E.~C.}\ \bibnamefont {Burks}}, \bibinfo {author}
  {\bibfnamefont {X.}~\bibnamefont {Che}}, \bibinfo {author} {\bibfnamefont
  {G.}~\bibnamefont {Yin}}, \bibinfo {author} {\bibfnamefont {J.}~\bibnamefont
  {Wang}}, \bibinfo {author} {\bibfnamefont {B.}~\bibnamefont {Lian}}, \bibinfo
  {author} {\bibfnamefont {Q.}~\bibnamefont {Zhou}}, \bibinfo {author}
  {\bibfnamefont {E.~S.}\ \bibnamefont {Choi}}, \bibinfo {author}
  {\bibfnamefont {K.}~\bibnamefont {Murata}}, \bibinfo {author} {\bibfnamefont
  {X.}~\bibnamefont {Kou}}, \bibinfo {author} {\bibfnamefont {Z.}~\bibnamefont
  {Chen}}, \bibinfo {author} {\bibfnamefont {T.}~\bibnamefont {Nie}}, \bibinfo
  {author} {\bibfnamefont {Q.}~\bibnamefont {Shao}}, \bibinfo {author}
  {\bibfnamefont {Y.}~\bibnamefont {Fan}}, \bibinfo {author} {\bibfnamefont
  {S.-C.}\ \bibnamefont {Zhang}}, \bibinfo {author} {\bibfnamefont
  {K.}~\bibnamefont {Liu}}, \bibinfo {author} {\bibfnamefont {J.}~\bibnamefont
  {Xia}},\ and\ \bibinfo {author} {\bibfnamefont {K.~L.}\ \bibnamefont
  {Wang}},\ }\bibfield  {title} {\bibinfo {title} {Chiral majorana fermion
  modes in a quantum anomalous hall insulator{\textendash}superconductor
  structure},\ }\href {https://doi.org/10.1126/science.aag2792} {\bibfield
  {journal} {\bibinfo  {journal} {Science}\ }\textbf {\bibinfo {volume}
  {357}},\ \bibinfo {pages} {294} (\bibinfo {year} {2017})}\BibitemShut
  {NoStop}%
\bibitem [{\citenamefont {Kayyalha}\ \emph {et~al.}(2020)\citenamefont
  {Kayyalha}, \citenamefont {Xiao}, \citenamefont {Zhang}, \citenamefont
  {Shin}, \citenamefont {Jiang}, \citenamefont {Wang}, \citenamefont {Zhao},
  \citenamefont {Xiao}, \citenamefont {Zhang}, \citenamefont {Fijalkowski},
  \citenamefont {Mandal}, \citenamefont {Winnerlein}, \citenamefont {Gould},
  \citenamefont {Li}, \citenamefont {Molenkamp}, \citenamefont {Chan},
  \citenamefont {Samarth},\ and\ \citenamefont {Chang}}]{kayyalha20}%
  \BibitemOpen
  \bibfield  {author} {\bibinfo {author} {\bibfnamefont {M.}~\bibnamefont
  {Kayyalha}}, \bibinfo {author} {\bibfnamefont {D.}~\bibnamefont {Xiao}},
  \bibinfo {author} {\bibfnamefont {R.}~\bibnamefont {Zhang}}, \bibinfo
  {author} {\bibfnamefont {J.}~\bibnamefont {Shin}}, \bibinfo {author}
  {\bibfnamefont {J.}~\bibnamefont {Jiang}}, \bibinfo {author} {\bibfnamefont
  {F.}~\bibnamefont {Wang}}, \bibinfo {author} {\bibfnamefont {Y.-F.}\
  \bibnamefont {Zhao}}, \bibinfo {author} {\bibfnamefont {R.}~\bibnamefont
  {Xiao}}, \bibinfo {author} {\bibfnamefont {L.}~\bibnamefont {Zhang}},
  \bibinfo {author} {\bibfnamefont {K.~M.}\ \bibnamefont {Fijalkowski}},
  \bibinfo {author} {\bibfnamefont {P.}~\bibnamefont {Mandal}}, \bibinfo
  {author} {\bibfnamefont {M.}~\bibnamefont {Winnerlein}}, \bibinfo {author}
  {\bibfnamefont {C.}~\bibnamefont {Gould}}, \bibinfo {author} {\bibfnamefont
  {Q.}~\bibnamefont {Li}}, \bibinfo {author} {\bibfnamefont {L.~W.}\
  \bibnamefont {Molenkamp}}, \bibinfo {author} {\bibfnamefont {M.~H.~W.}\
  \bibnamefont {Chan}}, \bibinfo {author} {\bibfnamefont {N.}~\bibnamefont
  {Samarth}},\ and\ \bibinfo {author} {\bibfnamefont {C.-Z.}\ \bibnamefont
  {Chang}},\ }\bibfield  {title} {\bibinfo {title} {Absence of evidence for
  chiral majorana modes in quantum anomalous hall-superconductor devices},\
  }\href {https://doi.org/10.1126/science.aax6361} {\bibfield  {journal}
  {\bibinfo  {journal} {Science}\ }\textbf {\bibinfo {volume} {367}},\ \bibinfo
  {pages} {64} (\bibinfo {year} {2020})}\BibitemShut {NoStop}%
\bibitem [{\citenamefont {Wang}\ \emph {et~al.}(2018)\citenamefont {Wang},
  \citenamefont {Kong}, \citenamefont {Fan}, \citenamefont {Chen},
  \citenamefont {Zhu}, \citenamefont {Liu}, \citenamefont {Cao}, \citenamefont
  {Sun}, \citenamefont {Du}, \citenamefont {Schneeloch}, \citenamefont {Zhong},
  \citenamefont {Gu}, \citenamefont {Fu}, \citenamefont {Ding},\ and\
  \citenamefont {Gao}}]{wang18}%
  \BibitemOpen
  \bibfield  {author} {\bibinfo {author} {\bibfnamefont {D.}~\bibnamefont
  {Wang}}, \bibinfo {author} {\bibfnamefont {L.}~\bibnamefont {Kong}}, \bibinfo
  {author} {\bibfnamefont {P.}~\bibnamefont {Fan}}, \bibinfo {author}
  {\bibfnamefont {H.}~\bibnamefont {Chen}}, \bibinfo {author} {\bibfnamefont
  {S.}~\bibnamefont {Zhu}}, \bibinfo {author} {\bibfnamefont {W.}~\bibnamefont
  {Liu}}, \bibinfo {author} {\bibfnamefont {L.}~\bibnamefont {Cao}}, \bibinfo
  {author} {\bibfnamefont {Y.}~\bibnamefont {Sun}}, \bibinfo {author}
  {\bibfnamefont {S.}~\bibnamefont {Du}}, \bibinfo {author} {\bibfnamefont
  {J.}~\bibnamefont {Schneeloch}}, \bibinfo {author} {\bibfnamefont
  {R.}~\bibnamefont {Zhong}}, \bibinfo {author} {\bibfnamefont
  {G.}~\bibnamefont {Gu}}, \bibinfo {author} {\bibfnamefont {L.}~\bibnamefont
  {Fu}}, \bibinfo {author} {\bibfnamefont {H.}~\bibnamefont {Ding}},\ and\
  \bibinfo {author} {\bibfnamefont {H.-J.}\ \bibnamefont {Gao}},\ }\bibfield
  {title} {\bibinfo {title} {Evidence for majorana bound states in an
  iron-based superconductor},\ }\href {https://doi.org/10.1126/science.aao1797}
  {\bibfield  {journal} {\bibinfo  {journal} {Science}\ }\textbf {\bibinfo
  {volume} {362}},\ \bibinfo {pages} {333} (\bibinfo {year}
  {2018})}\BibitemShut {NoStop}%
\bibitem [{\citenamefont {Huang}\ \emph {et~al.}(2018)\citenamefont {Huang},
  \citenamefont {Setiawan},\ and\ \citenamefont {Sau}}]{huang18}%
  \BibitemOpen
  \bibfield  {author} {\bibinfo {author} {\bibfnamefont {Y.}~\bibnamefont
  {Huang}}, \bibinfo {author} {\bibfnamefont {F.}~\bibnamefont {Setiawan}},\
  and\ \bibinfo {author} {\bibfnamefont {J.~D.}\ \bibnamefont {Sau}},\
  }\bibfield  {title} {\bibinfo {title} {Disorder-induced half-integer
  quantized conductance plateau in quantum anomalous hall
  insulator-superconductor structures},\ }\href
  {https://doi.org/10.1103/PhysRevB.97.100501} {\bibfield  {journal} {\bibinfo
  {journal} {Phys. Rev. B}\ }\textbf {\bibinfo {volume} {97}},\ \bibinfo
  {pages} {100501} (\bibinfo {year} {2018})}\BibitemShut {NoStop}%
\bibitem [{\citenamefont {Lian}\ \emph {et~al.}(2018)\citenamefont {Lian},
  \citenamefont {Wang}, \citenamefont {Sun}, \citenamefont {Vaezi},\ and\
  \citenamefont {Zhang}}]{lian18}%
  \BibitemOpen
  \bibfield  {author} {\bibinfo {author} {\bibfnamefont {B.}~\bibnamefont
  {Lian}}, \bibinfo {author} {\bibfnamefont {J.}~\bibnamefont {Wang}}, \bibinfo
  {author} {\bibfnamefont {X.-Q.}\ \bibnamefont {Sun}}, \bibinfo {author}
  {\bibfnamefont {A.}~\bibnamefont {Vaezi}},\ and\ \bibinfo {author}
  {\bibfnamefont {S.-C.}\ \bibnamefont {Zhang}},\ }\bibfield  {title} {\bibinfo
  {title} {Quantum phase transition of chiral majorana fermions in the presence
  of disorder},\ }\href {https://doi.org/10.1103/PhysRevB.97.125408} {\bibfield
   {journal} {\bibinfo  {journal} {Phys. Rev. B}\ }\textbf {\bibinfo {volume}
  {97}},\ \bibinfo {pages} {125408} (\bibinfo {year} {2018})}\BibitemShut
  {NoStop}%
\bibitem [{\citenamefont {Yamada}(2020)}]{yamada20anderson}%
  \BibitemOpen
  \bibfield  {author} {\bibinfo {author} {\bibfnamefont {M.~G.}\ \bibnamefont
  {Yamada}},\ }\bibfield  {title} {\bibinfo {title} {Anderson--kitaev spin
  liquid},\ }\href@noop {} {\bibfield  {journal} {\bibinfo  {journal} {npj
  Quantum Materials}\ }\textbf {\bibinfo {volume} {5}},\ \bibinfo {pages} {1}
  (\bibinfo {year} {2020})}\BibitemShut {NoStop}%
\bibitem [{\citenamefont {Altland}\ and\ \citenamefont
  {Zirnbauer}(1997)}]{altland1997}%
  \BibitemOpen
  \bibfield  {author} {\bibinfo {author} {\bibfnamefont {A.}~\bibnamefont
  {Altland}}\ and\ \bibinfo {author} {\bibfnamefont {M.~R.}\ \bibnamefont
  {Zirnbauer}},\ }\bibfield  {title} {\bibinfo {title} {Nonstandard symmetry
  classes in mesoscopic normal-superconducting hybrid structures},\ }\href
  {https://doi.org/10.1103/PhysRevB.55.1142} {\bibfield  {journal} {\bibinfo
  {journal} {Phys. Rev. B}\ }\textbf {\bibinfo {volume} {55}},\ \bibinfo
  {pages} {1142} (\bibinfo {year} {1997})}\BibitemShut {NoStop}%
\bibitem [{\citenamefont {Cho}\ and\ \citenamefont
  {Fisher}(1997)}]{cho1997criticality}%
  \BibitemOpen
  \bibfield  {author} {\bibinfo {author} {\bibfnamefont {S.}~\bibnamefont
  {Cho}}\ and\ \bibinfo {author} {\bibfnamefont {M.~P.~A.}\ \bibnamefont
  {Fisher}},\ }\bibfield  {title} {\bibinfo {title} {Criticality in the
  two-dimensional random-bond ising model},\ }\href
  {https://doi.org/10.1103/PhysRevB.55.1025} {\bibfield  {journal} {\bibinfo
  {journal} {Phys. Rev. B}\ }\textbf {\bibinfo {volume} {55}},\ \bibinfo
  {pages} {1025} (\bibinfo {year} {1997})}\BibitemShut {NoStop}%
\bibitem [{\citenamefont {Senthil}\ and\ \citenamefont
  {Fisher}(2000)}]{senthil2000quasiparticle}%
  \BibitemOpen
  \bibfield  {author} {\bibinfo {author} {\bibfnamefont {T.}~\bibnamefont
  {Senthil}}\ and\ \bibinfo {author} {\bibfnamefont {M.~P.~A.}\ \bibnamefont
  {Fisher}},\ }\bibfield  {title} {\bibinfo {title} {Quasiparticle localization
  in superconductors with spin-orbit scattering},\ }\href
  {https://doi.org/10.1103/PhysRevB.61.9690} {\bibfield  {journal} {\bibinfo
  {journal} {Phys. Rev. B}\ }\textbf {\bibinfo {volume} {61}},\ \bibinfo
  {pages} {9690} (\bibinfo {year} {2000})}\BibitemShut {NoStop}%
\bibitem [{\citenamefont {Bocquet}\ \emph {et~al.}(2000)\citenamefont
  {Bocquet}, \citenamefont {Serban},\ and\ \citenamefont
  {Zirnbauer}}]{bocquet00}%
  \BibitemOpen
  \bibfield  {author} {\bibinfo {author} {\bibfnamefont {M.}~\bibnamefont
  {Bocquet}}, \bibinfo {author} {\bibfnamefont {D.}~\bibnamefont {Serban}},\
  and\ \bibinfo {author} {\bibfnamefont {M.}~\bibnamefont {Zirnbauer}},\
  }\bibfield  {title} {\bibinfo {title} {Disordered 2d quasiparticles in class
  d: Dirac fermions with random mass, and dirty superconductors},\ }\href
  {https://doi.org/https://doi.org/10.1016/S0550-3213(00)00208-X} {\bibfield
  {journal} {\bibinfo  {journal} {Nuclear Physics B}\ }\textbf {\bibinfo
  {volume} {578}},\ \bibinfo {pages} {628} (\bibinfo {year}
  {2000})}\BibitemShut {NoStop}%
\bibitem [{\citenamefont {Chalker}\ \emph {et~al.}(2001)\citenamefont
  {Chalker}, \citenamefont {Read}, \citenamefont {Kagalovsky}, \citenamefont
  {Horovitz}, \citenamefont {Avishai},\ and\ \citenamefont
  {Ludwig}}]{chalker2001thermal}%
  \BibitemOpen
  \bibfield  {author} {\bibinfo {author} {\bibfnamefont {J.~T.}\ \bibnamefont
  {Chalker}}, \bibinfo {author} {\bibfnamefont {N.}~\bibnamefont {Read}},
  \bibinfo {author} {\bibfnamefont {V.}~\bibnamefont {Kagalovsky}}, \bibinfo
  {author} {\bibfnamefont {B.}~\bibnamefont {Horovitz}}, \bibinfo {author}
  {\bibfnamefont {Y.}~\bibnamefont {Avishai}},\ and\ \bibinfo {author}
  {\bibfnamefont {A.~W.~W.}\ \bibnamefont {Ludwig}},\ }\bibfield  {title}
  {\bibinfo {title} {Thermal metal in network models of a disordered
  two-dimensional superconductor},\ }\href
  {https://doi.org/10.1103/PhysRevB.65.012506} {\bibfield  {journal} {\bibinfo
  {journal} {Phys. Rev. B}\ }\textbf {\bibinfo {volume} {65}},\ \bibinfo
  {pages} {012506} (\bibinfo {year} {2001})}\BibitemShut {NoStop}%
\bibitem [{\citenamefont {Mildenberger}\ \emph {et~al.}(2006)\citenamefont
  {Mildenberger}, \citenamefont {Evers}, \citenamefont {Narayanan},
  \citenamefont {Mirlin},\ and\ \citenamefont {Damle}}]{mildenberger2006}%
  \BibitemOpen
  \bibfield  {author} {\bibinfo {author} {\bibfnamefont {A.}~\bibnamefont
  {Mildenberger}}, \bibinfo {author} {\bibfnamefont {F.}~\bibnamefont {Evers}},
  \bibinfo {author} {\bibfnamefont {R.}~\bibnamefont {Narayanan}}, \bibinfo
  {author} {\bibfnamefont {A.~D.}\ \bibnamefont {Mirlin}},\ and\ \bibinfo
  {author} {\bibfnamefont {K.}~\bibnamefont {Damle}},\ }\bibfield  {title}
  {\bibinfo {title} {Griffiths phase in the thermal quantum hall effect},\
  }\href {https://doi.org/10.1103/PhysRevB.73.121301} {\bibfield  {journal}
  {\bibinfo  {journal} {Phys. Rev. B}\ }\textbf {\bibinfo {volume} {73}},\
  \bibinfo {pages} {121301} (\bibinfo {year} {2006})}\BibitemShut {NoStop}%
\bibitem [{\citenamefont {Mildenberger}\ \emph {et~al.}(2007)\citenamefont
  {Mildenberger}, \citenamefont {Evers}, \citenamefont {Mirlin},\ and\
  \citenamefont {Chalker}}]{mildenberger2007density}%
  \BibitemOpen
  \bibfield  {author} {\bibinfo {author} {\bibfnamefont {A.}~\bibnamefont
  {Mildenberger}}, \bibinfo {author} {\bibfnamefont {F.}~\bibnamefont {Evers}},
  \bibinfo {author} {\bibfnamefont {A.~D.}\ \bibnamefont {Mirlin}},\ and\
  \bibinfo {author} {\bibfnamefont {J.~T.}\ \bibnamefont {Chalker}},\
  }\bibfield  {title} {\bibinfo {title} {Density of quasiparticle states for a
  two-dimensional disordered system: Metallic, insulating, and critical
  behavior in the class-d thermal quantum hall effect},\ }\href
  {https://doi.org/10.1103/PhysRevB.75.245321} {\bibfield  {journal} {\bibinfo
  {journal} {Phys. Rev. B}\ }\textbf {\bibinfo {volume} {75}},\ \bibinfo
  {pages} {245321} (\bibinfo {year} {2007})}\BibitemShut {NoStop}%
\bibitem [{\citenamefont {Evers}\ and\ \citenamefont
  {Mirlin}(2008)}]{evers2008}%
  \BibitemOpen
  \bibfield  {author} {\bibinfo {author} {\bibfnamefont {F.}~\bibnamefont
  {Evers}}\ and\ \bibinfo {author} {\bibfnamefont {A.~D.}\ \bibnamefont
  {Mirlin}},\ }\bibfield  {title} {\bibinfo {title} {Anderson transitions},\
  }\href {https://doi.org/10.1103/RevModPhys.80.1355} {\bibfield  {journal}
  {\bibinfo  {journal} {Rev. Mod. Phys.}\ }\textbf {\bibinfo {volume} {80}},\
  \bibinfo {pages} {1355} (\bibinfo {year} {2008})}\BibitemShut {NoStop}%
\bibitem [{\citenamefont {Kagalovsky}\ and\ \citenamefont
  {Nemirovsky}(2008)}]{kagalovsky2008universal}%
  \BibitemOpen
  \bibfield  {author} {\bibinfo {author} {\bibfnamefont {V.}~\bibnamefont
  {Kagalovsky}}\ and\ \bibinfo {author} {\bibfnamefont {D.}~\bibnamefont
  {Nemirovsky}},\ }\bibfield  {title} {\bibinfo {title} {Universal critical
  exponent in class d superconductors},\ }\href
  {https://doi.org/10.1103/PhysRevLett.101.127001} {\bibfield  {journal}
  {\bibinfo  {journal} {Phys. Rev. Lett.}\ }\textbf {\bibinfo {volume} {101}},\
  \bibinfo {pages} {127001} (\bibinfo {year} {2008})}\BibitemShut {NoStop}%
\bibitem [{\citenamefont {Kagalovsky}\ and\ \citenamefont
  {Nemirovsky}(2010)}]{kagalovsky2010critical}%
  \BibitemOpen
  \bibfield  {author} {\bibinfo {author} {\bibfnamefont {V.}~\bibnamefont
  {Kagalovsky}}\ and\ \bibinfo {author} {\bibfnamefont {D.}~\bibnamefont
  {Nemirovsky}},\ }\bibfield  {title} {\bibinfo {title} {Critical fixed points
  in class d superconductors},\ }\href
  {https://doi.org/10.1103/PhysRevB.81.033406} {\bibfield  {journal} {\bibinfo
  {journal} {Phys. Rev. B}\ }\textbf {\bibinfo {volume} {81}},\ \bibinfo
  {pages} {033406} (\bibinfo {year} {2010})}\BibitemShut {NoStop}%
\bibitem [{\citenamefont {Wimmer}\ \emph {et~al.}(2010)\citenamefont {Wimmer},
  \citenamefont {Akhmerov}, \citenamefont {Medvedyeva}, \citenamefont
  {Tworzyd\l{}o},\ and\ \citenamefont {Beenakker}}]{wimmer2010}%
  \BibitemOpen
  \bibfield  {author} {\bibinfo {author} {\bibfnamefont {M.}~\bibnamefont
  {Wimmer}}, \bibinfo {author} {\bibfnamefont {A.~R.}\ \bibnamefont
  {Akhmerov}}, \bibinfo {author} {\bibfnamefont {M.~V.}\ \bibnamefont
  {Medvedyeva}}, \bibinfo {author} {\bibfnamefont {J.}~\bibnamefont
  {Tworzyd\l{}o}},\ and\ \bibinfo {author} {\bibfnamefont {C.~W.~J.}\
  \bibnamefont {Beenakker}},\ }\bibfield  {title} {\bibinfo {title} {Majorana
  bound states without vortices in topological superconductors with
  electrostatic defects},\ }\href
  {https://doi.org/10.1103/PhysRevLett.105.046803} {\bibfield  {journal}
  {\bibinfo  {journal} {Phys. Rev. Lett.}\ }\textbf {\bibinfo {volume} {105}},\
  \bibinfo {pages} {046803} (\bibinfo {year} {2010})}\BibitemShut {NoStop}%
\bibitem [{\citenamefont {Medvedyeva}\ \emph {et~al.}(2010)\citenamefont
  {Medvedyeva}, \citenamefont {Tworzyd\l{}o},\ and\ \citenamefont
  {Beenakker}}]{medvedyeva2010}%
  \BibitemOpen
  \bibfield  {author} {\bibinfo {author} {\bibfnamefont {M.~V.}\ \bibnamefont
  {Medvedyeva}}, \bibinfo {author} {\bibfnamefont {J.}~\bibnamefont
  {Tworzyd\l{}o}},\ and\ \bibinfo {author} {\bibfnamefont {C.~W.~J.}\
  \bibnamefont {Beenakker}},\ }\bibfield  {title} {\bibinfo {title} {Effective
  mass and tricritical point for lattice fermions localized by a random mass},\
  }\href {https://doi.org/10.1103/PhysRevB.81.214203} {\bibfield  {journal}
  {\bibinfo  {journal} {Phys. Rev. B}\ }\textbf {\bibinfo {volume} {81}},\
  \bibinfo {pages} {214203} (\bibinfo {year} {2010})}\BibitemShut {NoStop}%
\bibitem [{\citenamefont {Laumann}\ \emph {et~al.}(2012)\citenamefont
  {Laumann}, \citenamefont {Ludwig}, \citenamefont {Huse},\ and\ \citenamefont
  {Trebst}}]{laumann2012}%
  \BibitemOpen
  \bibfield  {author} {\bibinfo {author} {\bibfnamefont {C.~R.}\ \bibnamefont
  {Laumann}}, \bibinfo {author} {\bibfnamefont {A.~W.~W.}\ \bibnamefont
  {Ludwig}}, \bibinfo {author} {\bibfnamefont {D.~A.}\ \bibnamefont {Huse}},\
  and\ \bibinfo {author} {\bibfnamefont {S.}~\bibnamefont {Trebst}},\
  }\bibfield  {title} {\bibinfo {title} {Disorder-induced majorana metal in
  interacting non-abelian anyon systems},\ }\href
  {https://doi.org/10.1103/PhysRevB.85.161301} {\bibfield  {journal} {\bibinfo
  {journal} {Phys. Rev. B}\ }\textbf {\bibinfo {volume} {85}},\ \bibinfo
  {pages} {161301} (\bibinfo {year} {2012})}\BibitemShut {NoStop}%
\bibitem [{\citenamefont {Yoshioka}\ \emph {et~al.}(2018)\citenamefont
  {Yoshioka}, \citenamefont {Akagi},\ and\ \citenamefont
  {Katsura}}]{yoshioka2018}%
  \BibitemOpen
  \bibfield  {author} {\bibinfo {author} {\bibfnamefont {N.}~\bibnamefont
  {Yoshioka}}, \bibinfo {author} {\bibfnamefont {Y.}~\bibnamefont {Akagi}},\
  and\ \bibinfo {author} {\bibfnamefont {H.}~\bibnamefont {Katsura}},\
  }\bibfield  {title} {\bibinfo {title} {Learning disordered topological phases
  by statistical recovery of symmetry},\ }\href
  {https://doi.org/10.1103/PhysRevB.97.205110} {\bibfield  {journal} {\bibinfo
  {journal} {Phys. Rev. B}\ }\textbf {\bibinfo {volume} {97}},\ \bibinfo
  {pages} {205110} (\bibinfo {year} {2018})}\BibitemShut {NoStop}%
\bibitem [{\citenamefont {Fulga}\ \emph {et~al.}(2020)\citenamefont {Fulga},
  \citenamefont {Oreg}, \citenamefont {Mirlin}, \citenamefont {Stern},\ and\
  \citenamefont {Mross}}]{fulga2020}%
  \BibitemOpen
  \bibfield  {author} {\bibinfo {author} {\bibfnamefont {I.~C.}\ \bibnamefont
  {Fulga}}, \bibinfo {author} {\bibfnamefont {Y.}~\bibnamefont {Oreg}},
  \bibinfo {author} {\bibfnamefont {A.~D.}\ \bibnamefont {Mirlin}}, \bibinfo
  {author} {\bibfnamefont {A.}~\bibnamefont {Stern}},\ and\ \bibinfo {author}
  {\bibfnamefont {D.~F.}\ \bibnamefont {Mross}},\ }\bibfield  {title} {\bibinfo
  {title} {Temperature enhancement of thermal hall conductance quantization},\
  }\href {https://doi.org/10.1103/PhysRevLett.125.236802} {\bibfield  {journal}
  {\bibinfo  {journal} {Phys. Rev. Lett.}\ }\textbf {\bibinfo {volume} {125}},\
  \bibinfo {pages} {236802} (\bibinfo {year} {2020})}\BibitemShut {NoStop}%
\bibitem [{\citenamefont {Gruzberg}\ \emph {et~al.}(2001)\citenamefont
  {Gruzberg}, \citenamefont {Read},\ and\ \citenamefont
  {Ludwig}}]{gruzberg2001random}%
  \BibitemOpen
  \bibfield  {author} {\bibinfo {author} {\bibfnamefont {I.~A.}\ \bibnamefont
  {Gruzberg}}, \bibinfo {author} {\bibfnamefont {N.}~\bibnamefont {Read}},\
  and\ \bibinfo {author} {\bibfnamefont {A.~W.~W.}\ \bibnamefont {Ludwig}},\
  }\bibfield  {title} {\bibinfo {title} {Random-bond ising model in two
  dimensions: The nishimori line and supersymmetry},\ }\href
  {https://doi.org/10.1103/PhysRevB.63.104422} {\bibfield  {journal} {\bibinfo
  {journal} {Phys. Rev. B}\ }\textbf {\bibinfo {volume} {63}},\ \bibinfo
  {pages} {104422} (\bibinfo {year} {2001})}\BibitemShut {NoStop}%
\bibitem [{\citenamefont {Chalker}\ and\ \citenamefont
  {Coddington}(1988)}]{chalker1988percolation}%
  \BibitemOpen
  \bibfield  {author} {\bibinfo {author} {\bibfnamefont {J.~T.}\ \bibnamefont
  {Chalker}}\ and\ \bibinfo {author} {\bibfnamefont {P.~D.}\ \bibnamefont
  {Coddington}},\ }\bibfield  {title} {\bibinfo {title} {Percolation, quantum
  tunnelling and the integer hall effect},\ }\href
  {https://doi.org/10.1088/0022-3719/21/14/008} {\bibfield  {journal} {\bibinfo
   {journal} {Journal of Physics C: Solid State Physics}\ }\textbf {\bibinfo
  {volume} {21}},\ \bibinfo {pages} {2665} (\bibinfo {year}
  {1988})}\BibitemShut {NoStop}%
\bibitem [{\citenamefont {Ludwig}\ \emph {et~al.}(1994)\citenamefont {Ludwig},
  \citenamefont {Fisher}, \citenamefont {Shankar},\ and\ \citenamefont
  {Grinstein}}]{ludwig1994integer}%
  \BibitemOpen
  \bibfield  {author} {\bibinfo {author} {\bibfnamefont {A.~W.~W.}\
  \bibnamefont {Ludwig}}, \bibinfo {author} {\bibfnamefont {M.~P.~A.}\
  \bibnamefont {Fisher}}, \bibinfo {author} {\bibfnamefont {R.}~\bibnamefont
  {Shankar}},\ and\ \bibinfo {author} {\bibfnamefont {G.}~\bibnamefont
  {Grinstein}},\ }\bibfield  {title} {\bibinfo {title} {Integer quantum hall
  transition: An alternative approach and exact results},\ }\href
  {https://doi.org/10.1103/PhysRevB.50.7526} {\bibfield  {journal} {\bibinfo
  {journal} {Phys. Rev. B}\ }\textbf {\bibinfo {volume} {50}},\ \bibinfo
  {pages} {7526} (\bibinfo {year} {1994})}\BibitemShut {NoStop}%
\bibitem [{\citenamefont {Gracey}(1991)}]{gracey1991}%
  \BibitemOpen
  \bibfield  {author} {\bibinfo {author} {\bibfnamefont {J.}~\bibnamefont
  {Gracey}},\ }\bibfield  {title} {\bibinfo {title} {Computation of the
  three-loop $\beta$-function of the $o(n)$ gross-neveu model in minimal
  subtraction},\ }\href
  {https://doi.org/https://doi.org/10.1016/0550-3213(91)90012-M} {\bibfield
  {journal} {\bibinfo  {journal} {Nuclear Physics B}\ }\textbf {\bibinfo
  {volume} {367}},\ \bibinfo {pages} {657} (\bibinfo {year}
  {1991})}\BibitemShut {NoStop}%
\bibitem [{\citenamefont {Gracey}\ \emph {et~al.}(2016)\citenamefont {Gracey},
  \citenamefont {Luthe},\ and\ \citenamefont {Schr\"oder}}]{gracey2016four}%
  \BibitemOpen
  \bibfield  {author} {\bibinfo {author} {\bibfnamefont {J.~A.}\ \bibnamefont
  {Gracey}}, \bibinfo {author} {\bibfnamefont {T.}~\bibnamefont {Luthe}},\ and\
  \bibinfo {author} {\bibfnamefont {Y.}~\bibnamefont {Schr\"oder}},\ }\bibfield
   {title} {\bibinfo {title} {Four loop renormalization of the gross-neveu
  model},\ }\href {https://doi.org/10.1103/PhysRevD.94.125028} {\bibfield
  {journal} {\bibinfo  {journal} {Phys. Rev. D}\ }\textbf {\bibinfo {volume}
  {94}},\ \bibinfo {pages} {125028} (\bibinfo {year} {2016})}\BibitemShut
  {NoStop}%
\bibitem [{\citenamefont {Choi}\ \emph {et~al.}(2017)\citenamefont {Choi},
  \citenamefont {Ryttov},\ and\ \citenamefont {Shrock}}]{choi17question}%
  \BibitemOpen
  \bibfield  {author} {\bibinfo {author} {\bibfnamefont {G.}~\bibnamefont
  {Choi}}, \bibinfo {author} {\bibfnamefont {T.~A.}\ \bibnamefont {Ryttov}},\
  and\ \bibinfo {author} {\bibfnamefont {R.}~\bibnamefont {Shrock}},\
  }\bibfield  {title} {\bibinfo {title} {Question of a possible infrared zero
  in the beta function of the finite-$n$ gross-neveu model},\ }\href
  {https://doi.org/10.1103/PhysRevD.95.025012} {\bibfield  {journal} {\bibinfo
  {journal} {Phys. Rev. D}\ }\textbf {\bibinfo {volume} {95}},\ \bibinfo
  {pages} {025012} (\bibinfo {year} {2017})}\BibitemShut {NoStop}%
\bibitem [{\citenamefont {Potter}\ and\ \citenamefont {Lee}(2010)}]{potter10}%
  \BibitemOpen
  \bibfield  {author} {\bibinfo {author} {\bibfnamefont {A.~C.}\ \bibnamefont
  {Potter}}\ and\ \bibinfo {author} {\bibfnamefont {P.~A.}\ \bibnamefont
  {Lee}},\ }\bibfield  {title} {\bibinfo {title} {Multichannel generalization
  of kitaev's majorana end states and a practical route to realize them in thin
  films},\ }\href {https://doi.org/10.1103/PhysRevLett.105.227003} {\bibfield
  {journal} {\bibinfo  {journal} {Phys. Rev. Lett.}\ }\textbf {\bibinfo
  {volume} {105}},\ \bibinfo {pages} {227003} (\bibinfo {year}
  {2010})}\BibitemShut {NoStop}%
\bibitem [{\citenamefont {Bernevig}(2013)}]{bernevig13}%
  \BibitemOpen
  \bibfield  {author} {\bibinfo {author} {\bibfnamefont {B.~A.}\ \bibnamefont
  {Bernevig}},\ }\href@noop {} {\emph {\bibinfo {title} {Topological insulators
  and topological superconductors}}}\ (\bibinfo  {publisher} {Princeton
  university press},\ \bibinfo {year} {2013})\BibitemShut {NoStop}%
\bibitem [{\citenamefont {Altland}\ and\ \citenamefont
  {Simons}(2010)}]{altland2010condensed}%
  \BibitemOpen
  \bibfield  {author} {\bibinfo {author} {\bibfnamefont {A.}~\bibnamefont
  {Altland}}\ and\ \bibinfo {author} {\bibfnamefont {B.~D.}\ \bibnamefont
  {Simons}},\ }\href@noop {} {\emph {\bibinfo {title} {Condensed matter field
  theory}}}\ (\bibinfo  {publisher} {Cambridge university press},\ \bibinfo
  {year} {2010})\BibitemShut {NoStop}%
\bibitem [{\citenamefont {Aharony}\ and\ \citenamefont
  {Narovlansky}(2018)}]{aharony2018renormalization}%
  \BibitemOpen
  \bibfield  {author} {\bibinfo {author} {\bibfnamefont {O.}~\bibnamefont
  {Aharony}}\ and\ \bibinfo {author} {\bibfnamefont {V.}~\bibnamefont
  {Narovlansky}},\ }\bibfield  {title} {\bibinfo {title} {Renormalization group
  flow in field theories with quenched disorder},\ }\href
  {https://doi.org/10.1103/PhysRevD.98.045012} {\bibfield  {journal} {\bibinfo
  {journal} {Phys. Rev. D}\ }\textbf {\bibinfo {volume} {98}},\ \bibinfo
  {pages} {045012} (\bibinfo {year} {2018})}\BibitemShut {NoStop}%
\bibitem [{\citenamefont {Peskin}\ and\ \citenamefont
  {Schroeder}(1995)}]{peskin-schroeder}%
  \BibitemOpen
  \bibfield  {author} {\bibinfo {author} {\bibfnamefont {M.~E.}\ \bibnamefont
  {Peskin}}\ and\ \bibinfo {author} {\bibfnamefont {D.~V.}\ \bibnamefont
  {Schroeder}},\ }\href@noop {} {\emph {\bibinfo {title} {Introduction to
  Quantum Field Theory}}}\ (\bibinfo  {publisher} {Westview},\ \bibinfo {year}
  {1995})\BibitemShut {NoStop}%
\bibitem [{\citenamefont {Amit}\ and\ \citenamefont
  {Martin-Mayor}(2005)}]{amit}%
  \BibitemOpen
  \bibfield  {author} {\bibinfo {author} {\bibfnamefont {D.~J.}\ \bibnamefont
  {Amit}}\ and\ \bibinfo {author} {\bibfnamefont {V.}~\bibnamefont
  {Martin-Mayor}},\ }\href@noop {} {\emph {\bibinfo {title} {Field Theory,
  Renormalization Group and Critical Phenomena}}}\ (\bibinfo  {publisher}
  {World Scientific},\ \bibinfo {year} {2005})\BibitemShut {NoStop}%
\bibitem [{\citenamefont {Bondi}\ \emph {et~al.}(1990)\citenamefont {Bondi},
  \citenamefont {Curci}, \citenamefont {Paffuti},\ and\ \citenamefont
  {Rossi}}]{bondi1990a}%
  \BibitemOpen
  \bibfield  {author} {\bibinfo {author} {\bibfnamefont {A.}~\bibnamefont
  {Bondi}}, \bibinfo {author} {\bibfnamefont {G.}~\bibnamefont {Curci}},
  \bibinfo {author} {\bibfnamefont {G.}~\bibnamefont {Paffuti}},\ and\ \bibinfo
  {author} {\bibfnamefont {P.}~\bibnamefont {Rossi}},\ }\bibfield  {title}
  {\bibinfo {title} {Metric and central charge in the perturbative approach to
  two dimensional fermionic models},\ }\href
  {https://doi.org/https://doi.org/10.1016/0003-4916(90)90380-7} {\bibfield
  {journal} {\bibinfo  {journal} {Annals of Physics}\ }\textbf {\bibinfo
  {volume} {199}},\ \bibinfo {pages} {268} (\bibinfo {year}
  {1990})}\BibitemShut {NoStop}%
\bibitem [{\citenamefont {Schuessler}\ \emph {et~al.}(2009)\citenamefont
  {Schuessler}, \citenamefont {Ostrovsky}, \citenamefont {Gornyi},\ and\
  \citenamefont {Mirlin}}]{schuessler09analytic}%
  \BibitemOpen
  \bibfield  {author} {\bibinfo {author} {\bibfnamefont {A.}~\bibnamefont
  {Schuessler}}, \bibinfo {author} {\bibfnamefont {P.~M.}\ \bibnamefont
  {Ostrovsky}}, \bibinfo {author} {\bibfnamefont {I.~V.}\ \bibnamefont
  {Gornyi}},\ and\ \bibinfo {author} {\bibfnamefont {A.~D.}\ \bibnamefont
  {Mirlin}},\ }\bibfield  {title} {\bibinfo {title} {Analytic theory of
  ballistic transport in disordered graphene},\ }\href
  {https://doi.org/10.1103/PhysRevB.79.075405} {\bibfield  {journal} {\bibinfo
  {journal} {Phys. Rev. B}\ }\textbf {\bibinfo {volume} {79}},\ \bibinfo
  {pages} {075405} (\bibinfo {year} {2009})}\BibitemShut {NoStop}%
\bibitem [{\citenamefont {Roy}\ and\ \citenamefont
  {Das~Sarma}(2014)}]{roy2014diffusive}%
  \BibitemOpen
  \bibfield  {author} {\bibinfo {author} {\bibfnamefont {B.}~\bibnamefont
  {Roy}}\ and\ \bibinfo {author} {\bibfnamefont {S.}~\bibnamefont
  {Das~Sarma}},\ }\bibfield  {title} {\bibinfo {title} {Diffusive quantum
  criticality in three-dimensional disordered dirac semimetals},\ }\href
  {https://doi.org/10.1103/PhysRevB.90.241112} {\bibfield  {journal} {\bibinfo
  {journal} {Phys. Rev. B}\ }\textbf {\bibinfo {volume} {90}},\ \bibinfo
  {pages} {241112} (\bibinfo {year} {2014})}\BibitemShut {NoStop}%
\bibitem [{\citenamefont {Syzranov}\ \emph {et~al.}(2016)\citenamefont
  {Syzranov}, \citenamefont {Ostrovsky}, \citenamefont {Gurarie},\ and\
  \citenamefont {Radzihovsky}}]{syzranov2016critical}%
  \BibitemOpen
  \bibfield  {author} {\bibinfo {author} {\bibfnamefont {S.~V.}\ \bibnamefont
  {Syzranov}}, \bibinfo {author} {\bibfnamefont {P.~M.}\ \bibnamefont
  {Ostrovsky}}, \bibinfo {author} {\bibfnamefont {V.}~\bibnamefont {Gurarie}},\
  and\ \bibinfo {author} {\bibfnamefont {L.}~\bibnamefont {Radzihovsky}},\
  }\bibfield  {title} {\bibinfo {title} {Critical exponents at the
  unconventional disorder-driven transition in a weyl semimetal},\ }\href
  {https://doi.org/10.1103/PhysRevB.93.155113} {\bibfield  {journal} {\bibinfo
  {journal} {Phys. Rev. B}\ }\textbf {\bibinfo {volume} {93}},\ \bibinfo
  {pages} {155113} (\bibinfo {year} {2016})}\BibitemShut {NoStop}%
\bibitem [{\citenamefont {Syzranov}\ \emph {et~al.}(2015)\citenamefont
  {Syzranov}, \citenamefont {Gurarie},\ and\ \citenamefont
  {Radzihovsky}}]{syzranov2015unconventional}%
  \BibitemOpen
  \bibfield  {author} {\bibinfo {author} {\bibfnamefont {S.~V.}\ \bibnamefont
  {Syzranov}}, \bibinfo {author} {\bibfnamefont {V.}~\bibnamefont {Gurarie}},\
  and\ \bibinfo {author} {\bibfnamefont {L.}~\bibnamefont {Radzihovsky}},\
  }\bibfield  {title} {\bibinfo {title} {Unconventional localization transition
  in high dimensions},\ }\href {https://doi.org/10.1103/PhysRevB.91.035133}
  {\bibfield  {journal} {\bibinfo  {journal} {Phys. Rev. B}\ }\textbf {\bibinfo
  {volume} {91}},\ \bibinfo {pages} {035133} (\bibinfo {year}
  {2015})}\BibitemShut {NoStop}%
\bibitem [{\citenamefont {Schwartz}(2014)}]{schwartz2014quantum}%
  \BibitemOpen
  \bibfield  {author} {\bibinfo {author} {\bibfnamefont {M.~D.}\ \bibnamefont
  {Schwartz}},\ }\href@noop {} {\emph {\bibinfo {title} {Quantum field theory
  and the standard model}}}\ (\bibinfo  {publisher} {Cambridge University
  Press},\ \bibinfo {year} {2014})\BibitemShut {NoStop}%
\bibitem [{\citenamefont {Chayes}\ \emph {et~al.}(1986)\citenamefont {Chayes},
  \citenamefont {Chayes}, \citenamefont {Fisher},\ and\ \citenamefont
  {Spencer}}]{chayes1986finite}%
  \BibitemOpen
  \bibfield  {author} {\bibinfo {author} {\bibfnamefont {J.~T.}\ \bibnamefont
  {Chayes}}, \bibinfo {author} {\bibfnamefont {L.}~\bibnamefont {Chayes}},
  \bibinfo {author} {\bibfnamefont {D.~S.}\ \bibnamefont {Fisher}},\ and\
  \bibinfo {author} {\bibfnamefont {T.}~\bibnamefont {Spencer}},\ }\bibfield
  {title} {\bibinfo {title} {Finite-size scaling and correlation lengths for
  disordered systems},\ }\href {https://doi.org/10.1103/PhysRevLett.57.2999}
  {\bibfield  {journal} {\bibinfo  {journal} {Phys. Rev. Lett.}\ }\textbf
  {\bibinfo {volume} {57}},\ \bibinfo {pages} {2999} (\bibinfo {year}
  {1986})}\BibitemShut {NoStop}%
\bibitem [{\citenamefont {Dudka}\ \emph {et~al.}(2016)\citenamefont {Dudka},
  \citenamefont {Fedorenko}, \citenamefont {Blavatska},\ and\ \citenamefont
  {Holovatch}}]{dudka2016}%
  \BibitemOpen
  \bibfield  {author} {\bibinfo {author} {\bibfnamefont {M.}~\bibnamefont
  {Dudka}}, \bibinfo {author} {\bibfnamefont {A.~A.}\ \bibnamefont
  {Fedorenko}}, \bibinfo {author} {\bibfnamefont {V.}~\bibnamefont
  {Blavatska}},\ and\ \bibinfo {author} {\bibfnamefont {Y.}~\bibnamefont
  {Holovatch}},\ }\bibfield  {title} {\bibinfo {title} {Critical behavior of
  the two-dimensional ising model with long-range correlated disorder},\ }\href
  {https://doi.org/10.1103/PhysRevB.93.224422} {\bibfield  {journal} {\bibinfo
  {journal} {Phys. Rev. B}\ }\textbf {\bibinfo {volume} {93}},\ \bibinfo
  {pages} {224422} (\bibinfo {year} {2016})}\BibitemShut {NoStop}%
\bibitem [{\citenamefont {Roy}\ and\ \citenamefont
  {Das~Sarma}(2016)}]{roy2016erratum}%
  \BibitemOpen
  \bibfield  {author} {\bibinfo {author} {\bibfnamefont {B.}~\bibnamefont
  {Roy}}\ and\ \bibinfo {author} {\bibfnamefont {S.}~\bibnamefont
  {Das~Sarma}},\ }\bibfield  {title} {\bibinfo {title} {Erratum: Diffusive
  quantum criticality in three-dimensional disordered dirac semimetals [phys.
  rev. b 90, 241112(r) (2014)]},\ }\href
  {https://doi.org/10.1103/PhysRevB.93.119911} {\bibfield  {journal} {\bibinfo
  {journal} {Phys. Rev. B}\ }\textbf {\bibinfo {volume} {93}},\ \bibinfo
  {pages} {119911} (\bibinfo {year} {2016})}\BibitemShut {NoStop}%
\bibitem [{\citenamefont {Gross}\ and\ \citenamefont {Neveu}(1974)}]{gross74}%
  \BibitemOpen
  \bibfield  {author} {\bibinfo {author} {\bibfnamefont {D.~J.}\ \bibnamefont
  {Gross}}\ and\ \bibinfo {author} {\bibfnamefont {A.}~\bibnamefont {Neveu}},\
  }\bibfield  {title} {\bibinfo {title} {Dynamical symmetry breaking in
  asymptotically free field theories},\ }\href
  {https://doi.org/10.1103/PhysRevD.10.3235} {\bibfield  {journal} {\bibinfo
  {journal} {Phys. Rev. D}\ }\textbf {\bibinfo {volume} {10}},\ \bibinfo
  {pages} {3235} (\bibinfo {year} {1974})}\BibitemShut {NoStop}%
\bibitem [{\citenamefont {Louvet}\ \emph {et~al.}(2016)\citenamefont {Louvet},
  \citenamefont {Carpentier},\ and\ \citenamefont
  {Fedorenko}}]{louvet2016disorder}%
  \BibitemOpen
  \bibfield  {author} {\bibinfo {author} {\bibfnamefont {T.}~\bibnamefont
  {Louvet}}, \bibinfo {author} {\bibfnamefont {D.}~\bibnamefont {Carpentier}},\
  and\ \bibinfo {author} {\bibfnamefont {A.~A.}\ \bibnamefont {Fedorenko}},\
  }\bibfield  {title} {\bibinfo {title} {On the disorder-driven quantum
  transition in three-dimensional relativistic metals},\ }\href
  {https://doi.org/10.1103/PhysRevB.94.220201} {\bibfield  {journal} {\bibinfo
  {journal} {Phys. Rev. B}\ }\textbf {\bibinfo {volume} {94}},\ \bibinfo
  {pages} {220201} (\bibinfo {year} {2016})}\BibitemShut {NoStop}%
\end{thebibliography}%

\end{document}